\documentclass[fleqn,usenatbib]{mnras}

\usepackage{newtxtext,newtxmath}

\usepackage[T1]{fontenc}
\usepackage{ae,aecompl}
\let\cite=\citep

\usepackage{graphicx}	
\usepackage{amsmath}	
\usepackage{amssymb}	

\usepackage{caption}
\usepackage{subcaption}
\usepackage{chngcntr}

\title[Impact of baryons and neutrinos on lensing peaks]{The impact of baryonic physics and massive neutrinos on weak lensing peak statistics}

\author[M. Fong et al.]{Matthew Fong,$^{1}$\thanks{E-mail: matthew.fong@utdallas.edu (MF)}
Miyoung Choi,$^{1}$\thanks{E-mail: miyoung.choi@utdallas.edu (MC)}
Victoria Catlett,$^{1}$
Brandyn Lee,$^{1}$
Austin Peel,$^{2}$
\newauthor
Rachel Bowyer,$^{3}$
Lindsay J. King,$^{1}$\thanks{E-mail: lindsay.king@utdallas.edu (LJK)}
Ian G. McCarthy$^{4}$
\\
$^{1}$Physics Department, The University of Texas at Dallas, 800 W. Campbell Road, Richardson, TX 75080, USA\\
$^{2}$AIM, CEA, CNRS, Universit{\'e} Paris-Saclay, Universit{\'e} Paris Diderot, Sorbonne Paris Cit{\'e}, F-91191 Gif-sur-Yvette, France\\
$^{3}$Department of Astrophysical and Planetary Sciences, 391 UCB, University of Colorado, Boulder, CO 80309-0391, USA\\
$^{4}$Astrophysics Research Institute, Liverpool John Moores University, 146 Brownlow Hill, Liverpool L3 5RF, UK
}

\date{Accepted XXX. Received YYY; in original form ZZZ}

\pubyear{2019}

\begin{document}
\label{firstpage}
\pagerange{\pageref{firstpage}--\pageref{lastpage}}
\maketitle

\begin{abstract}
We study the impact of baryonic processes and massive neutrinos on weak lensing peak statistics that can be used to constrain cosmological parameters. We use the BAHAMAS suite of cosmological simulations, which self-consistently include baryonic processes and the effect of massive neutrino free-streaming on the evolution of structure formation. We construct synthetic weak lensing catalogues by ray-tracing through light-cones, and use the aperture mass statistic for the analysis. The peaks detected on the maps reflect the cumulative signal from massive bound objects and general large-scale structure. We present the first study of weak lensing peaks in simulations that include both baryonic physics and massive neutrinos (summed neutrino mass $M_{\nu} =$ 0.06, 0.12, 0.24, and 0.48 eV assuming normal hierarchy), so that the uncertainty due to physics beyond the gravity of dark matter can be factored into constraints on cosmological models. Assuming a fiducial model of baryonic physics, we also investigate the correlation between peaks and massive haloes, over a range of summed neutrino mass values. As higher neutrino mass tends to suppress the formation of massive structures in the Universe, the halo mass function and lensing peak counts are therefore modified as a function of $M_{\nu}$. Over most of the S/N range, the impact of fiducial baryonic physics is greater (less) than neutrinos for 0.06 and 0.12 (0.24 and 0.48) eV models. Both baryonic physics and massive neutrinos should be accounted for when deriving cosmological parameters from weak lensing observations.

\end{abstract}




\section{Introduction}

Galaxy clusters and large-scale structure (LSS) provide a powerful laboratory to study the Universe
\citep[e.g.][]{Bond1980, Blumenthal1984, Voit2005, Allen2011, Kravtsov2012}.
Measurements of LSS help constrain cosmological parameters, independent from observations of the cosmic microwave background (CMB) and other probes. 

Agreement between various astrophysical probes has provided strong evidence for a concordance cosmology, the $\Lambda$CDM model ($\Lambda$ Cold Dark Matter).
However, recent high-precision measurements have suggested a tension in some of the parameter estimates. For example, some authors have determined that local measurements of Hubble's constant $H_{0}$ ($73.48 \pm 1.66$ km/s/Mpc, \citealt{Riess2018}) disagree with the value derived from the joint analysis of CMB and Baryon Acoustic Oscillations (BAO) ($67.4\pm0.5$ km/s/Mpc, \citealt{Planck2018}).
However some analyses have suggested that the discrepancy between the measurements is of low significance. For example, \citet{Feeney2018} developed a Bayesian hierarchical model of the distance ladder that finds a local $H_0$ value nearly identical to the \textit{Planck} CMB measurement.
Another tension that has been found by some studies is between measurements of $\Omega_{\rm m}$ and of $\sigma_{8}$, the present-day matter density of the Universe and the density perturbation amplitude on $8h^{-1}$Mpc scale (see for example the discussion in \citealt{Hildebrandt2017,McCarthy2018, Planck2018}), where $h$ is $H_0$ scaled by $100$ km/s/Mpc. 

Traditionally, simulations of cosmological structure formation have considered only collisionless gravitational dynamics.
However, with the increase in computational capabilities, some large-volume simulations have now been carried out to model LSS for various cosmologies and using various prescriptions for baryonic physics \citep[e.g.][]{Dolag2009, Schaye2010, Vogelsberger2013, LeBrun2014, Schaye2015, Dubois2014, McCarthy2017, Pillepich2018}.
These and other simulations have shown that baryonic physics affects the total matter power spectrum \citep[e.g.][]{Semboloni2011, vanDaalen2011, Schneider2015}, 1-point lensing statistics \citep{Castro2018}, the halo mass function \citep[e.g.][]{Sawala2013, Velliscig2014, Cusworth2014} and galaxy cluster density profiles and mass estimation \citep[e.g.][]{Schaller2015, Mummery2017, Henson2017, Lee2018}. 
For example, on galaxy and cluster scales baryonic feedback produces an outward pressure that acts against the infall of matter, resulting in a shallower inner density profile, corresponding to a lower concentration of mass \citep{Mummery2017}. 
For the ongoing $1400$ deg$^2$ Subaru Hyper Suprime-Cam survey, \citet{Osato2015ApJ...806..186O} estimate that when the peak statistics, or other lensing statistics are used alone, the parameter bias due to baryonic physics is
not significant with most of the biased parameters lying within the 
$1\sigma$ error. However this would not be the case for a survey covering a larger fraction of the sky.
These works have illustrated that the addition of baryons can have a significant impact on our inference of cosmological parameters.

Neutrinos are the most ubiquitous subatomic particle in the Universe.
Massive neutrinos act like a form of hot dark matter. The high speeds of neutrinos at early times tend to resist gravitational collapse and to impede the growth of structure, while cold dark matter facilitates structure formation.
Theoretical and observational studies of the impact of neutrinos on cosmological structure formation have been carried out by e.g. \citet{Hu1998, Bashinsky2004, Hannestad2006, Gratton2008, Namikawa2010, Lahav2010, Bird2012, Wagner2012, Costanzi2013, Villaescusa-Navarro2014, Castorina2014, Roncarelli2015, Mummery2017, Moscardini2018, Hagstotz2018}.
However, their mass is not yet known, and the relevance of massive neutrinos to structure formation and to astrophysical observables is an open question. 
\citet{Lesgourgues2006} found that the three active neutrino species have a summed mass of at least $0.06$ eV for normal or inverted hierarchies, by studying atmospheric and solar oscillation experiments.  
In their fiducial analysis, \citet{PlanckXIII} adopt a value of $M_\nu = 0.06$ eV.  The CMB data itself can be used to constrain the summed neutrino mass and, when combined with external BAO constraints, \citet{PlanckXIII} set an upper limit of $M_\nu < 0.21$ eV.  However, the derived upper limit is sensitive to the treatment of internal tensions in the primary CMB data \citep[e.g.][]{Addison2016, Ashdown2017} and, when this is factored in, values of up to 0.4 eV are potentially compatible with the data \citep[e.g.][]{Valentino2017, McCarthy2018, Poulin2018}.

Some studies find that the aforementioned tension in cosmological parameter measurements can potentially be remedied with the inclusion of massive neutrinos \citep[e.g.][]{Wyman2014, Battye2014, McCarthy2018}. \citet{Mummery2017} used cosmo-OWLS (the OverWhelmingly Large Simulations; \citealt{LeBrun2014}) and BAHAMAS (BAryons and HAloes of MAssive Systems; \citealt{McCarthy2018}) to study how baryonic physics and neutrinos impact the halo mass function, mass density profiles of haloes, the halo mass-concentration relation, and the clustering properties of haloes.

The impact of baryonic physics and massive neutrinos is thought to be significant and has been considered as a systematic in various works on weak lensing (WL) statistics with dark matter only simulations and observational surveys \citep[e.g.][]{Yang2013, Hildebrandt2017, Martinet2018}. 
In this work we are motivated by the studies above that suggest the importance of physics beyond the gravity associated with cold dark matter. 

The statistics of the peaks on weak gravitational lensing maps has been shown to be a powerful probe of cosmology and massive galaxy clusters \citep[e.g.][]{Kruse1999, Kruse2000, Jain2000, Dietrich2010, Maturi2010, Kratochvil2010, Fan2010, Yang2011, Hamana2012, Martinet2015, Lin2015, Kacprzak2016, Liu2016a, Liu2016b, Peel2017, Peel2018, Martinet2018, Li2019, Shan2018, Davies2019}. 
Weak lensing peaks arise from massive structures such as galaxy clusters but also from the LSS of the Universe. Thus, the peak statistics contain information about the Universe on both non-linear and linear scales \citep[e.g.][]{Yang2011, Liu2016a, Martinet2018}. 

In this paper, we estimate the impact that baryons and massive neutrinos have on the counts of weak gravitational lensing peaks. 
The cosmological hydrodynamical simulations that we use in this work is BAHAMAS, a suite that includes the effects of massive neutrinos and for which the efficiencies of stellar and Active Galactic Nuclei (AGN) feedback have been carefully calibrated to match the observed baryon fractions of massive systems \citep{McCarthy2017}. 
For each run we use light-cones from BAHAMAS and generate sets of synthetic weak lensing surveys with different source redshift distributions and source number densities for galaxies from which the weak lensing signal is measured. 
The synthetic surveys used in this work are based on the Kilo Degree Survey (KiDS, \citealt{Hildebrandt2017, deJong2017}), and on Deep Ground Based and Deep Spaced Based survey characteristics, such as expected for the Large Synoptic Survey Telescope (LSST, \citealt{Chang2013}) and \textit{Euclid} \citep[e.g.][]{Laureijs2011, Amendola2018} and for the \textit{Hubble Space Telescope} (\textit{HST}, e.g. the weak lensing observations in \citealt{King2016}) respectively. We consider surveys of the same size area as each other ($625$ deg$^{2}$).
We determine peaks on maps of the aperture mass statistic \citep{Schneider1996}, as applied and adapted in e.g. \citet{Schneider1998, Schirmer2007, Maturi2010}.

The impact of baryonic physics on weak lensing peak statistics has been studied \citep[e.g.][]{Osato2015ApJ...806..186O, Castro2018} as well as the impact of summed neutrino mass \citep[e.g.][]{Peel2018, Liu2019, Li2019}, but never their combination as explored in this paper.
We explore the impact of a range of different baryonic physics prescriptions on the statistics of the signal-to-noise (S/N) peaks, compared with the statistics of the peaks in dark matter only simulations with collisionless dynamics. For a fiducial baryonic physics model, we explore the impact of massive neutrinos; the lower bound of the summed neutrino mass in BAHAMAS is taken from \citet{Lesgourgues2006}, with additional runs increasing the summed neutrino mass by factors of 2 up to $0.48$ eV. 
Using a fiducial baryonic prescription outlined in Section \ref{sec:BAHAMAS}, we also examine the correlation between high S/N peaks and massive galaxy clusters and assess our ability to detect clusters using aperture mass peaks. 
As far as we know, our work is the first systematic study of weak lensing peaks using simulations which incorporate massive neutrinos in concert with baryonic physics. A consideration of a broad range of cosmological models has been explored for S/N peak distributions \citep[e.g.][]{Yang2011, Liu2016a, Liu2016b, Martinet2018, Peel2018} but these simulations and other works do not include both the impact of summed neutrino mass and baryonic physics. In this paper we focus on baryonic physics and massive neutrinos and restrict our consideration to the two sets of cosmologies of BAHAMAS based on the WMAP 9-yr mission \citep[\textit{WMAP}~9,][]{Hinshaw2013} and the Planck 2015 mission \citep[\textit{Planck}~2015,][]{PlanckXVI} parameters.

The remainder of the paper is organised as follows:
in Section \ref{sec:gravitationalLensing} we introduce weak gravitational lensing. 
In Section \ref{sec:BAHAMAS} we outline the BAHAMAS simulation and how we generate synthetic lensing catalogues. 
In Section \ref{sec:apertureMass} we introduce how we quantify weak lensing signatures using the aperture mass statistic and S/N peaks.
In Section \ref{s:results} we present the results on the impact of baryonic physics and summed neutrino mass on the S/N peak statistics. We also consider how summed neutrino mass affects the cluster mass function, mass of individual galaxy clusters, and the correlation between peak detection and cluster mass.
In Section \ref{s:discussions} we present a summary of our findings and conclude.

\section{Gravitational Lensing}
\label{sec:gravitationalLensing}

The light bundles from distant background sources are distorted as they pass through the gravitational fields of massive objects and the general LSS. This distortion can be determined in General Relativity, where light bundles follow null geodesics of the space-time metric that is distorted by mass. 
Cosmological simulations of structure formation for particular cosmological models provide a description of the distribution of mass in the Universe. In this section we introduce concepts of gravitational lensing relevant to the analysis of cosmological simulations.

The distorted appearance of the weakly lensed images of distant sources can be described by lensing convergence and shear. The weak lensing calculations here assume the so-called ``Born approximation", where the paths of light rays are approximated as straight lines in comoving coordinates. This has been shown to be accurate for weak lensing \citep[e.g.][]{Bartelmann2001, White2004}. In a given cosmological model, the weak lensing convergence at a particular angular position, $\kappa(\theta)$, depends on the spatial distribution of mass density fluctuations, $\delta$, in the Universe, and the redshift distribution of sources that are being lensed by the fluctuations, $p_{s}(z)$:
\begin{equation}
\kappa(\theta) = \frac{3\Omega_\mathrm{m} H_0^2}{2c^2} \int_0^{\chi(z_\mathrm{max})} (1+z) s(\chi) \delta(\chi, \theta) d\chi,
\label{eq:simulationConvergence}
\end{equation} 
where $c$ is the speed of light, 
$\chi$ is the cosmological comoving distance, and $s(\chi)$ is a lensing kernel defined as:
\begin{equation}
s(\chi) = \chi (z) \int_z^{z_\mathrm{max}} p_s(z') \left( \frac{\chi(z') - \chi(z))}{\chi(z')} \right) dz'.
\label{eq:lensingKernel}
\end{equation}
The lensing kernel depends on the source redshift probability distribution, $p_{s}(z)$, where the maximum source redshift is $z_\mathrm{max}$ and the distribution is normalised to $1$. The light-cones we use from BAHAMAS extend out to $z_\mathrm{max}=3$. The lensing convergence results in an isotropic distortion of a lensed source.

Given a map of the weak lensing convergence, the complex shear, $\gamma=\gamma_1 + i \gamma_2$, can be obtained using Fourier transform techniques since both the convergence and shear can be written as linear combinations of second derivatives of the lensing potential. Following e.g. \citet{Clowe2004}:
\begin{equation}
\tilde{\gamma}  =  \left(\frac{\hat{k}_1^2-\hat{k}_2^2}{\hat{k}_1^2+\hat{k}_2^2}\tilde{\kappa}, \frac{2\hat{k}_1\hat{k}_2}{\hat{k}_1^2+\hat{k}_2^2}\tilde{\kappa}\right),
\label{eq:fftmethod}
\end{equation}
where $\tilde{\gamma}$ and $\tilde{\kappa}$ are the Fourier transforms of the complex shear and scalar convergence, and $\hat{k}$ are wave vectors in Fourier space. The inverse Fourier transform of
$\tilde{\gamma}$ yields $\gamma$, and the complex reduced shear $g$ is given by $\gamma/(1-\kappa)$ which can also be written as:
\begin{equation}
    g = 
    g_1 + i g_2 = 
    |g|e^{2i\phi},
    \label{eq:g}
\end{equation}
where $\phi$ is the phase angle of the distortion and $|g|$ is related to the strength of the shear. The shear or reduced shear results in an anisotropic distortion of the lensed source.

Maps of the convergence and reduced shear are determined from the mass distributions in cosmological simulations as detailed in the next section.\footnote{The publicly available BAHAMAS convergence and shear maps can be found at \url{http://www.astro.ljmu.ac.uk/~igm/BAHAMAS/}}
These maps can then be used to determine the shapes of lensed galaxies.
The shapes and orientation angles of the unlensed and lensed sources can be expressed as complex ellipticities $\epsilon^{s}$ and $\epsilon$ respectively. 
The phase of a complex ellipticity is twice the orientation angle of the galaxy on the sky and the modulus of the complex ellipticity is related to the axis ratio through $\frac{1-b/a}{1+b/a}$, where $a$ and $b$ are the major and minor axes. 

The intrinsic shape of a galaxy can be expressed as:
\begin{equation}
    \epsilon^{s} = |\epsilon^{s}|e^{2i\phi^{s}} \equiv \epsilon_1^s + i \epsilon_2^s,
\end{equation}
where $\epsilon_1^s$ and $\epsilon_2^s$ are complex components of the source ellipticity $|\epsilon^{s}| \cos (2 \phi^{s})$ and $|\epsilon^{s}| \sin (2 \phi^{s})$ respectively and $\phi^{s}$ is the orientation of the source galaxy.

If a background source galaxy is circular, or $\epsilon^{s} = 0$, then the lensed galaxy ellipticity will be exactly $\epsilon = g$. In general, for non-critical lenses, the lensed and unlensed ellipticities are related by \citep[e.g.]{Bartelmann1996, Schneider1998, Schneider2005}: 
\begin{equation}
    \epsilon = \frac{\epsilon^{s} + g}{1 + g^{*}\epsilon^{s}} \equiv \epsilon_1 + i\epsilon_2,
    \label{eq:ellipticity}
\end{equation}
where * denotes taking the complex conjugate.

\section{Synthetic Lensing Catalogues: BAHAMAS Simulations}
\label{sec:BAHAMAS}
We use light-cones extracted from BAHAMAS \citep{McCarthy2017}.
Motivated by the fact that neutrinos are massive and by the tension between cosmological parameter estimates from the LSS and the primary CMB, BAHAMAS allows for a range of non-zero neutrino masses \citep{McCarthy2018}. 
BAHAMAS is the only suite of cosmological hydrodynamical simulations that have been explicitly calibrated on the baryon fractions of collapsed systems. This guarantees that the response of the redistribution of total matter due to baryonic physics is broadly correct \citep[see Table 1 of][]{McCarthy2017}.
In this section we will discuss the different runs with varying cosmology and neutrino mass that we use. We also discuss the creation of synthetic light-cones and weak lensing galaxy catalogues used in our study.

\subsection{Cosmology, Neutrino Mass, and Baryonic Physics}
\label{sec:Cosmologies}

The initial conditions for BAHAMAS are based on the cosmological parameters derived from the cosmic microwave background missions \textit{WMAP}~9 and \textit{Planck}~2015, using the six-parameter standard $\Lambda$CDM model.
\citet{McCarthy2018} generated a suite of cosmological simulations that vary the baryonic physics model, the summed mass of neutrinos, and the background cosmology.

To explore the role of baryonic physics, we use the collisionless dynamics (DMONLY) run, the fiducial calibrated baryonic physics model (i.e., the model has been calibrated to reproduce the baryon fractions of groups and clusters), as well as two variations of the baryonic physics model where the efficiency of feedback from AGN was raised (AGN High) and lowered (AGN low) to approximately bracket the observed baryon fractions of groups and clusters.

For the summed mass of neutrinos, \citet{McCarthy2018} ran simulations with $M_\nu$ = 0.06, 0.12, 0.24, and 0.48 eV.  This was done both in the context of WMAP9 and Planck 2015 cosmologies \citep[see][for details]{McCarthy2018}. The present-day neutrino density of each run is $\Omega_\nu = \frac{\rho_\nu^0}{\rho_{\rm c}^0} = M_\nu / (93.14 \ {\rm eV}\,h^2)$, using the present day number density of neutrinos given by \citet{Lesgourgues2012}, where $\rho_\nu$ and $\rho_{\rm c}^0$ are the neutrino and critical densities today. When massive neutrinos were added in the \textit{WMAP}~9-based cosmology, the cold dark matter density was reduced in order to keep a flat geometry, or $\Omega_{b} + \Omega_\mathrm{cdm} +  \Omega_\nu +  \Omega_\Lambda = 1$. 
For the ``\textit{Planck}~2015$/A_{Lens}-based$" simulations used in this paper, the Markov Chains of \citet{PlanckCollaboration2016} corresponding to the ``CMB+BAO+CMB lensing" with marginalization over $A_{Lens}$ (the amplitude of the CMB lensing power spectrum) was used \citep[see Figure 2 of][]{McCarthy2018}. Cosmological parameter sets were selected that have summed neutrino mass within $\Delta M_{\nu} = 0.02$ eV of the target value, and the weighted mean of the other important cosmological parameters is taken from the Markov Chains.
By selecting parameter values in this way, it ensures that the selected cosmologies are consistent with the CMB+BAO constraints and again preserving flatness, i.e. $\Omega_{\rm{K}} = 0$. Note that there is an underlying fiducial baryonic physics for every model with non-zero $M_\nu$. The details of the different runs are presented in Table \ref{tab:cosmological_parameters}.

\begin{table*}
	\centering
	\caption{Cosmological parameter values for 12 suites of the BAHAMAS simulations \citep{McCarthy2017,McCarthy2018}. Adjustments on the summed neutrino mass, baryonic matter fractions, AGN feedback temperatures, $\sigma_8$ values, and changes in $S_8$ are given in this table. The columns are: (1) The summed mass of the 3 active neutrino species (we adopt a normal hierarchy for the individual masses). Note that there is underlying fiducial baryonic physics for every non-zero $M_\nu$ model; (2) the logarithm to base 10 of the AGN feedback temperature defined by \citet{McCarthy2018}; (3) the total matter density; (4) present-day baryon density; (5) present-day dark matter density; (6) present-day neutrino density, computed as $\Omega_\nu = M_\nu / (93.14 \ {\rm eV}\,h^2)$; (7) present-day (linearly-evolved) amplitude of the matter power spectrum on a scale of 8 $\mathrm{Mpc}\,h^{-1}$ (note that we use $A_s$ rather than $\sigma_8$ to compute the power spectrum used for the initial conditions, thus the initial conditions are CMB normalised); (8) $S_8 = \sigma_8 \sqrt{\Omega_\mathrm{m} / 0.3}$.}
	\label{tab:cosmological_parameters}
	\begin{tabular}{lccccccc}
		\hline 
        (1)        & (2)        & (3)               & (4)                & (5)              & (6)      & (7)              &  (8) \\
		$M_\nu$(eV)   & $\log(\Delta T_{\rm heat} [$K$])$ & $\Omega_\mathrm{m}$ & $\Omega_b $  & $\Omega_{\rm cdm} $ &  $\Omega_\nu$ & $\sigma_8$ & $S_8$ \\
        \hline
		{\bf WMAP9-based}\\        
		0 (DMONLY) & - & 0.2793 & 0.0463 & 0.2330 & 0.0 & 0.8211 & 0.7923\\
        0 (Low AGN) & ${7.6}$ & 0.2793 & 0.0463 & 0.2330 & 0.0 & 0.8211 & 0.7923\\
0 (Fiducial AGN) & $7.8$ & 0.2793 & 0.0463 & 0.2330 & 0.0 & 0.8211 & 0.7923\\
        0 (High AGN) & ${8.0}$ & 0.2793 & 0.0463 & 0.2330 & 0.0 & 0.8211 & 0.7923\\
        \\
        0.06 & ${7.8}$ & 0.2793 & 0.0463 & 0.2317 & 0.0013 & 0.8069 & 0.7786\\
        0.12 & ${7.8}$ & 0.2793 & 0.0463 & 0.2304 & 0.0026 & 0.7924 & 0.7646\\
        0.24 & ${7.8}$ & 0.2793 & 0.0463 & 0.2277 & 0.0053 & 0.7600 & 0.7333\\
        0.48 & ${7.8}$ & 0.2793 & 0.0463 & 0.2225 & 0.0105 & 0.7001 & 0.6755\\
        \\
		{\bf \textit{Planck}~2015/$A_{\rm Lens}$-based}\\        
		0.06 & ${7.8}$ & 0.3067 & 0.0482 & 0.2571 & 0.0014 & 0.8085 & 0.8175\\
        0.12 & ${7.8}$ & 0.3091 & 0.0488 & 0.2574 & 0.0029 & 0.7943 & 0.8063\\
        0.24 & ${7.8}$ & 0.3129 & 0.0496 & 0.2576 & 0.0057 & 0.7664 & 0.7827\\
        0.48 & ${7.8}$ & 0.3197 & 0.0513 & 0.2567 & 0.0117 & 0.7030 & 0.7257\\
        \hline
        
	\end{tabular}
\end{table*}

\subsection{Light-Cones and Synthetic Lensing Catalogues}
\label{sec:lightCones}
The simulation boxes of BAHAMAS are $400$ $\mathrm{Mpc}\,h^{-1}$ per comoving side with each box containing $2 \times 1024^{3}$ particles. In order to construct BAHAMAS light-cones, \citet{McCarthy2018} saved particle data at snapshots from $z = 3$ to today. There are 15 snapshots at $z =$ 0.0, 0.125, 0.25, 0.375, 0.5, 0.75, 1.0, 1.25, 1.5, 1.75, 2.0, 2.25, 2.5, 2.75, and 3.0. The snapshots were then randomly oriented and translated, and slices of $5 \times 5$ deg$^{2}$ (at pixel resolution of 10 arcseconds) were taken from the snapshots. A total of 25 randomisations of rotations and translations of the 15 snapshots were performed, the same for each cosmology and prescription for baryonic physics and neutrino mass, so that cosmic variance does not play a role when comparing light-cones across different runs. Here we consider $25$ light-cones with a total area of $625$ deg$^{2}$.

The light-cones extracted from the simulations then give us ideal convergence maps, following Section \ref{sec:gravitationalLensing}, integrating the spatial matter overdensities $\delta(\chi, \theta)$ through the discrete slices along the line of sight with the kernel containing the source redshift distribution $p_s(z)$, as in Equation \ref{eq:simulationConvergence} and \ref{eq:lensingKernel}. We adopt two different redshift distributions, one for the 450 deg$^2$ Kilo Degree Survey \citep[KiDS-450,][]{Hildebrandt2017} and one predicted for the LSST survey \citep{Chang2013}. The convergence maps for the former were created by \citet{McCarthy2018} and the convergence maps for the latter were constructed for this paper. Shear maps, and reduced shear maps, are constructed from convergence maps using Equation \ref{eq:fftmethod} and \ref{eq:g}. Synthetic catalogues of weakly lensed galaxies are generated using the reduced shear maps from the simulations. 

We populate the maps with unlensed galaxies that are randomly placed on the sky, with number density appropriate for observations with a particular survey. The moduli of the complex ellipticities (related to the axis ratios) are drawn from a Gaussian distribution with $\sigma_{|\epsilon^{s}|} = 0.25$ and zero mean. Note that this value of 0.25 is for the modulus of $\epsilon^{s}$ rather than per component, hence the value is slightly lower than measured in observations \citep[see e.g. Figure A3 of][]{Schrabback2018}. The galaxy orientation angles, $\phi^s$, are randomly assigned between 0 and $\pi$. At the position of a source galaxy, the reduced shear is extracted at the nearest pixel of the BAHAMAS maps. The observed ellipticity is determined using Equation \ref{eq:ellipticity}.
 
In this paper we roughly base one of the 625 deg$^{2}$ synthetic surveys on \citet{Hildebrandt2017}, where they analyse $\sim450$ deg$^{2}$ of imaging data from KiDS. In KiDS the effective number density of galaxies is $8.53$ gal/arcmin$^{2}$ (the number density of galaxies was determined in \citealt{Hildebrandt2017} by using the method proposed by \citealt{Heymans2012}), and the observed ellipticity dispersion is $\sigma_{\epsilon} \approx 0.29$ per component. For our first synthetic survey we use an effective source number density of $9$ gal/arcmin$^{2}$.

We also consider synthetic surveys of the same $625$ deg$^{2}$ area, but with convergence (shear and reduced shear) maps constructed using the source redshift distribution from \citet{Chang2013}, estimated for the deeper upcoming LSST. Both LSST and $\textit{Euclid}$ \citep{Amendola2018,Laureijs2011} surveys have an expected effective galaxy source number density $n_{\rm{eff}} = 30$ gal/arcmin$^{2}$.
We also consider $n_{\rm{eff}} = 60$ gal/arcmin$^{2}$ when we create our synthetic lensing catalogues, to reflect deeper space based characteristics, such as the \textit{HST} \citep[see for example the source number density in][]{King2016}. 
Note that we use the same redshift distribution for this even deeper survey, but increase the effective number density of sources from which the shear can be measured. This neglects the contribution from some higher redshift galaxies which carry a more significant lensing signal than lower redshift ones. The different surveys with the effective number density of 9, 30, and 60 gal/arcmin$^2$ will be referred to as KiDS, Deep Ground Based or Space Based (DGB/SB), and Deep Space Based (DSB) respectively. 

Similar to \citet{Martinet2018}, on the $625$ deg$^2$ we generate 5 synthetic catalogues of source ellipticities at the same positions but with random shapes per suite (as in Table \ref{tab:cosmological_parameters}) to make sure that the simulations are not biased to one particular realisation of shape noise.

\section{Aperture Mass Peaks}
\label{sec:apertureMass}

In this section we outline the aperture mass statistic \citep{Schneider1996} that we use to measure the weak lensing signal, and how we map the aperture mass over the synthetic surveys. Finally we describe how we determine the weak lensing peaks based on the aperture mass maps.

\subsection{Aperture Mass Statistic}
\label{sec:apertureMassStatistic}
We can take advantage of the fact that the average shape of unlensed galaxies is circular for a large enough sample, or $\langle \epsilon^{s}\rangle  = 0$. So if we take an ensemble of lensed galaxy shapes over a small patch of sky, we can recover the reduced shear, $\langle \epsilon \rangle  = g$ (see Equation \ref{eq:ellipticity}).

An important tool is the aperture mass \citep{Schneider1996}, which can be used to map dark and luminous matter. 
Although not employed in this paper, the variance of the aperture mass as a function of aperture scale can be used to constrain cosmological parameters using cosmic shear \citep[e.g. see][and references therein]{Schneider2005}. The aperture mass statistic has been used on wide field surveys to identify massive objects, e.g. \citet{Hetterscheidt2005}. Various cluster detection methods that incorporate tomographic redshift information for the source galaxies and different shapes of filter function have also been developed \citep[e.g.][]{Schirmer2007, Maturi2010}. For example, \citet{Hennawi2005} considered the efficiency of cluster detection using the aperture mass statistic and also including tomographic information and optimal filtering.

The aperture mass, $M_{\rm{ap}}$, is constructed by integrating the weighted convergence within an aperture:
\begin{equation}
    M_{\rm{ap}}(\vec{\theta_0}) = \int d^2\theta \, U(\vec{\theta} - \vec{\theta}_0) \, \kappa(\vec{\theta}),
    \label{eq:apertureMassKappa}
\end{equation}
where $\theta_{0}$ is the 2D location of the aperture centre, and $U$ is a weight function that is compensated within the filter radius, smoothly goes to zero at a finite radius, and is zero outside of that radius. It should also be localised in Fourier space with no oscillatory behaviour in the power spectrum \citep{Leonard2012}.
Since the convergence and the shear are related (see Section \ref{sec:gravitationalLensing}), the aperture mass can be expressed as
\begin{equation}
    M_{\rm{ap}}(\vec{\theta}_0) = \int d^2\theta \, Q(\vec{\theta} - \vec{\theta}_0) \, \gamma_t(\vec{\theta},\vec{\theta}_0)\,,
    \label{eq:apertureMassIntegral}
\end{equation}
where the shear weight function or filter function is $Q(\theta) = \frac{2}{\theta^2} \int_0^\theta d\theta' \theta' U(\theta') - U(\theta)$ and the tangential shear is $\gamma_{t} \approx g_{t}$ in the weak lensing limit.

In practice for real or synthetic observations of lensed galaxies we express the integral in Equation \ref{eq:apertureMassIntegral} as a sum over discrete galaxies. The aperture mass then becomes the weighted sum over the tangential ellipticities: 
\begin{equation}
    M_{\rm{ap}}(\vec{\theta}_0) = \frac{1}{n_{\rm{gal}}} \sum_i^{N_{\rm{gal}}} Q(\vec{\theta_i} - \vec{\theta}_0) \, \epsilon_t(\vec{\theta}_i, \vec{\theta}_0)\,,
    \label{eq:apertureMass}
\end{equation}
where $n_{\rm{gal}}$ is the number density of observed galaxies inside the aperture, $N_{\rm{gal}}$ is the total number of galaxies inside the aperture, and $\epsilon_{t}$ is the observed tangential ellipticity of a galaxy expressed as:
\begin{equation}
    \epsilon_{t} (\vec{\theta}, \vec{\theta}_0) = -[\epsilon_1(\vec{\theta}) \cos (2 \varphi (\vec{\theta},\vec{\theta_0})) + \epsilon_2(\vec{\theta}) \sin (2 \varphi (\vec{\theta},\vec{\theta_0}))],
\end{equation}
where $\varphi (\vec{\theta},\vec{\theta_0})$ is the angular position with respect to the centre of the aperture.

\begin{figure}
    \includegraphics[width=\columnwidth]{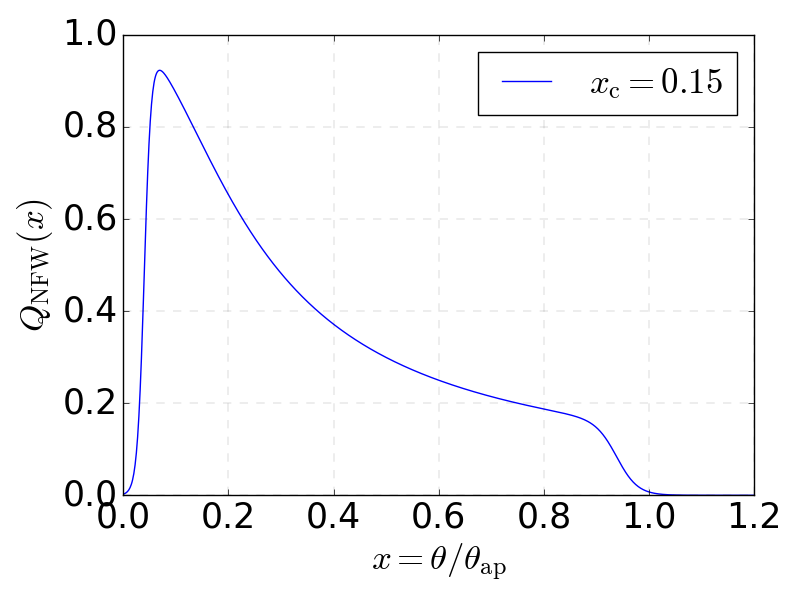}
    \caption{$Q_{\rm{NFW}}(x)$ is the aperture mass filter function used in this work (Equation \ref{eq:NFWFilterFunction}), taken from \protect\citet{Schirmer2007}.
    The aperture mass calculated in Equation \ref{eq:apertureMass} cuts off at the filter size, $\theta = \theta_{\rm{ap}}$ and the falloff towards the centre would down-weight a bright central galaxy in a galaxy cluster and the strong lensing regime.
    }
    \label{fig:Qnfw}
\end{figure}

Filter functions are optimised for different applications (see for example \citealt{Maturi2010}).
In this work we use a filter function that is optimised to detect NFW haloes taken from \citet{Schirmer2007}:
\begin{equation}
    Q_{\rm{NFW}} (x) = \frac{1}{1 + e^{6-150x} + e^{-47+50x}} \frac{\tanh (x/x_{\rm c})}{x/x_{\rm c}},
    \label{eq:NFWFilterFunction}
\end{equation} 
where $x=\theta / \theta_{\rm{ap}}$ is the angular distance from the aperture centre $\theta_0$ scaled by the filter size $\theta_{\rm{ap}}$. 
$x_{\rm c}$ is analogous to the halo concentration in the NFW profile, and it was empirically set to $x_{\rm c}=0.15$ in \citet{Hetterscheidt2005}. 
This is consistent with \citet{Martinet2018}. Figure \ref{fig:Qnfw} is a plot of the filter function, $Q_{\rm{NFW}}(x)$ (Equation \ref{eq:NFWFilterFunction}), used in this work. Note that the filter function down-weights galaxies towards the centre of the aperture, which excludes contamination by the presence of a bright central galaxy in a cluster and the strong lensing regime. Furthermore there is a possibility that the impact of different baryonic physics models can be mitigated due to the down-weight or exclusion of the central region in the filter. Compared to DMONLY, free-streaming neutrinos suppress the median halo density profile roughly independent of the distance to the centre, while baryonic physics effects depend more strongly on the distance \citep[see][]{Mummery2017}. The filter function is truncated at $\theta = \theta_{\rm{ap}}$ and the aperture mass is calculated within the filter size, $\theta_{\rm{ap}}$. 

\subsection{S/N Peaks}
\label{sec:snrPeaks}
\begin{figure*}
    \centering
    \includegraphics[width=\textwidth]{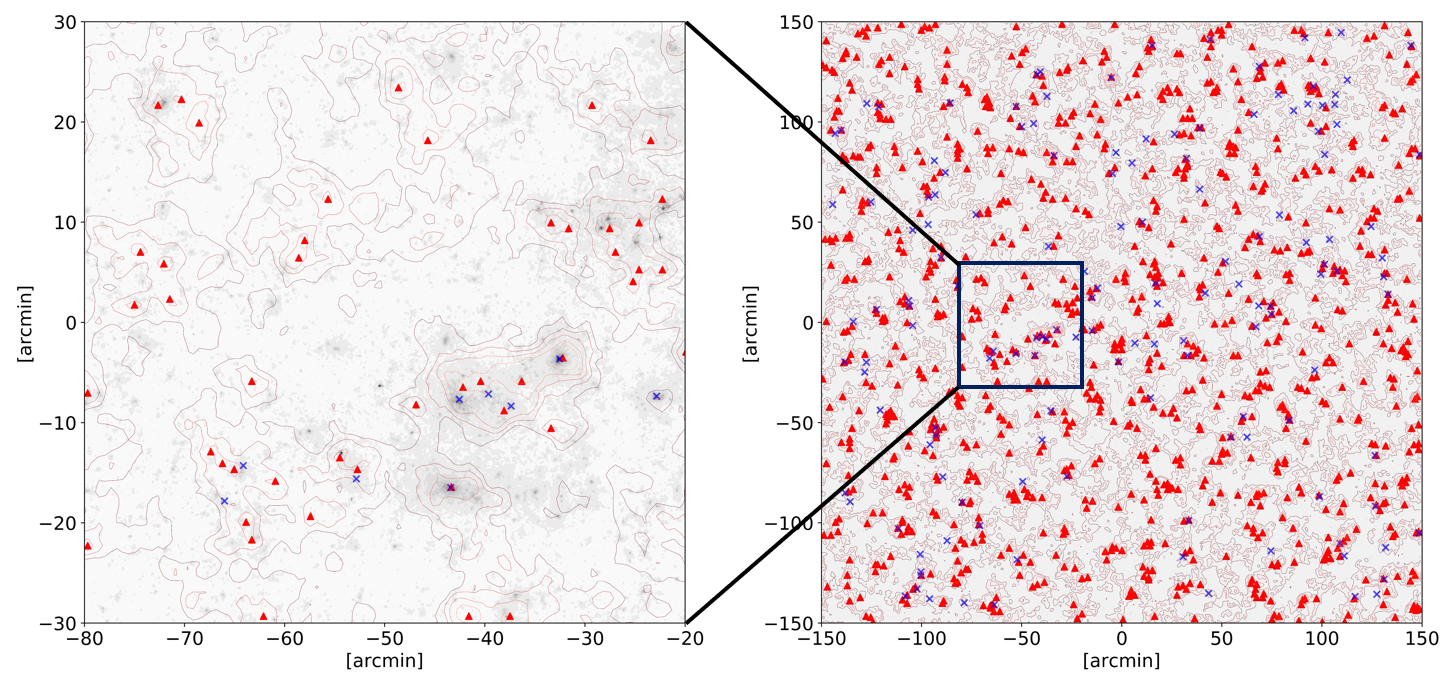}
    \caption{A 60 $\times$ 60 arcmin$^2$ sub-region of a 5 $\times$ 5 deg$^2$ aperture mass map for source number density 9 gal/arcmin$^{2}$. The aperture size is 12.5 arcmin, and the S/N contour lines are plotted from 0 with an increment of 1. The field in grey scale in the background is the light-cone convergence map, and the cross marks are the locations of clusters from the BAHAMAS friends-of-friends catalogues, where the lowest mass cut is $10^{14} \rm{M_\odot}$ and the upper limit redshift cut for the clusters is at z=0.5. The triangles are the peak locations above S/N=2.}
    \label{fig:aperture_map}
\end{figure*}

To calculate the S/N ratio for the aperture mass statistic we proceed as in \citet{Martinet2018}, where the standard deviation of the aperture mass in the absence of shear is given by: 
\begin{equation}
    \sigma(M_{\rm{ap}}(\vec{\theta}_0)) = \frac{1}{ \sqrt{2}n_{\rm{gal}}} \left( \sum_i^{N_{\rm{gal}}} |\epsilon(\vec{\theta}_i)|^2 Q^2(\vec{\theta}_i - \vec{\theta}_0) \right)^{\frac{1}{2}}.
\end{equation}
The S/N ratio for an aperture measurement is given by:
\begin{equation}
    \frac{S}{N}(\vec{\theta}_0) = \frac{\sqrt{2} \sum_i^{N_{\rm{gal}}} Q(\vec{\theta}_i - \vec{\theta}_0) \epsilon_t(\vec{\theta}_i,\vec{\theta}_0)}{\sqrt{\sum_i^{N_{\rm{gal}}} |\epsilon(\vec{\theta}_i)|^2 Q^2(\vec{\theta}_i - \vec{\theta}_0)}}.
    \label{SNR}
\end{equation}
We use publicly available software, developed by \citet{Bard2012}, which implements GPU computing for fast calculation of the aperture mass. This algorithm uses the positions and shapes of galaxies in this calculation.
For a compromise between computational efficiency and optimised resolution for peak detection, the grid resolution of the aperture maps (where the apertures are centred) is set to 512 $\times$ 512 pixels which is 0.5859 arcmin per pixel, similar to \citet{Martinet2018}. The algorithm scans the grid points over the synthetic weak lensing catalogues and returns the aperture mass, variance, and S/N values. 
We ran the code on the synthetic data from Section \ref{sec:lightCones} using filter sizes $\theta_{\rm{ap}} = 8.0, 10.0, 12.5,$ and $15.0$ arcmin. Figure \ref{fig:aperture_map} shows a sample of the direct overlay of the S/N contours (lower limit of S/N at 0 and contours increase in increments of 1) on top of the BAHAMAS light-cone convergence map (grey scale) which is produced by integrating the weighted line-of-sight overdensities according to Equation \ref{eq:simulationConvergence}. 
The cross marks are the locations of galaxy clusters taken directly from the BAHAMAS Friends-of-Friends (FoF) catalogues.
The triangles are the S/N peak locations (S/N $ > 2$) obtained by a peak detection Python script which examines if each pixel of the S/N map has a higher value than the 8 nearest neighboring pixels. We focus on the analysis of the statistics of these peaks in our work.

To quantify the difference in the weak lensing peak counts for varying survey characteristics, and for simulation runs with different baryonic physics and summed neutrino mass, we take the ratio of the weak lensing peak distributions with respect to the reference models. We use five different shape noise realisations for each simulation run, consistent with \citet{Martinet2018}, in order to avoid potential bias to one particular noise realisation. Varying only shape noise neglects sample variance which is partially accounted for by using several different line-of-sights through the simulations, in that we use $25$ light-cones.

Note that \citet{Martinet2018} used dark matter only simulations to investigate the impact of a much wider range of cosmologies (particularly considering $\Omega_\mathrm{m}$ and $\sigma_8$) on peak counts for comparison with KiDS data. In this work, as noted in the introduction, we focus on investigating the impact of baryonic processes (including star formation, AGN feedback and supernova feedback) and massive neutrinos on the peak statistics.

\section{Results}
\label{s:results}

In Subsection \ref{sub:miyoungResults} we present the results of our studies on the impact of baryonic physics and summed neutrino mass on the weak lensing peak statistics. Then in Subsection \ref{sec:highSNR} we present the results on how summed neutrino mass affects the correlation between higher S/N peaks and massive clusters. In both Subsections we consider synthetic surveys with different source redshift distributions and effective number density of sources from which lensing shear can be measured. These synthetic surveys are referred to as KiDS, DGB/SB, and DSB, as outlined in Section \ref{sec:lightCones}.

\subsection{Dependence of Weak Lensing Peak Counts on Baryonic Physics and Summed Neutrino Mass}
\label{sub:miyoungResults}

\begin{figure}
     \centering
     \begin{subfigure}[b]{\columnwidth}
         \centering
         \includegraphics[width=\columnwidth]{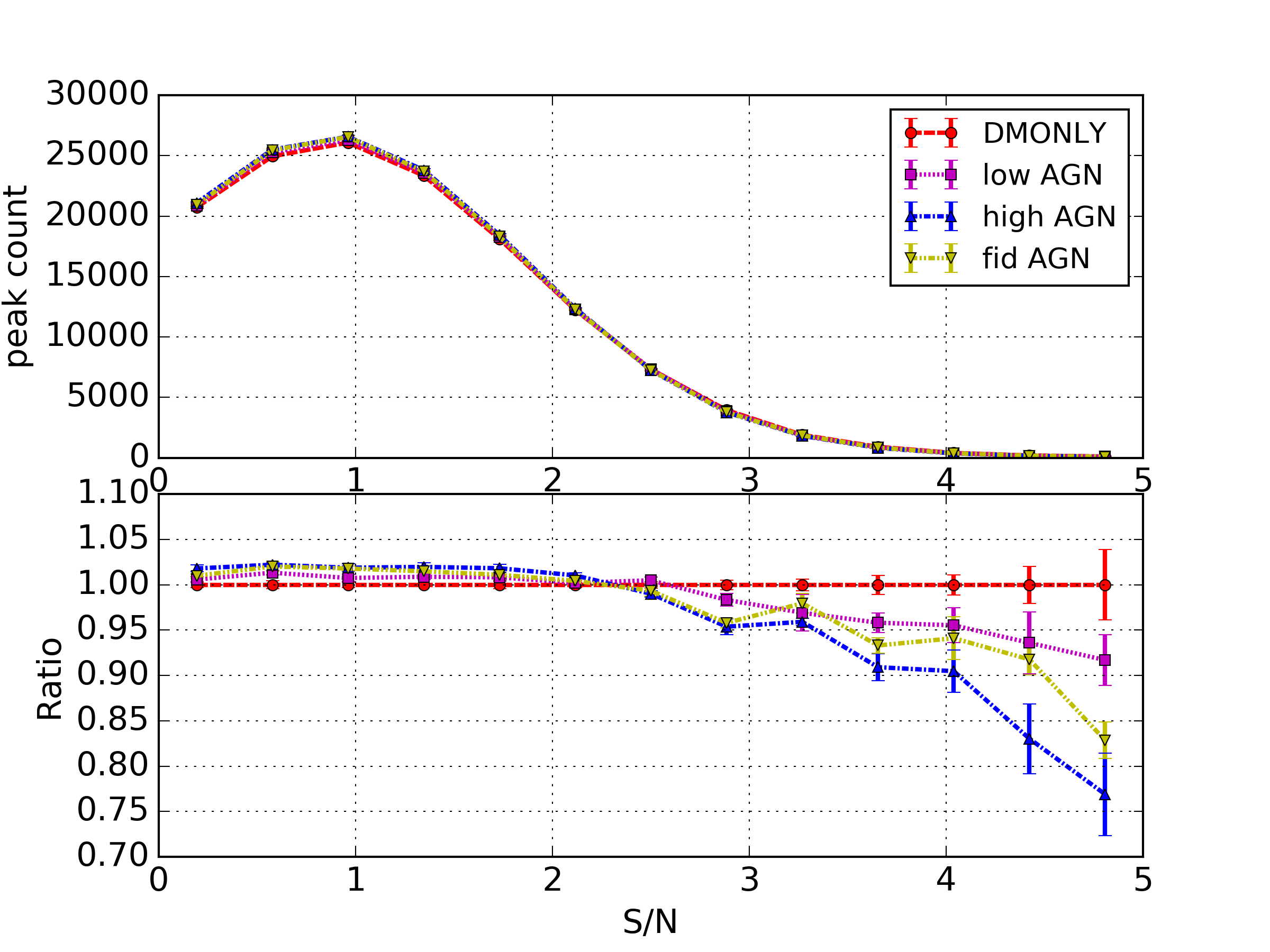}
         \caption{}
         \label{fig:AGNs_KiDS}
     \end{subfigure}
     \hfill
     \begin{subfigure}[b]{\columnwidth}
         \centering
         \includegraphics[width=\columnwidth]{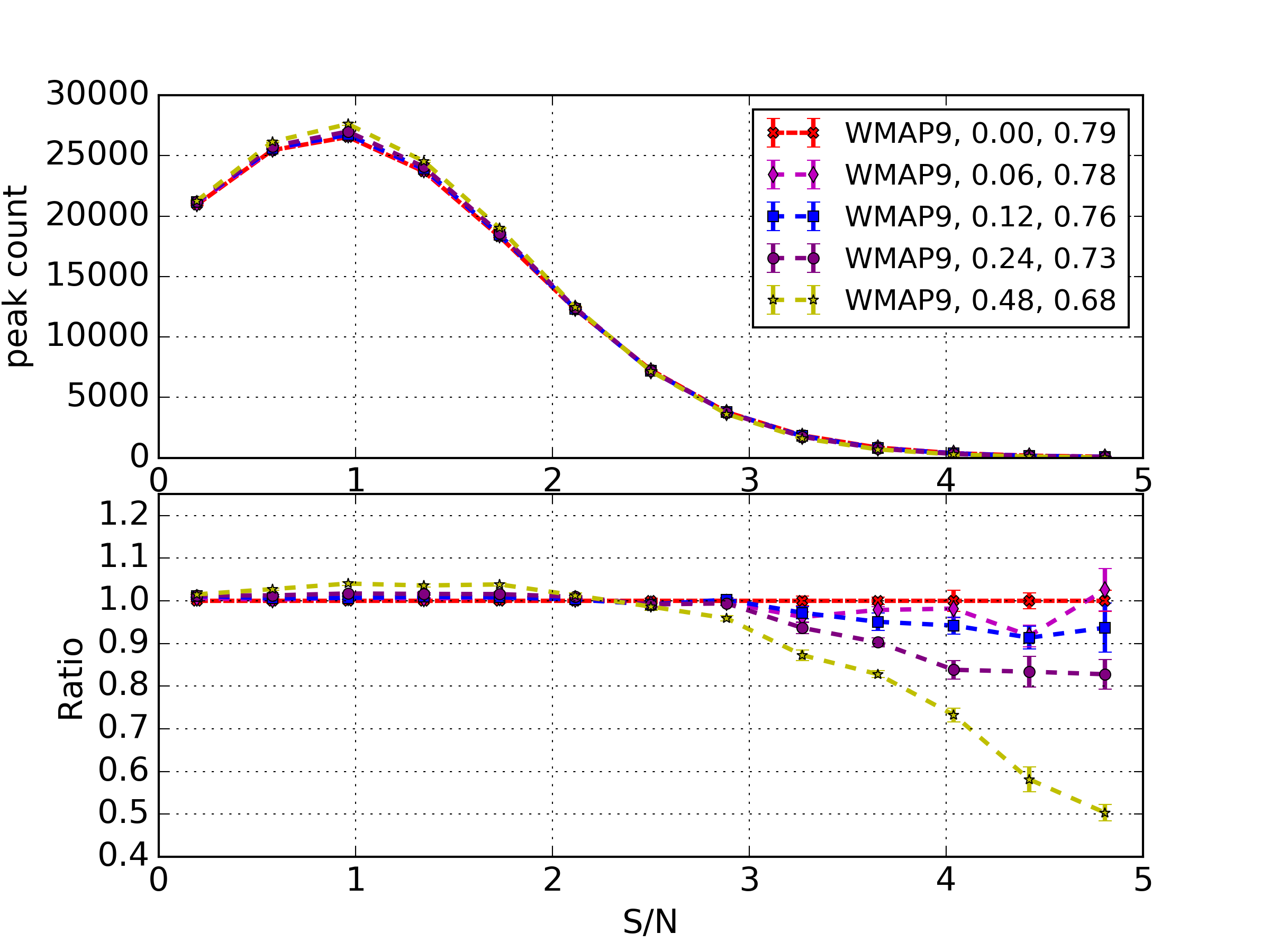}
         \caption{}
         \label{fig:Mnus_KiDS}
     \end{subfigure}
     \hfill
      \begin{subfigure}[b]{\columnwidth}
         \centering
         \includegraphics[width=\columnwidth]{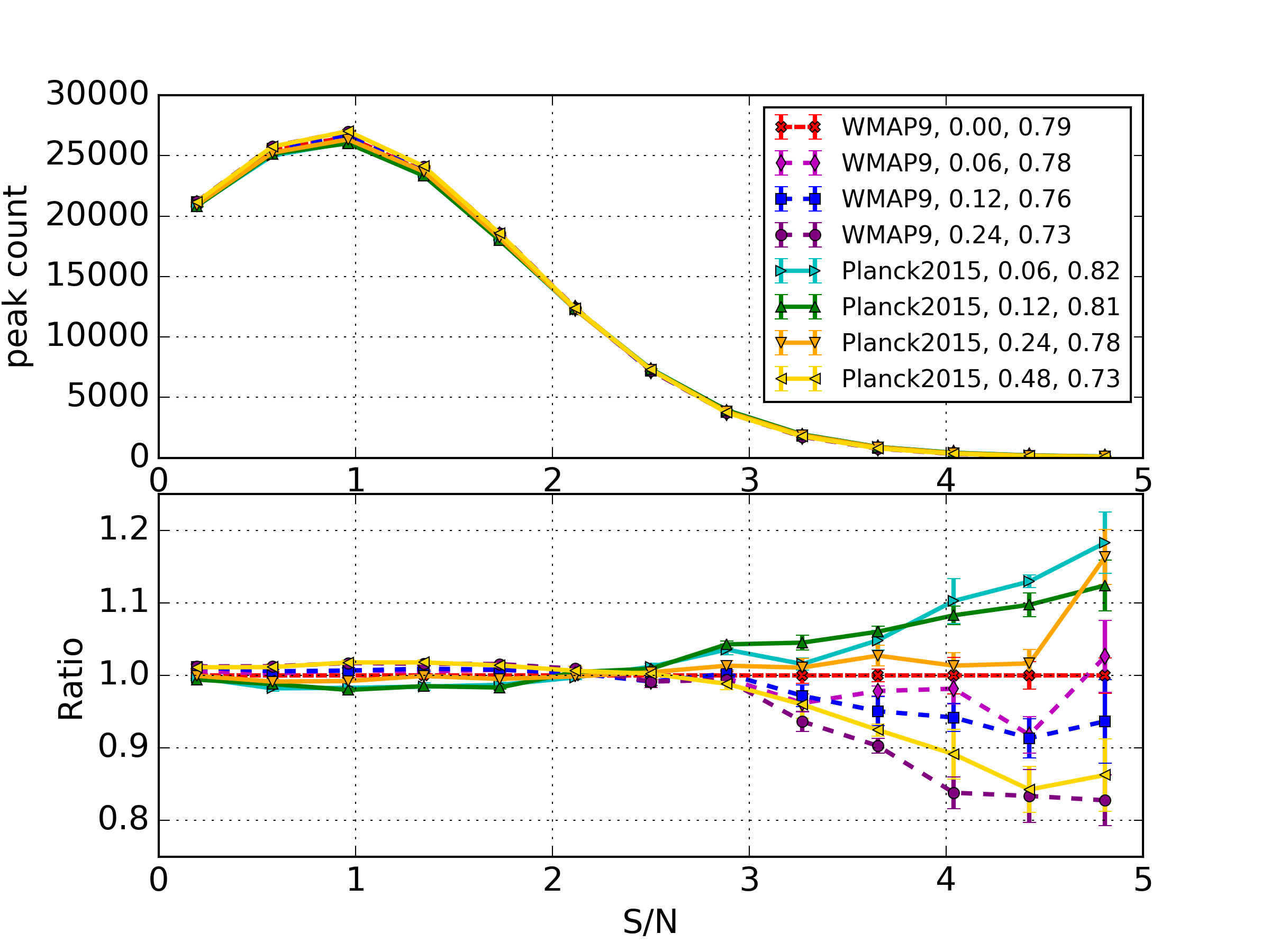}
         \caption{}
         \label{fig:Planck_KiDS}
     \end{subfigure}
        \vspace*{-7mm}
        \caption{
        The impact of (a) baryonic physics, (b) massive neutrinos, and (c) \textit{WMAP}~9 and \textit{Planck}~2015 cosmologies on the weak lensing peak statistics with aperture filter size of 12.5 arcmin all with source number density of 9 gal/arcmin$^2$. The bottom panels show the ratio, taking the ratio with the \textit{WMAP}~9 DMONLY results for subplot (a) and \textit{WMAP}~9 with fiducial AGN feedback for subplots (b) and (c). 
        Note that we exclude the \textit{WMAP}~9 $M_{\nu}$ = 0.48 eV model in (c) as it is well outside of the range of the \textit{Planck}~2015 models. All of the models in (b) and (c) have fiducial baryonic physics. The error bars show the variance of the five different shape noise realisations. In (b) and (c) the legend entries are Cosmology, $M_{\nu}$ (eV), and $S_{8}$. Note that the bins are spaced linearly in S/N from 0 to 5 in 13 bins with the markers indicating the bin centres.}
        \label{fig:result_peak_KiDS}
\end{figure}

\begin{figure}
     \centering
     \begin{subfigure}[b]{\columnwidth}
         \centering
         \includegraphics[width=\columnwidth]{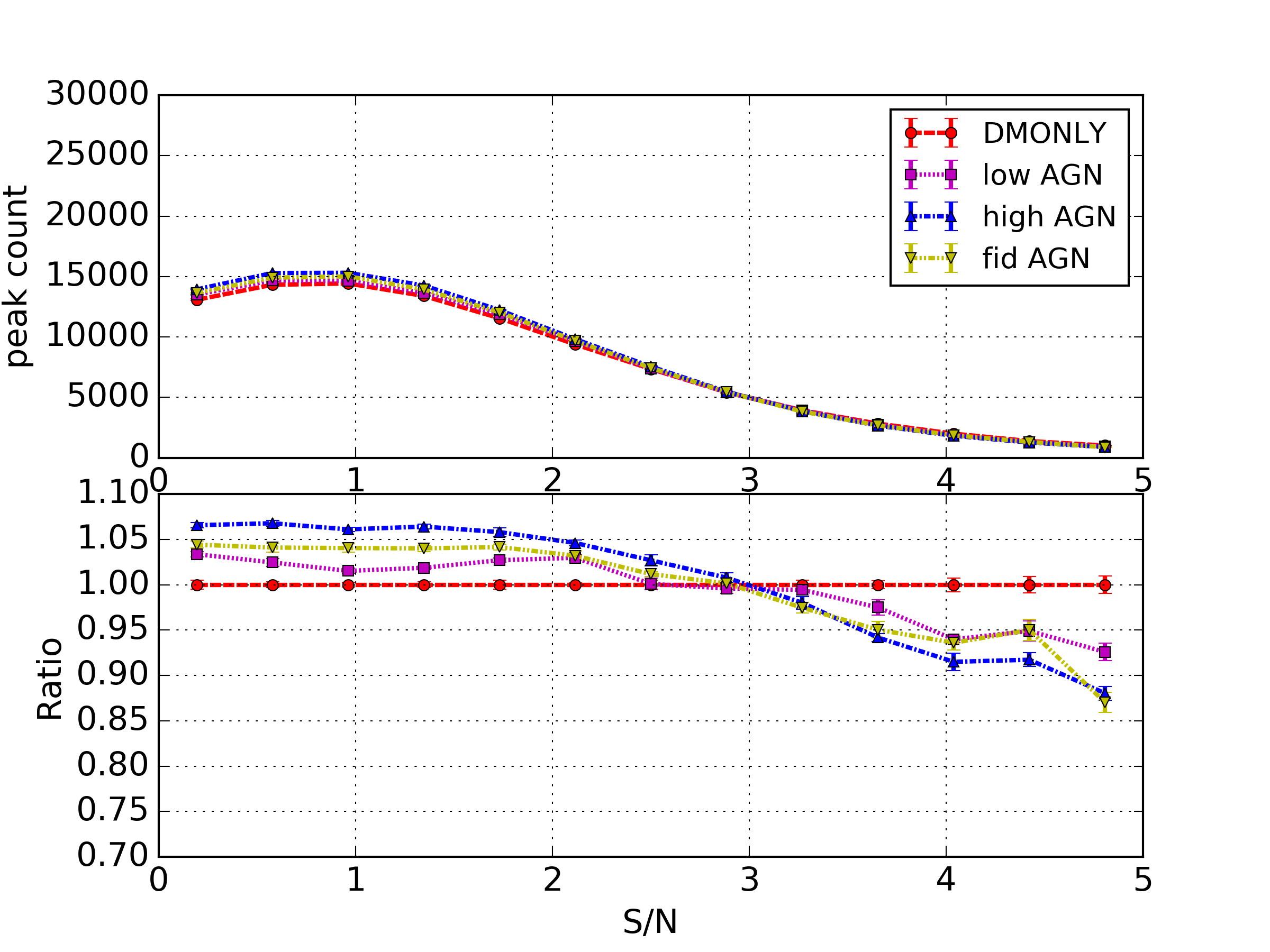}
         \caption{}
         \label{fig:AGNs_LSST}
     \end{subfigure}
     \hfill
     \begin{subfigure}[b]{\columnwidth}
         \centering
         \includegraphics[width=\columnwidth]{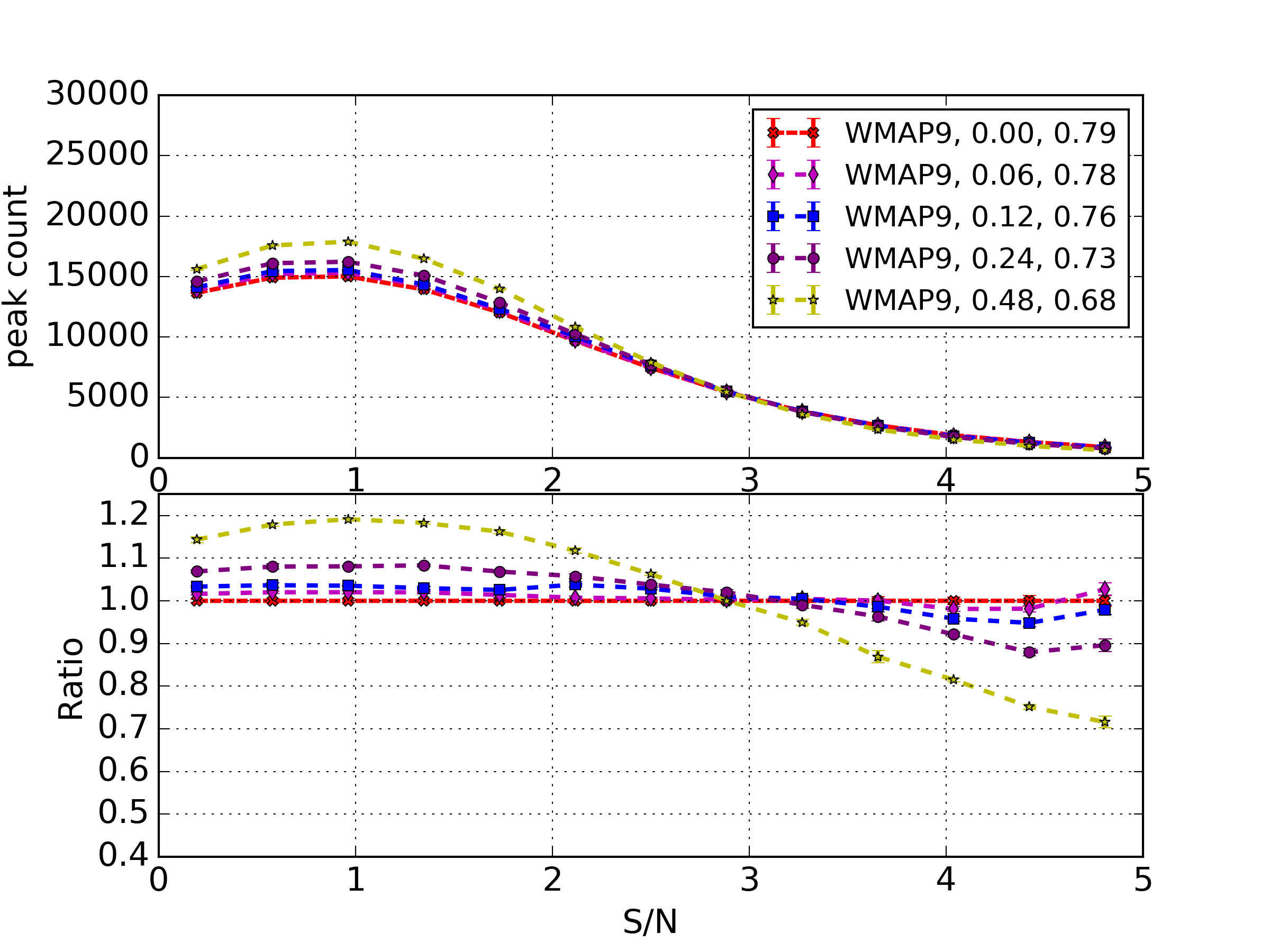}
         \caption{}
         \label{fig:Mnus_LSST}
     \end{subfigure}
     \hfill
     \begin{subfigure}[b]{\columnwidth}
         \centering
         \includegraphics[width=\columnwidth]{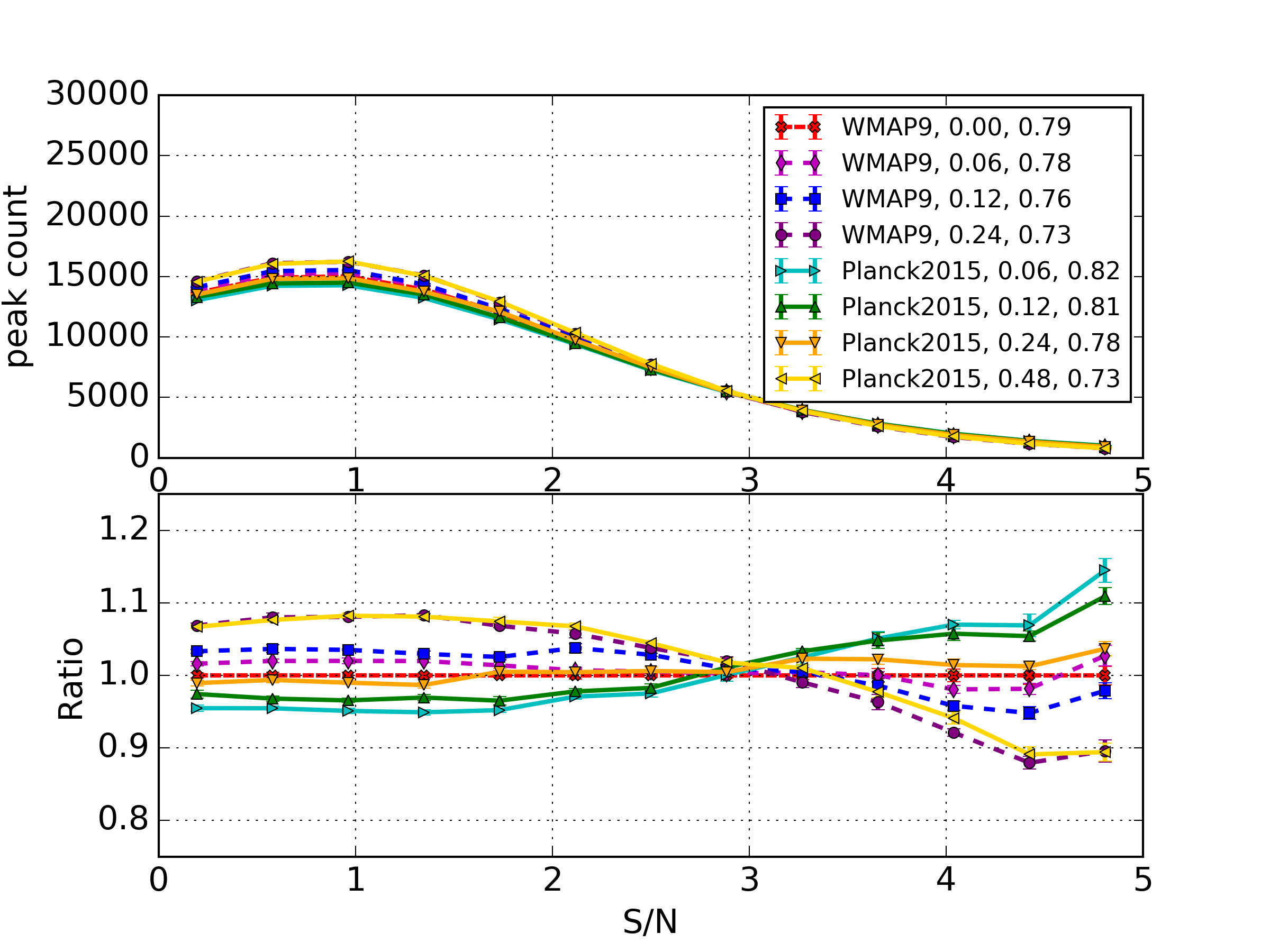}
         \caption{}
         \label{fig:Planck_LSST}
     \end{subfigure}
        \vspace*{-7mm}
        \caption{Comparison of the WL peak distributions with the higher source number density (30 gal/arcmin$^2$) for DGB/SB surveys. The other parameters such as the intrinsic ellipticity dispersion and the suites of simulations are the same as in Figure \ref{fig:result_peak_KiDS}. Note that we exclude the \textit{WMAP}~9 $M_{\nu}$ = 0.48 eV model in (c) as it is well outside of the range of the \textit{Planck}~2015 models. All of the models in (b) and (c) have fiducial baryonic physics. The error bars show the variance of the five different shape noise realisations. In (b) and (c) the legend entries are Cosmology, $M_{\nu}$ (eV), and $S_{8}$. Note that the bins are spaced linearly in S/N with the markers indicating the bin centres.
        }
        \label{fig:result_peak_LSST}
\end{figure}
Figures \ref{fig:result_peak_KiDS} and \ref{fig:result_peak_LSST} show the impact of the different baryonic physics models and summed neutrino mass in \textit{WMAP}~9 and \textit{Planck}~2015 cosmologies for the KiDS and DGB/SB survey respectively. The filter size is fixed at 12.5 arcmin for consistency with \citet{Martinet2018}. 
In the top panels of each subfigure the S/N peak distributions are presented, and the bottom panels show the ratio of the distributions with respect to the reference model. Note that the bins are spaced linearly in S/N from 0 to 5 in 13 bins, with the markers indicating the bin centres.

Figures \ref{fig:AGNs_KiDS} and \ref{fig:AGNs_LSST} show the impact of baryonic processes on the WL peak statistics with different background source number densities of 9 and 30 gal/arcmin$^2$ respectively. 
The bottom panels show the ratio of the S/N distributions for different AGN feedback strengths with respect to the DMONLY case. Note that in this case massive neutrinos are not included. 
In Figure \ref{fig:AGNs_KiDS}, with $n_{\rm eff} = 9$ gal/arcmin$^2$, the low S/N peaks show a modest boost from baryons resulting in only a few percent deviation from DMONLY. 
However in Figure \ref{fig:AGNs_LSST}, with $n_{\rm eff} = 30$ gal/arcmin$^2$, the low S/N peak counts tend to be boosted by up to about $6$ percent in comparison with DMONLY. 
According to the discussion in \citet{Martinet2018}, much of the constraining power of peak counts for cosmological models comes from the lower S/N range. Our results indicate that when baryonic processes are accounted for in the error budget, care should be taken to calibrate the impact of baryons as a function of the source number density. Effects of different aperture filter sizes on the S/N distributions is discussed in Section \ref{sec:highSNR} and more specific examples are shown in the Appendix.

In DMONLY simulations higher S/N peak values are more likely to be produced by more massive clusters, although there is an additional complication due to the dependence of peak height on cluster redshift for a given source population \citep[e.g. Figure 17 of][]{Hamana2004}.
The impact of baryons is more significant at higher S/N, where higher peak values are more likely to arise from massive haloes \citep{Hennawi2005}. The massive haloes are more likely to have greater AGN feedback strength that cause the mass density profiles to be flatter and less concentrated \citep{Mummery2017}, thereby returning lower S/N values than the DMONLY case and having a larger effect on the high S/N peaks. 
In Table \ref{tab:cosmological_parameters} the $S_8$ values do not vary with AGN temperature, yet these plots show that increasing AGN temperature changes the peak distribution similar to decreasing $S_8$ \citep[see][]{Martinet2018}. 
In agreement with previous work \citep[e.g.][]{Osato2015ApJ...806..186O}, this implies that estimating cosmological parameters when using cosmological simulations without baryonic processes can lead to a bias. 
This is also true for the summed neutrino mass, as discussed below.
Additionally, the deviation among the higher S/N peak counts with different AGN feedback strength is hard to resolve, even though the error bars due to scatter from the different shape noise realisations are tighter than those for the lower number density of sources (as anticipated).

The middle panels, Figures \ref{fig:Mnus_KiDS} and \ref{fig:Mnus_LSST} (9 and 30 gal/arcmin$^2$ respectively), are the S/N peak distributions for different summed neutrino mass. Note that all models here include the underlying fiducial baryonic physics prescription. The S/N peak counts show that there is a negative (positive) difference for the higher (lower) S/N peaks for all of the $M_{\nu}$ models (with the exception of the last S/N peak bin in the 0.06 eV model, where the statistics in this bin are poor). Furthermore increasing $M_{\nu}$ suppresses (boosts) the higher (lower) S/N peaks. This is expected as free-streaming neutrinos impede the growth of LSS and therefore have a larger impact on more massive haloes, suppressing the high end of the halo mass function more significantly with increasing neutrino mass \citep[e.g.][]{Costanzi2013, Mummery2017}.

Note the similarity of the suppression (boosting) in high (low) S/N peak counts due to AGN feedback and massive neutrinos. 
This can make differentiating baryonic physics and massive neutrinos based purely on S/N peak counts quite difficult. \citet{Mummery2017} shows that baryonic physics and summed neutrino mass impact on the halo mass function and halo clustering and matter clustering (or matter power spectrum, P(k)), independently. That is, the separate effects of baryonic physics and summed neutrino mass can be multiplied to estimate the combined effect within a few percent error. Since the weak lensing peak statistics depends on the mass function and on the convergence (Equation \ref{eq:apertureMassKappa}), the S/N peak counts should behave similarly. 
Furthermore, because baryonic physics and summed neutrino mass alter the halo mass function and matter clustering with a different dependence on redshift \citep[see Figures 1 and 13 of][]{Mummery2017}, this can potentially be used to distinguish and quantify the effects when considering S/N peak counts in tomographic redshift bins.
The redshift dependence of the peak counts could be explored within the framework of the halo model using modifications to haloes to account for baryonic physics and modelling the impact on their formation histories as a result of massive neutrinos \citep[e.g.][]{Hamana2004, Massara2014, Semboloni2011, Mead2016}.
Another possibility is to harness the differential impact of baryonic physics and massive neutrinos on halo mass density profiles, as shown in \citet{Mummery2017}. Choosing different filter shapes and sizes should then have an impact on the detection of peaks as a function of baryonic physics and massive neutrinos. 
For example, the optimisation of S/N for specific objects depends on the filter and its parameters. In this work we use a filter optimised for NFW haloes \citep[Equation \ref{eq:NFWFilterFunction} as given by][]{Schirmer2007}. Filters can also be adjusted to better study haloes of particular masses and redshifts \citep[see e.g.][]{Hetterscheidt2005}.
A detailed treatment of these topics is beyond the scope of this paper and will be considered in future work.

The impact of \textit{WMAP}~9 and \textit{Planck}~2015 cosmologies for $M_{\nu}$ = 0.00, 0.06, 0.12, 0.24, and 0.48 eV summed neutrino mass models (all with fiducial baryonic physics) are compared with \textit{WMAP}~9 $M_{\nu}$ = 0.00 eV, as a reference model, in Figures \ref{fig:Planck_KiDS} and \ref{fig:Planck_LSST}, for 9 and 30 gal/arcmin$^2$ respectively. Note that we exclude the \textit{WMAP}~9 $M_{\nu}$ = 0.48 eV model as it is well outside of the range of the \textit{Planck}~2015 models (see Figures \ref{fig:Mnus_KiDS} and \ref{fig:Mnus_LSST}). 
The peak counts with the \textit{Planck}~2015 cosmology for the $M_{\nu}$ 0.06 to 0.24 eV models also show a positive (negative) difference for the higher (lower) S/N peaks, with respect to the \textit{WMAP}~9 $M_{\nu}$ = 0.00 eV reference model. Furthermore the \textit{Planck}~2015 $M_{\nu}$ = 0.24 and 0.48 eV models clearly bracket the \textit{WMAP}~9 $M_{\nu}$ = 0.00, 0.06, and 0.12 eV models for both KiDS and DGB/SB.
This suggests degeneracy between summed neutrino mass and other cosmological parameters in the framework of fiducial baryonic physics. Importantly the S/N peak counts are roughly ordered with respect to the $S_8$ values ($S_8 = \sigma_8 \sqrt{\Omega_\mathrm{m} / 0.3}$, see Table \ref{tab:cosmological_parameters}), in agreement with \citet{Martinet2018}.
The BAHAMAS simulations were carried out with initial amplitude of density fluctuations normalised to the CMB, rather than normalised to have the same $\sigma_{8}$ today. However, using simulations with massive neutrinos and dark matter \citet{Costanzi2013} found that when simulations were normalised to have the same $\sigma_{8}$ today (with the same value of $\Omega_{\rm m}$), even for a very high value of summed neutrino mass the halo mass function is very close to that of a dark matter only simulation, out to $z=1$.

For DGB/SB the weak lensing peak counts with $M_\nu=0.12$ eV for \textit{WMAP}~9 and \textit{Planck}~2015 have an absolute maximum relative difference of $\sim$5 and $\sim$10 percent respectively (up to S/N of 5) compared with the \textit{WMAP}~9 zero neutrino mass model. 
As can be seen from Figure \ref{fig:Planck_LSST}, the differences between the peak counts for suites with summed neutrino masses 0.06 and 0.12 eV inside the \textit{WMAP}~9 and \textit{Planck}~2015 cosmologies are smaller than the differences between the counts across the cosmologies.
However for higher summed neutrino mass, models across cosmologies but with similar $S_8$ values have peak counts that are difficult to distinguish. For example the \textit{WMAP}~9 $M_\nu=$0.00 (0.24) eV model has similar peak counts to the \textit{Planck}~2015 $M_\nu=$0.24 (0.48) eV model.

\begin{figure*}
     \centering
     \begin{subfigure}[b]{\columnwidth}
         \centering
         \includegraphics[width=\columnwidth]{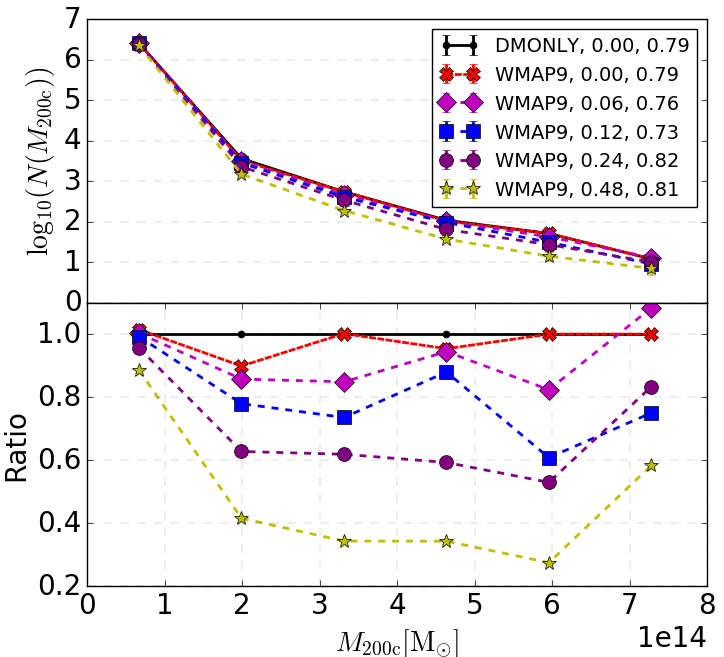}
         \label{fig:HMF_varyNeutrinos}
     \end{subfigure}
     \hfill
     \begin{subfigure}[b]{\columnwidth}
         \centering
         \includegraphics[width=\columnwidth]{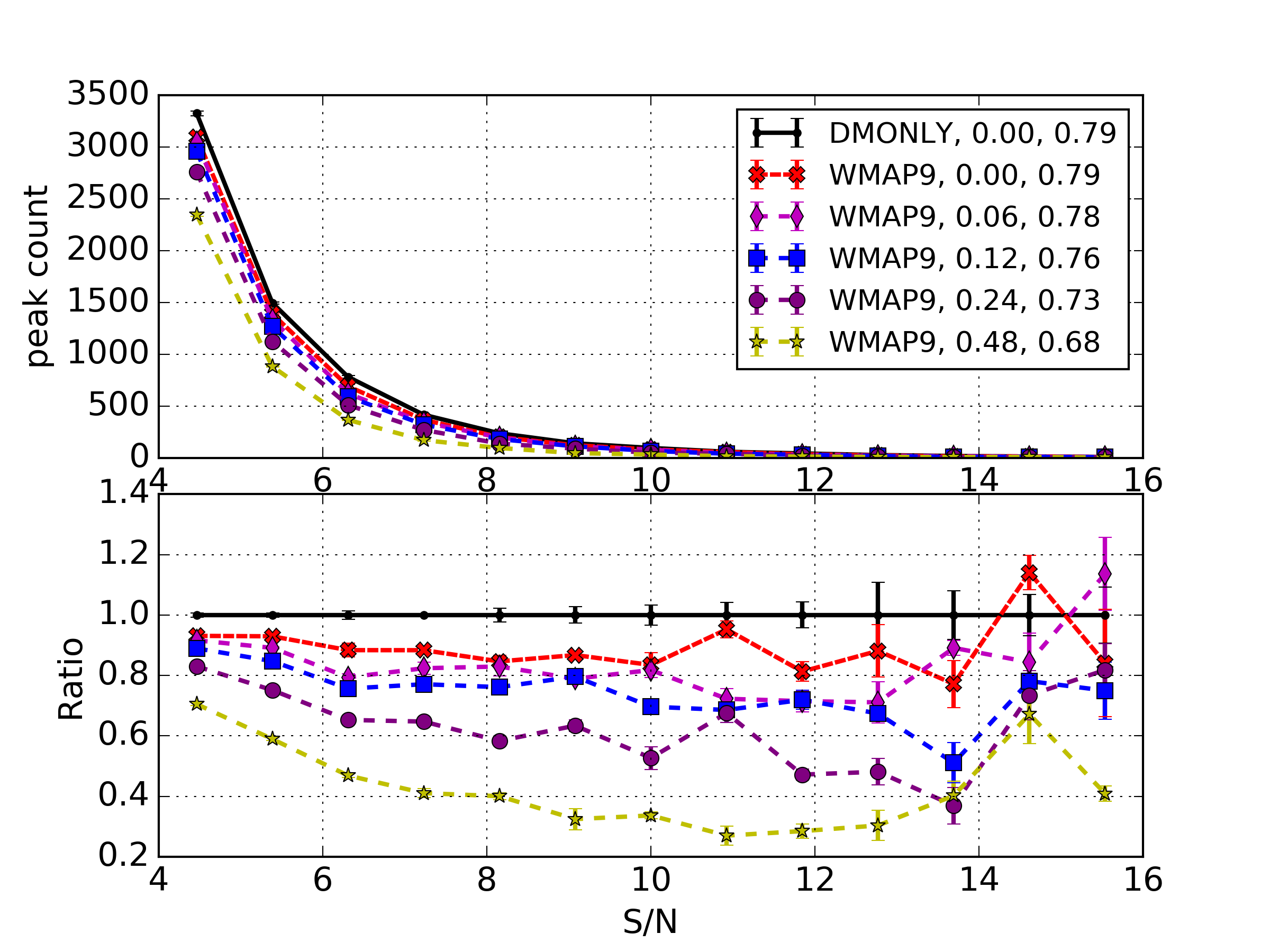}
         \label{fig:SNR_varyNeutrinos}
     \end{subfigure}
        \caption{Impact of summed neutrino mass with underlying fiducial baryonic physics  ($M_\nu$ = 0.00, 0.06, 0.12, 0.24, and 0.48 eV models, all with fiducial baryonic physics, see Table \ref{tab:cosmological_parameters}). The left panel shows the halo mass function (plotted for redshifts out to $z=0.9$) and the right panel shows the S/N peak count distributions, compared with the DMONLY model (collisionless dynamics). This is shown for the DGB/SB (30 gal/arcmin$^2$) survey based on \textit{WMAP}~9 cosmology. The legend is labelled as Cosmology, $M_\nu$ (eV), and $S_{8}$. The comparisons in the bottom panels are with respect to the DMONLY (collisionless dynamics) model. The error bars show the estimated error using Poisson noise ($\sqrt{N(M_{200c}}$) on the left panel and the variance of the five different shape noise realisations on the right panel. In the figures we extend the S/N from 4 to 16 in 13 bins in order to assess the impact on very high S/N peaks, which are more likely produced by higher mass haloes. 
        }
        \label{fig:neutrinoImpact} 
\end{figure*}
Figure 1 in \citet{Mummery2017} shows that the halo mass function is well represented by a product of the separate impacts of baryonic physics (even with extreme AGN feedback strengths) and summed neutrino mass. The Figure shows that the baryonic physics prescription has a larger impact in the $10^{13} - 10^{14} \rm{M_{\odot}}$ range, but becomes more like DMONLY for the high mass range. That is because the AGN feedback ejects matter from the centre of galaxy clusters, but as the cluster mass becomes larger the ejected matter is more likely to stay bound with the cluster. The figure also shows that increasing summed neutrino mass tends to suppress the halo mass function, with greater effect for higher mass bins.
The combined effects can be seen in the left panel of Figure \ref{fig:neutrinoImpact} (plotted for redshifts out to $z=0.9$\footnote{The distribution amplitudes grew for higher redshift ranges, but the ratios remained mostly the same.}). The error bars are estimated with the Poisson noise ($\sqrt{N(M_{200c}}$). Note that the error bars are small.
The impact of summed neutrino mass, all with fiducial baryonic physics, on the halo mass function is compared with the DMONLY model (collisionless dynamics). As the summed neutrino mass increases, the halo mass function becomes further suppressed toward the middle of the distribution. But the distribution is more similar to DMONLY for higher mass bins. This reflects the independent behaviour described in \citet{Mummery2017}, where the separate impacts are multiplicative.

In the right panel of Figure \ref{fig:neutrinoImpact} we show how summed neutrino mass with fiducial baryonic physics impacts on the S/N peaks, over a wide range of S/N values. 
In the figures we extend the S/N from 4 to 16 in 13 bins in order to assess the impact on very high S/N peaks.
The high S/N values are more likely produced by massive clusters and should reflect the behaviour seen in the halo mass function (left panel of Figure \ref{fig:neutrinoImpact}). 
This shows how increasing summed neutrino mass with fiducial baryonic physics suppresses the S/N peaks, but the peak counts turn over and tends toward DMONLY for the higher S/N bins. This can be explained by the combined impact of baryonic physics and summed neutrino mass on the halo mass function. Note that the amplitudes of the peak counts are different from Figure \ref{fig:Mnus_LSST} due to wider binning and that the bins are spaced linearly in S/N with the markers indicating the bin centres.

\begin{figure*}
     \centering
     \begin{subfigure}[b]{\columnwidth}
         \centering
         \includegraphics[width=\columnwidth]{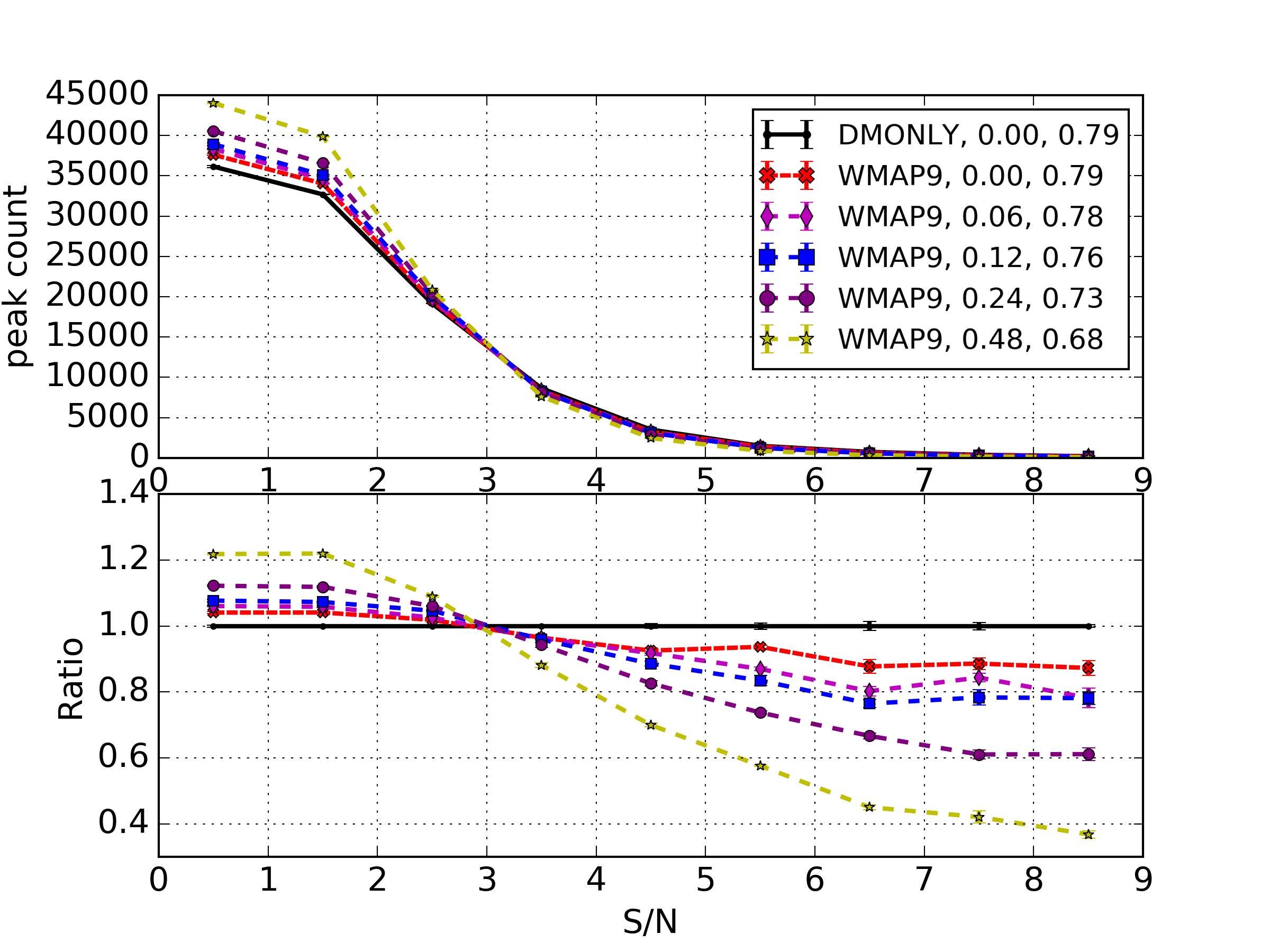}
         \caption{30 gal/arcmin$^2$}
         \label{fig:Mnus_30gals}
     \end{subfigure}
     \hfill
     \begin{subfigure}[b]{\columnwidth}
         \centering
         \includegraphics[width=\columnwidth]{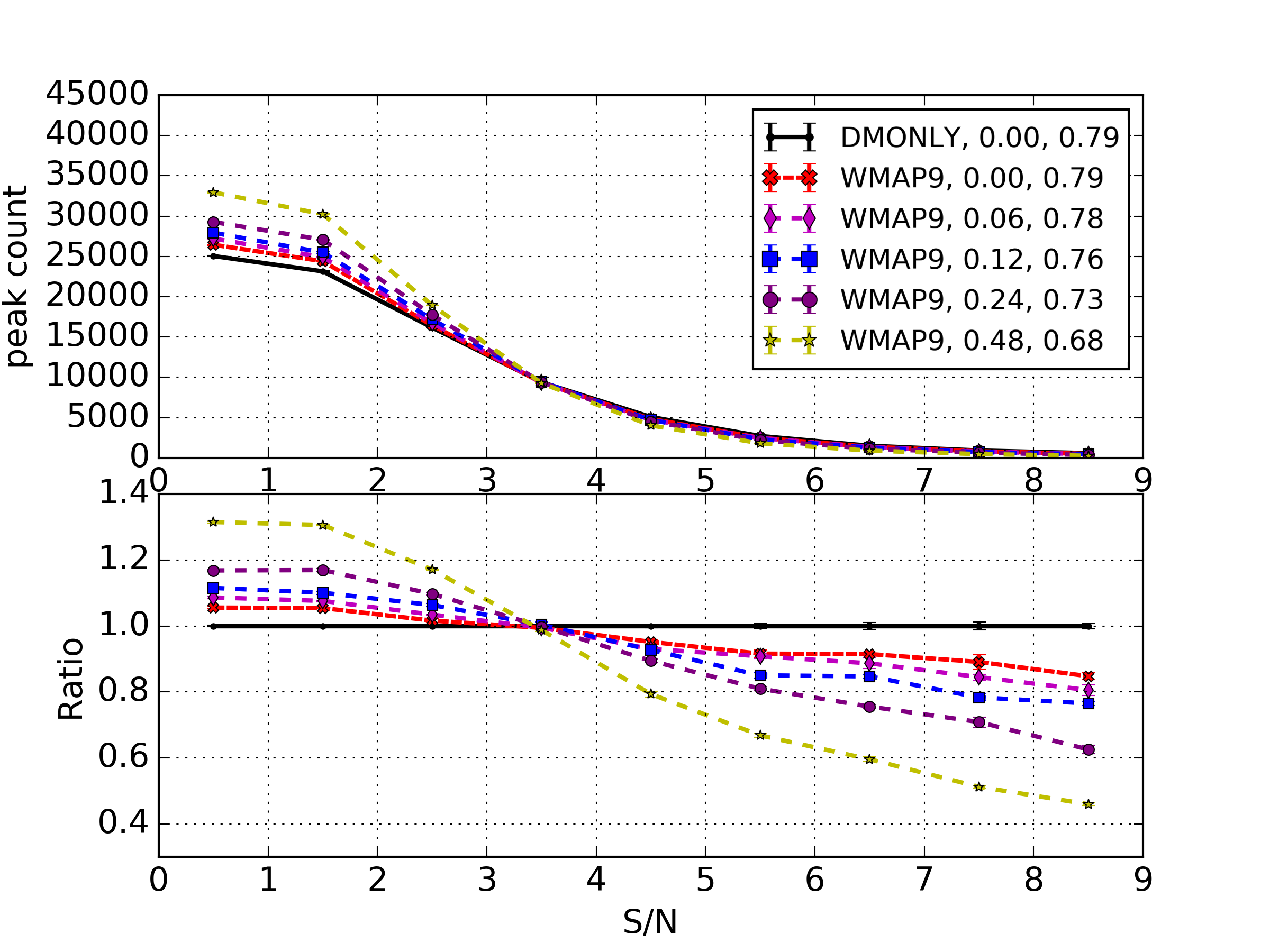}
         \caption{60 gal/arcmin$^2$}
         \label{fig:Mnus_60gals}
     \end{subfigure}
        \caption{Impact of baryonic physics and neutrino mass, $M_\nu$ = 0.00, 0.06, 0.12, 0.24, and 0.48 eV models, all with fiducial baryonic physics, on the S/N peak distributions from DGB/SB (30 gal/arcmin$^2$) and DSB (60 gal/arcmin$^2$) surveys based on \textit{WMAP}~9 cosmology. The legend goes as Cosmology, $M_{\nu}$ (eV), and $S_{8}$. The comparisons in the bottom panels are with respect to the DMONLY (collisionless dynamics) model with zero neutrino mass. The error bars show the variance of the five different shape noise realisations. In this figure we extend the S/N from 0 to 10 in 9 bins in order to assess the impact at high S/N. 
        Note that the amplitudes of the peak counts are different from Figures \ref{fig:Mnus_KiDS} and \ref{fig:Mnus_LSST} due to wider binning. 
        }
        \label{fig:snrmax10} 
\end{figure*}

Figure \ref{fig:snrmax10} illustrates how the numbers of the peak counts change with source number densities of 30 and 60 gal/arcmin$^2$, DGB/SB and DSB survey weak lensing characteristics respectively (both use the same redshift distribution when constructing the lensing convergence). In this figure we extend the S/N from 0 to 10 in 9 bins, with the markers indicating the bin centres, in order to assess the impact at high S/N. 
Note that the amplitudes of the peak counts are different from Figures \ref{fig:Mnus_KiDS} and \ref{fig:Mnus_LSST} due to wider binning. 
The peak counts of these models are compared with those of DMONLY. Higher source number density tends to offset the impact of shape noise, suppressing the number of lower S/N peak counts and boosting the higher S/N peak counts, while also decreasing the scatter in peak counts from field to field. 
This can be explained by generating a shape-noise-only field, where the field-of-view is populated by galaxies with ellipticities drawn from a Gaussian distribution (with appropriate number density) and with no foreground structure \citep[as in the noise-only fields in][]{Martinet2018}. Adding real foreground structure from the light-cones increases (decreases) the number of higher (lower) S/N values. Our results highlight that accounting for the impact of baryonic physics and massive neutrinos in observational surveys must take the source number density into consideration. 
Furthermore a given real structure is more likely to have a higher corresponding S/N value with increasing source number density. We discuss this further in Section \ref{sec:highSNR}. 

\begin{figure}
    \centering
    \includegraphics[width=\columnwidth]{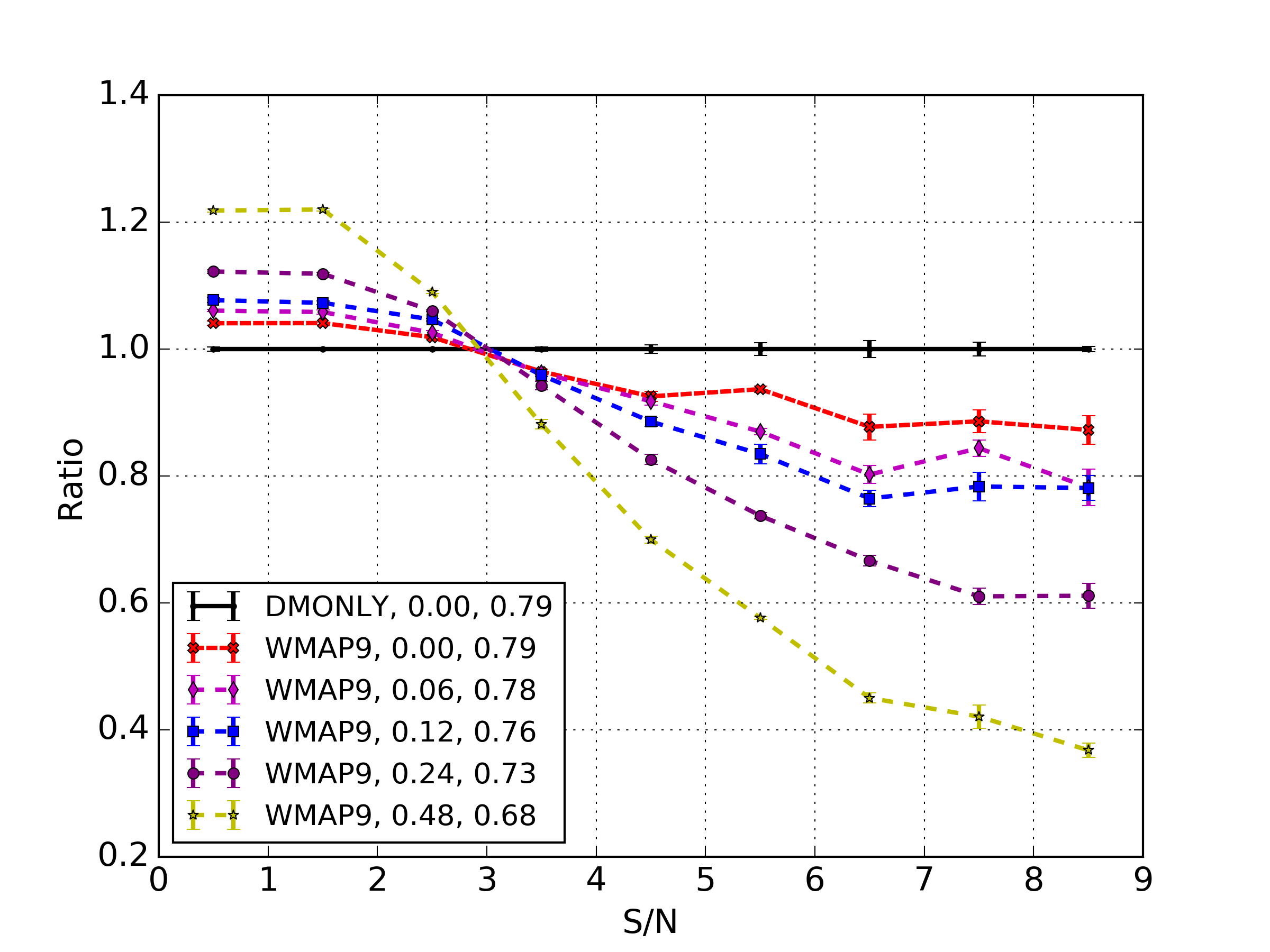}
    \caption{Relative percentage differences in peak counts with respect to the DMONLY peak count distribution (Figure \ref{fig:Mnus_30gals}). For the DGB/SB survey, five models ($M_{\nu} =$ 0.00, 0.06, 0.12, 0.24, and 0.48 eV, all with fiducial baryonic physics) of the \textit{WMAP}~9 cosmology are compared with the DMONLY model in order to determine percentage differences in each interval of S/N. The legend is labelled as Cosmology, $M_\nu$ (eV), and $S_{8}$.}
    \label{fig:percentDiff}
\end{figure}

\begin{table}
    \caption{The relative percentage differences in S/N peak counts for the DGB/SB survey (these values are taken from Figures \ref{fig:Mnus_30gals} and \ref{fig:percentDiff}). The values are in units of percent (\%) and compare the five different simulations ($M_{\nu}=$ 0.0, 0.06, 0.12, 0.24, and 0.48 eV, all with fiducial baryonic physics) per S/N bin with the DMONLY simulation.}
    \resizebox{\columnwidth}{!}{
    \begin{tabular}{c|c|c|c|c|c}
        \hline
         & \multicolumn{5}{c | }{$M_\nu$ [eV]} \\
         S/N & 0.00 & 0.06 & 0.12 & 0.24 & 0.48\\
         \hline
         0-1 & 4.1 & 6.1 & 7.7 & 12.2 & 21.8 \\
         1-2 & 4.1 & 5.8 & 7.3 & 11.9 & 22.0 \\
         2-3 & 1.9 & 2.6 & 4.6 & 6.0 & 9.0 \\
         3-4 & -3.6 &   -3.8 &   -4.1 & -5.7 & -11.9 \\
         4-5 & -7.4 &   -8.2 &  -11.4 & -17.4 & -30.0 \\
         5-6 & -6.3 &  -13.0 &  -16.5 & -26.3 & -42.4 \\
         6-7 & -12.3 &  -19.8 &  -23.5 & -33.3 & -53.0 \\
         7-8 & -11.4 &   -15.6 &  -21.7 & -39.0 & -56.0 \\
         8-9 & -12.7 &  -21.8 &  -21.9 & -38.9 & -63.2 \\
         9-10 & -16.2 &  -15.4 &  -26.8 & -42.3 & -67.5 \\
         \hline
    \end{tabular}
    \label{tab:varyMnu_SNRpercentDiff}
    }
\end{table}

Figure \ref{fig:percentDiff} (corresponding values listed in Table \ref{tab:varyMnu_SNRpercentDiff}) compares the peak counts in \textit{WMAP}~9 $M_{\nu}=$ 0.0, 0.06, 0.12, 0.24, and 0.48 eV models (all with fiducial baryonic physics) to DMONLY, for DGB/SB data. 
Assuming that baryonic physics and massive neutrinos act independently \citep{Mummery2017}: at lower summed neutrino mass, 0.06 and 0.12 eV, the impact of fiducual baryonic physics on the peak counts tends to be greater than that of massive neutrinos up to S/N $\sim5$; at higher summed neutrino mass, 0.24 and 0.48 eV, the presence of massive neutrinos tends to be more important than fiducial baryonic physics. These conclusions are also sensitive to the source number density and intrinsic galaxy ellipticity dispersion.

\subsection{Results on High S/N Weak Lensing Peaks}
\label{sec:highSNR}
In the previous subsection we compared our results to the DMONLY model for consistency with \citet{Mummery2017}, where it was shown that baryons and neutrinos impact on the halo mass function independently and that their effects are multiplicative. In this subsection we focus on the high S/N peaks for the \textit{Planck}~2015 cosmology with fiducial baryonic physics and varying neutrino mass, where the implementation of neutrinos in the simulations impact on other cosmological parameters other than $\Omega_{\rm cdm}$.
We discuss: 
(1) How summed neutrino mass affects the high mass end of the halo mass function and also give some examples of the impact on the mass of specific clusters.
(2) S/N peak dependence on survey characteristics and noise.
(3) How filter size impacts the detection of high S/N peaks.
(4) Using cluster positions to find nearby S/N peaks to study the correlation between cluster mass and S/N peaks. 
This would be analogous to targeting clusters (for example known clusters or clusters selected using another technique) and measuring their aperture mass signal.
(5) Using S/N peak locations to find nearby clusters, analogous to carrying out a blind weak lensing survey.

We use the FoF catalogues, containing halo information within the light-cones, from the simulation selecting objects with masses $M_{200c}\geq10^{14} \rm{M_\odot}$ and $z \leq 0.9$. Clusters are identified by running a FoF algorithm on the full simulation snapshot data. Spherical overdensity masses for the FoF groups are calculated using the SUBFIND algorithm. An overdensity of 200 with respect to the critical density at the snapshot redshift of the FoF group is used (i.e., $M_{\rm 200c}$).

In the top panel of the left panel of Figure \ref{fig:neutrinoImpact} we plot the number counts of clusters for each neutrino mass suite for the \textit{WMAP}~9 cosmology. 
For the bottom panel we take the ratios with respect to the $M_{\nu}=$ 0.06 eV number counts. 
The left panel of Figure \ref{fig:neutrinoImpact} shows that when the summed neutrino mass increases there are progressively fewer massive haloes of a particular mass. Comparing the number counts to the $M_{\nu}=$ 0.06 eV model, the trends with increasing summed neutrino mass are similar for both \textit{WMAP}~9 and \textit{Planck}~2015 cosmologies (not plotted here).
These results are consistent with the findings of e.g., \citet{Costanzi2013, Castorina2014, Mummery2017, Hagstotz2018}.

The reason for this trend is that neutrinos can free-stream out of overdense regions which inhibits the growth of structure. Increasing the mass in the neutrino component means that a larger fraction of the total mass can free-stream out.
We also studied the impact of summed neutrino mass on cluster shapes by analysing their moment of inertia tensors, finding that their axis ratios have no significant change but the overall sizes are altered. This will be a subject of a future paper.

\begin{figure*}
    \includegraphics[width=\columnwidth]{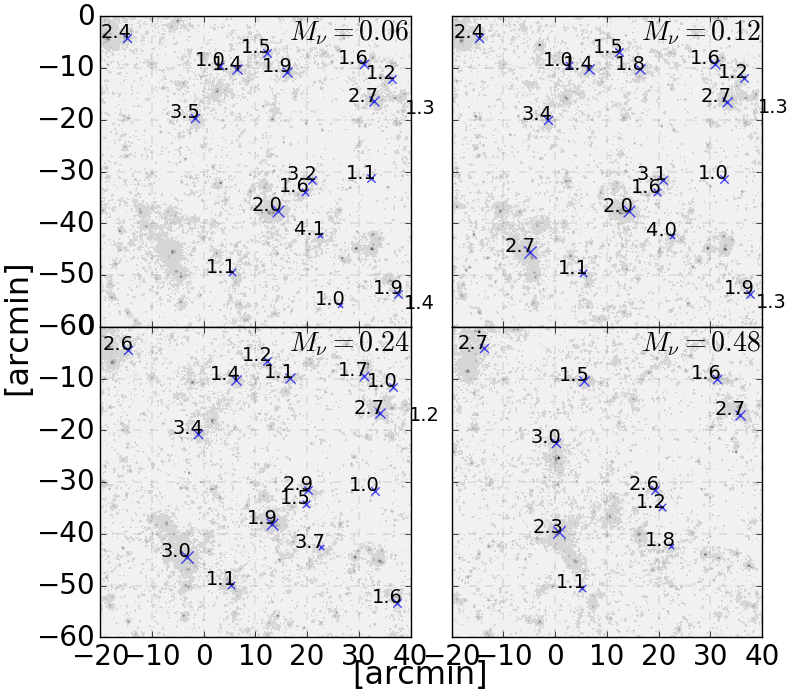}
    \caption{The locations of clusters on the convergence maps for different summed neutrino mass. This is the same field-of-view for four different simulations in the \textit{Planck}~2015 cosmology with $n_{\rm{eff}} = 9$ gal/arcmin$^{2}$. The four panels have $M_{\nu}$ = 0.06, 0.12, 0.24, and 0.48 eV for the top left, top right, bottom left, and bottom right panels respectively. The grey scale is the convergence map and $\times$s are the cluster locations. The smaller the $\times$ the higher the cluster redshift. The values next to the $\times$s are the cluster masses in units of $10^{14}\,\rm{M_\odot}$.
    }
    \label{fig:plotAllDiffMnu}
\end{figure*}
Figure \ref{fig:plotAllDiffMnu} shows the spatial distribution of massive clusters in a $1 \times 1$ deg$^{2}$ field-of-view for four different simulations, where they only differ by neutrino mass. 
The four panels are $M_{\nu}$ = 0.06, 0.12, 0.24, and 0.48 eV for the top left, top right, bottom left, and bottom right panels respectively. 
The grey scale is the convergence map and $\times$s are the cluster locations. 
The smaller the $\times$ the higher the cluster's redshift. 
The values next to the $\times$s are the cluster masses in units of 
$10^{14}\,\rm{M_{\odot}}$.
This shows that with increasing summed neutrino mass most cluster masses decrease 
(see the left panel of Figure \ref{fig:neutrinoImpact}), though not all. Note that we only show cluster masses $M_{200c} \geq 10^{14}\,\rm{M_{\odot}}$, and some clusters drop below this limit.


\subsubsection{Noise realisations}
\label{sub:noiserealisations}
We study a total of five shape noise realisations of weak lensing on $625$ deg$^{2}$, for the \textit{Planck}~2015 cosmology with $M_\nu = 0.06$ eV. As described above, each noise realisation has a different set of random seeds for intrinsic galaxy ellipticities (see Section \ref{sec:lightCones}). The galaxy positions are the same for each run. In this section we consider cluster-mass objects and high peak values (S/N $ \geq 3$).

\begin{figure*}
     \centering
     \begin{subfigure}[b]{\columnwidth}
         \centering
         \includegraphics[width=\columnwidth]{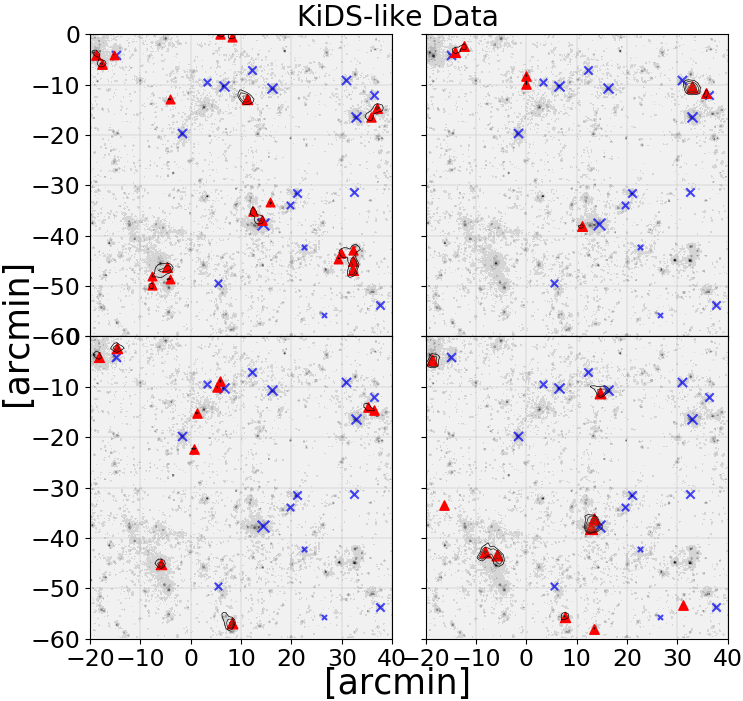}
         \caption{9 gal/arcmin$^2$}
         \label{fig:plotAllDiffRuns}
     \end{subfigure}
     \hfill
     \begin{subfigure}[b]{\columnwidth}
         \centering
         \includegraphics[width=\columnwidth]{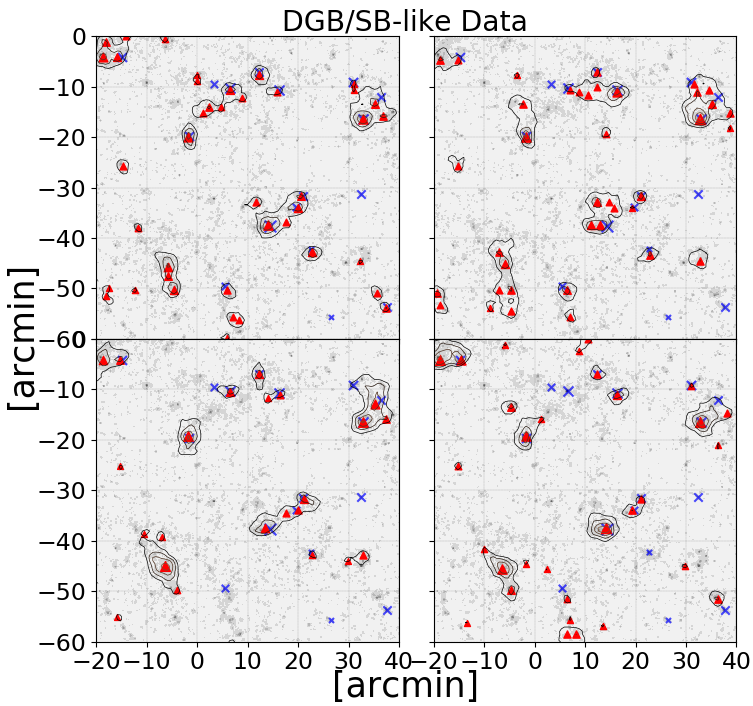}
         \caption{30 gal/arcmin$^2$}
         \label{fig:LSST_plotAllDiffRuns}
     \end{subfigure}
        \caption{The correspondence between the locations of S/N peaks and massive dark matter haloes for four sets of shape noise realisations for KiDS and DGB/SB surveys. 
        The grey scale is the true noiseless convergence map of the simulation. The contours and triangles are the aperture mass S/N and S/N peaks ($\geq 3$) respectively. These are calculated for source shape noise $\sigma_{|\epsilon|} = 0.25$. The $\times$ markers are the true cluster locations from the halo finder catalogue. 
        A smaller triangle or $\times$ corresponds to a lower S/N peak value or a higher cluster redshift, respectively. All four panels show the same field-of-view ($1 \times 1$ deg$^{2}$) with the \textit{Planck}~2015 cosmology and $M_{\nu}$ = 0.06 eV. 
        }
        \label{fig:survey_plotAllDiffRuns}
\end{figure*}
For KiDS (DGB/SB) data with $n_{\rm{eff}} =$ 9 (30) gal/arcmin$^{2}$ Figure \ref{fig:plotAllDiffRuns} (Figure \ref{fig:LSST_plotAllDiffRuns}) shows the impact of different shape noise realisations on the aperture mass maps and S/N peaks. 
Each panel shows the same $1 \times 1$ deg$^{2}$ field-of-view.
The grey scale and $\times$ markers are the convergence maps and cluster locations respectively. 
The line contours and triangles are the aperture mass S/N and S/N peaks ($\geq 3$) respectively.
A smaller triangle or $\times$ corresponds to a lower S/N peak value or a higher cluster redshift, respectively.
$\times$ markers and the convergence maps do not change between noise realisations because they are taken directly from the BAHAMAS simulation.
Figures \ref{fig:plotAllDiffRuns} and \ref{fig:LSST_plotAllDiffRuns} 
show that aperture mass S/N maps and therefore peak locations depend heavily on the noise realisation, but are more consistent for a higher effective number density of source galaxies.
The match between S/N peaks and cluster locations is in rough agreement with \citet{Hamana2004, Yang2011, Liu2016a, Martinet2018}, where peaks are not always associated with clusters.


\subsection{Filter Sizes}
\label{sec:filterSizes}
Even though we use a fiducial filter size of 12.5 arcmin \citep[see][]{Martinet2018} throughout most of this paper, in this subsection we briefly illustrate the well-studied impact of filter size \citep[e.g.][]{Hetterscheidt2005, Schirmer2007, Martinet2018}. We focus on the 0.06 eV summed neutrino mass model. 
In practice for a real observational survey the filter size and form would be adjusted depending on the noise properties of the survey and the science goals.


\begin{figure}
    \includegraphics[width=\columnwidth]{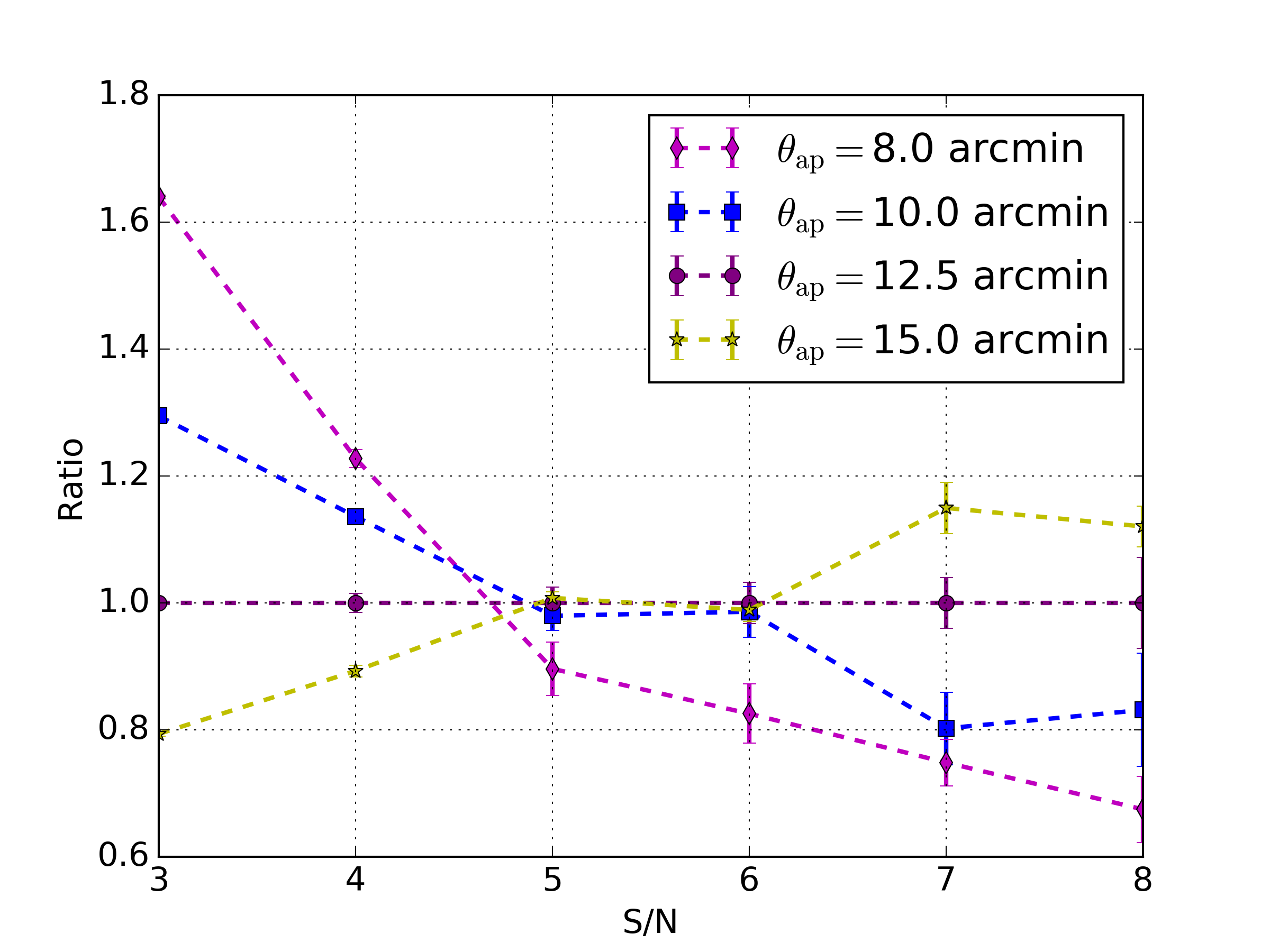}
    \includegraphics[width=\columnwidth]{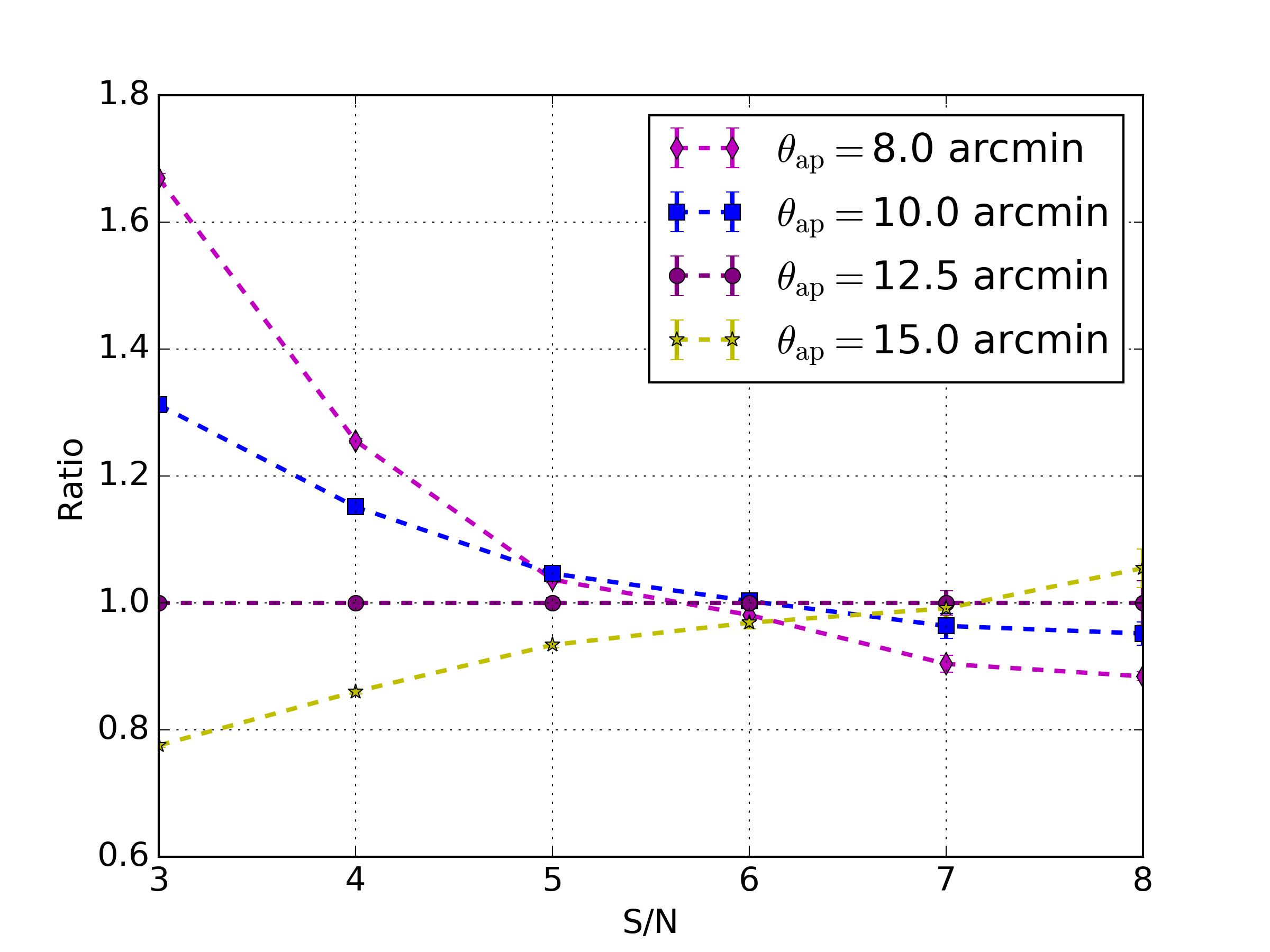}
    \caption{The ratio of the S/N distributions for various aperture sizes for KiDS and DGB/SB, top and bottom panels respectively, with respect to $\theta_{\rm{ap}} = 12.5$ arcmin. The lines represent $\theta_{\rm{ap}} = 8.0, 10.0, 12.5,$ and $15.0$ arcmin.
    The error bars show the variance of the five different shape noise realisations. The model used here was \textit{Planck}~2015 cosmology, with fiducial baryonic physics and 0.06 eV summed neutrino mass.
    }
    \label{fig:varyFilterSizeAllConesCountSNRs}
\end{figure}
Figure \ref{fig:varyFilterSizeAllConesCountSNRs} shows how the S/N peak distribution varies with filter size. The lines represent $\theta_{\rm{ap}} = 8.0, 10.0, 12.5$ and $15.0$ arcmin.
The panels show the ratio of the S/N distributions relative to $\theta_{\rm{ap}} = 12.5$ arcmin. The model used here was \textit{Planck}~2015 cosmology, with fiducial baryonic physics and 0.06 eV summed neutrino mass.
This shows that increasing aperture size suppresses the S/N peak counts for lower S/N values while boosting counts for higher values. The larger filter sizes tend to be more sensitive to larger structures (contributing more to the higher S/N values) and less sensitive to smaller structures (suppressing lower S/N counts). 
In other words, the larger filters smooth shape noise, reducing the number of low signal-to-noise peaks.
This is consistent with works which have sought to optimise the aperture mass filter functions or have employed other optimal filtering techniques \citep[e.g.][]{Hennawi2005, Schirmer2007, Maturi2010}.
\citet{Martinet2018} found that a filter size of $12.5$ arcmin maximises the number of peaks above a S/N of $3$ for the KiDS data. 
However our synthetic catalogues are not constructed in exactly the same manner. 
For example we have a different intrinsic shape noise and the optimal filter size can also change as a function of number density (see Figures \ref{fig:Mnus_30gals} and \ref{fig:Mnus_60gals}).

Neutrinos and baryons have different effects on the core of haloes which could be discriminated through a filter size and shape optimization. For examples of different filter sizes see the Appendix.


\subsection{Summed Neutrino Mass and S/N vs. Cluster Mass}
\label{sec:summedNeutrinoMass}
In Section \ref{s:results} we discussed how summed neutrino mass impacts the S/N peak distribution, in particular we found that increasing summed neutrino mass suppresses the high S/N peaks. 
In the beginning of this section (Section \ref{sec:highSNR}) we discussed how increasing summed neutrino mass tends to decrease the masses of clusters, consistent with the authors mentioned above.
In this subsection we study the relationship between S/N peak values and cluster mass, and the dependence on summed neutrino mass.

First we study the correlation between S/N peaks and cluster locations.
Note that given the redshift distribution of KiDS, it is not likely that clusters with redshifts above $z=0.5$ or so will be easily detected with lensing, 
since the number density of background galaxies from which the shear can be estimated is low \citep{Martinet2018}.
We also limit the lower bound S/N peak values to $3$ for consistency with the regime dominated by clusters rather than LSS in a KiDS-like survey \citep{Martinet2018}.

The clusters and any associated S/N peak locations can then be used to investigate how well S/N peaks trace cluster mass. 
We do this by taking circular areas of radius 
$2.0$ arcmin 
centred on cluster centres to identify nearby S/N peaks. 
Note that there may not be any S/N peaks within the circles (see Figure \ref{fig:survey_plotAllDiffRuns}).
When multiple S/N peaks are enclosed, we allow for two options: either choose the closest S/N peak or the highest value enclosed. 
In this paper we will show only the results for choosing the closest S/N peak to a cluster,\footnote{When choosing the highest S/N peaks some points in the plot of S/N against $M_{200c}$ are shifted to higher S/N values, but the distributions for choosing by closest and highest are mostly the same.} and allow a S/N peak to be chosen by multiple clusters.

\begin{figure*}
     \centering
     \begin{subfigure}[b]{\columnwidth}
         \centering
         \includegraphics[width=\columnwidth]{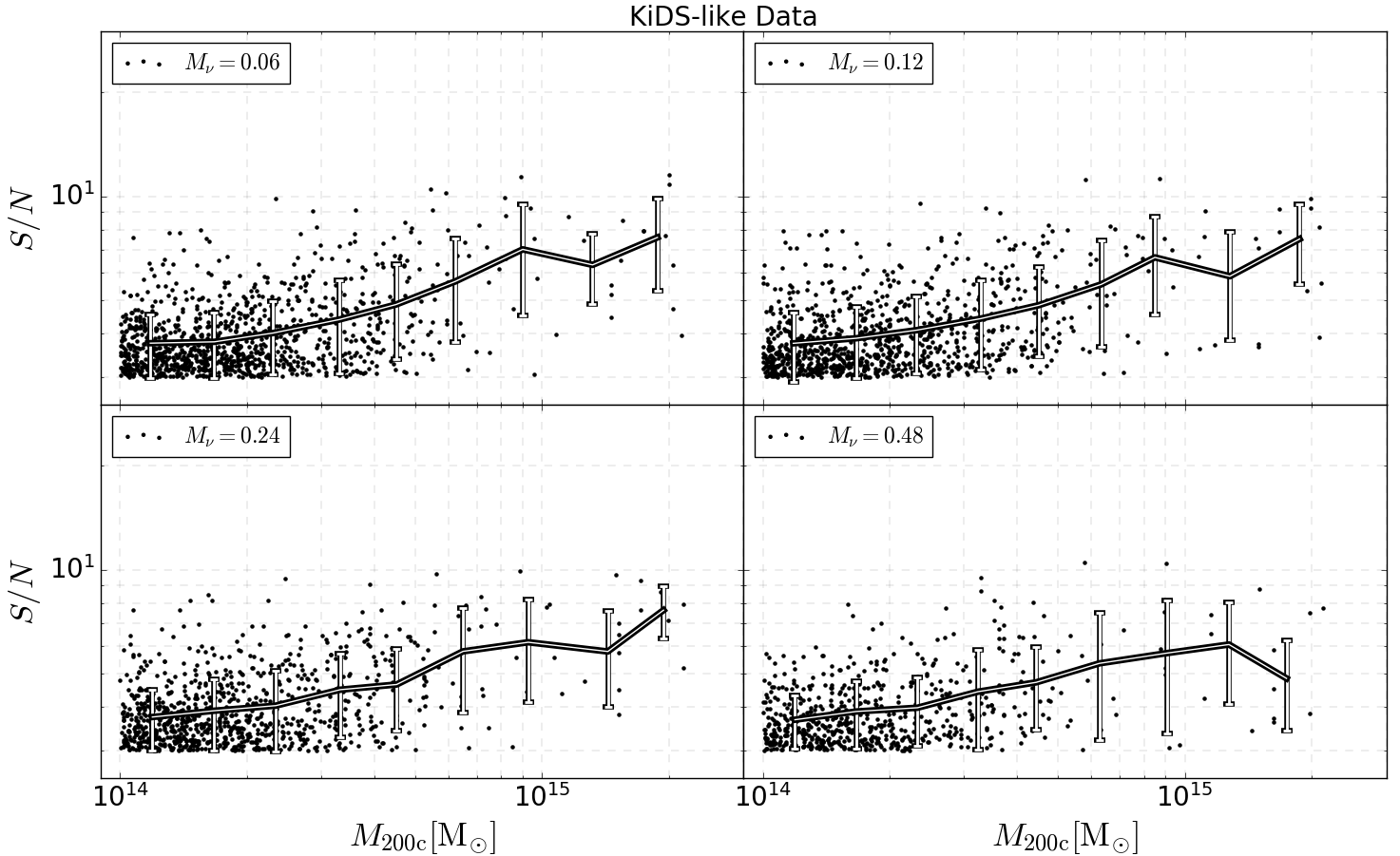}
         \caption{9 gal/arcmin$^2$}
         \label{fig:MvsSNR}
     \end{subfigure}
     \hfill
     \begin{subfigure}[b]{\columnwidth}
         \centering
         \includegraphics[width=\columnwidth]{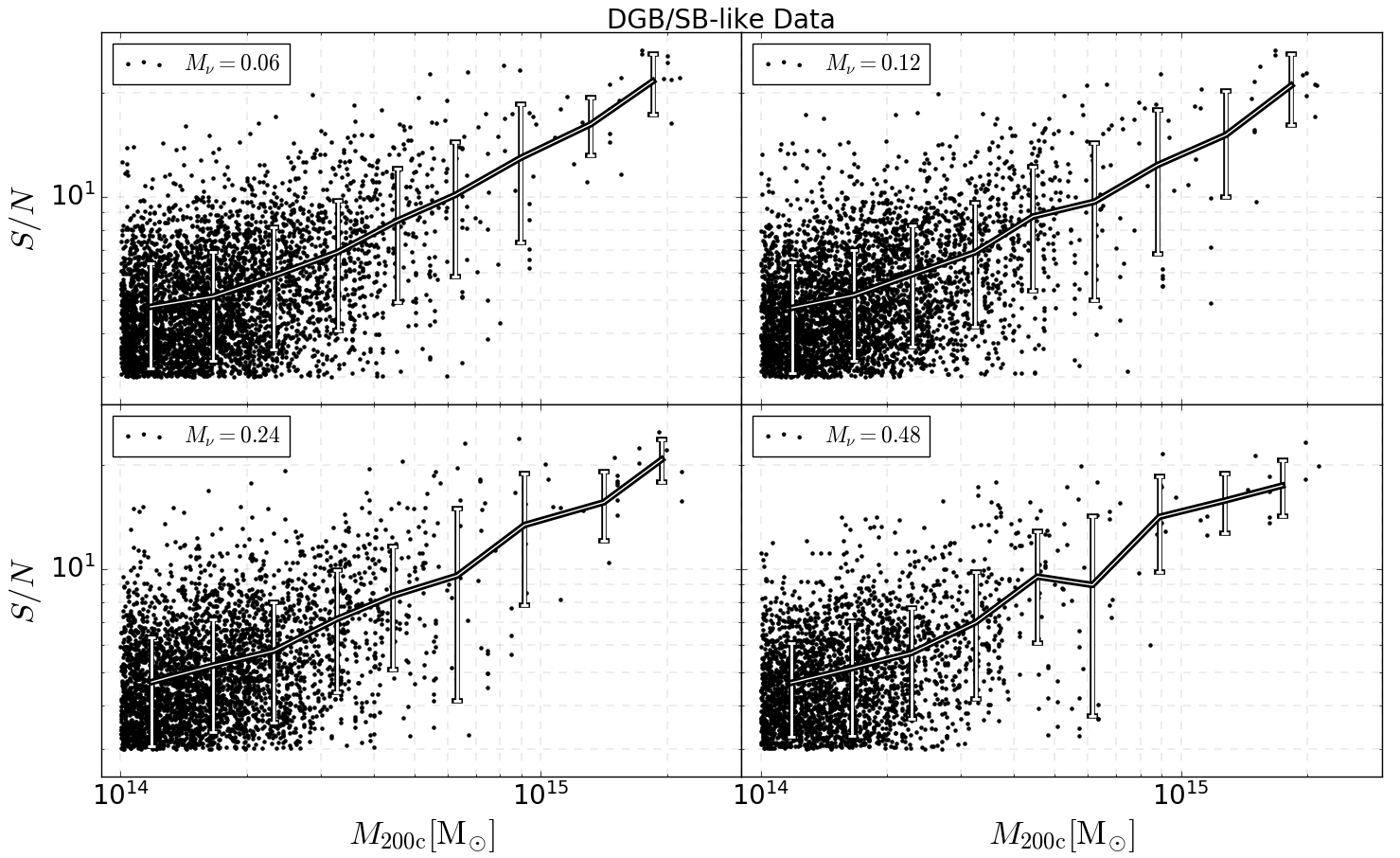}
         \caption{30 gal/arcmin$^2$}
         \label{fig:LSST_MvsSNR}
     \end{subfigure}
        \caption{The S/N peaks vs. $M_{200c}$ for different summed neutrino mass for KiDS and DGB/SB (left and right panels respectively). 
        Here we use known galaxy cluster centres and find the nearest S/N peak within $2.0$ arcmin.
        The different sub-panels correspond to different summed neutrino mass $M_{\nu}$ = 0.06, 0.12, 0.24, and 0.48 eV. 
        The points are the S/N values plotted against cluster mass, the line shows the mean S/N in logarithmic mass bins, and the error bars show the variance (note that there is a limit at S/N = 3). 
        The Pearson Correlation Coefficient (PCC) is calculated for each sub-panel (see text for details).
        The PCC values for each sub-panel are ($M_{\nu}$, PCC), for KiDS: (0.06 eV, 0.48), (0.12 eV, 0.44), (0.24 eV, 0.45), (0.48 eV, 0.37); and for DGB/SB: (0.06 eV, 0.58), (0.12 eV, 0.57), (0.24 eV, 0.55), (0.48 eV, 0.56).
        }
        \label{fig:survey_MvsSNR}
\end{figure*}
In the case of assuming known cluster locations and choosing the closest S/N peak, in 
Figure \ref{fig:survey_MvsSNR} we plot S/N peak values vs. $M_{200c}$ for the four different summed neutrino masses, $M_{\nu}$ = 0.06, 0.12, 0.24, and 0.48 eV, with the left and right panels showing the results for KiDS and DGB/SB respectively.
The line shows the mean and the error bars show the variance of S/N inside the mass bins (note that there is a limit at S/N = 3, which impacts on the error bars for the lower mass halo bins).
Note that in these panels the clusters are not segregated into redshift bins. This means that the lensing efficiencies are not accounted for when comparing S/N peaks to cluster mass.
The Pearson Correlation Coefficient (PCC) measures the linear relationship between two data sets, where the value can range from $-1$ to $+1$. PCC = 0 means there is no linear correlation and $+1$ ($-1$) means there is a positive (negative) linear relationship.
For KiDS the PCC values for each sub-panel are ($M_{\nu}$, PCC): (0.06 eV, 0.48), (0.12 eV, 0.44), (0.24 eV, 0.45), (0.48 eV, 0.37); and for DGB/SB: (0.06 eV, 0.58), (0.12 eV, 0.57), (0.24 eV, 0.55), (0.48 eV, 0.56).

To check the noise properties we have also taken the galaxy cluster positions from the FoF catalogue and compared these with random positions generated to have the same number as the S/N peaks and assigned S/N values from the distribution shown in Figure \ref{fig:MvsSNR}, for each synthetic survey and summed neutrino mass. Overall the PCC values have means $\sim0$ for both KiDS and DGB/SB survey characteristics.

A deeper survey gives a higher S/N peak value for a given cluster. The correlation between the peak values and mass are not significantly changed as a function of neutrino mass, even though the high end of the S/N peak distribution and the cluster mass function become suppressed with increasing summed neutrino mass (see Figures \ref{fig:Mnus_LSST}, \ref{fig:Mnus_KiDS}, and \ref{fig:neutrinoImpact}). In addition the shape of the filter function will give rise to different weights for lensed galaxies at different distances from the clusters' centres where their tangential distortions will depend on the mass density profile.
Even though the profiles may differ as a function of summed neutrino mass, our results indicate that although corresponding clusters might have lower masses in a higher summed neutrino mass simulation, the values of the S/N are also decreasing when the same filter is used in the measurement. Note that the simulations in these figures all have the same underlying fiducial baryonic physics. 
As discussed in \citet{Mummery2017} the impact of baryonic physics and summed neutrino mass on the mean halo density profiles are different on different scales. 
Accordingly, optimising filter parameters that account for the impacts of baryonic physics and summed neutrino mass on S/N peak measurements is an interesting question which would require a detailed investigation outside the scope of this work.

\begin{figure*}
     \centering
     \begin{subfigure}[b]{\columnwidth}
         \centering
         \includegraphics[width=\columnwidth]{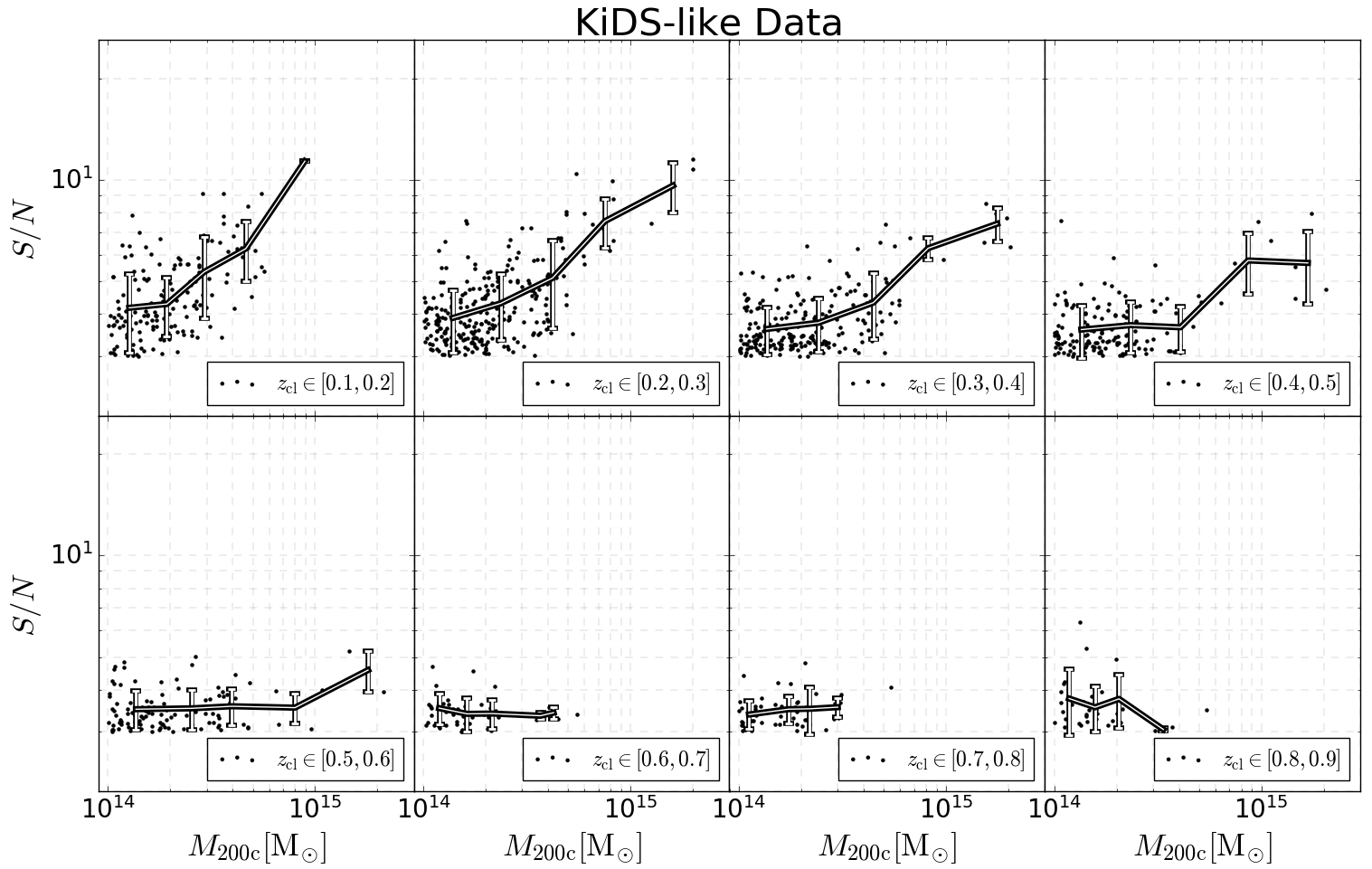}
         \caption{9 gal/arcmin$^2$}
         \label{fig:KiDS_MvsSNR_binZ}
     \end{subfigure}
     \hfill
     \begin{subfigure}[b]{\columnwidth}
         \centering
         \includegraphics[width=\columnwidth]{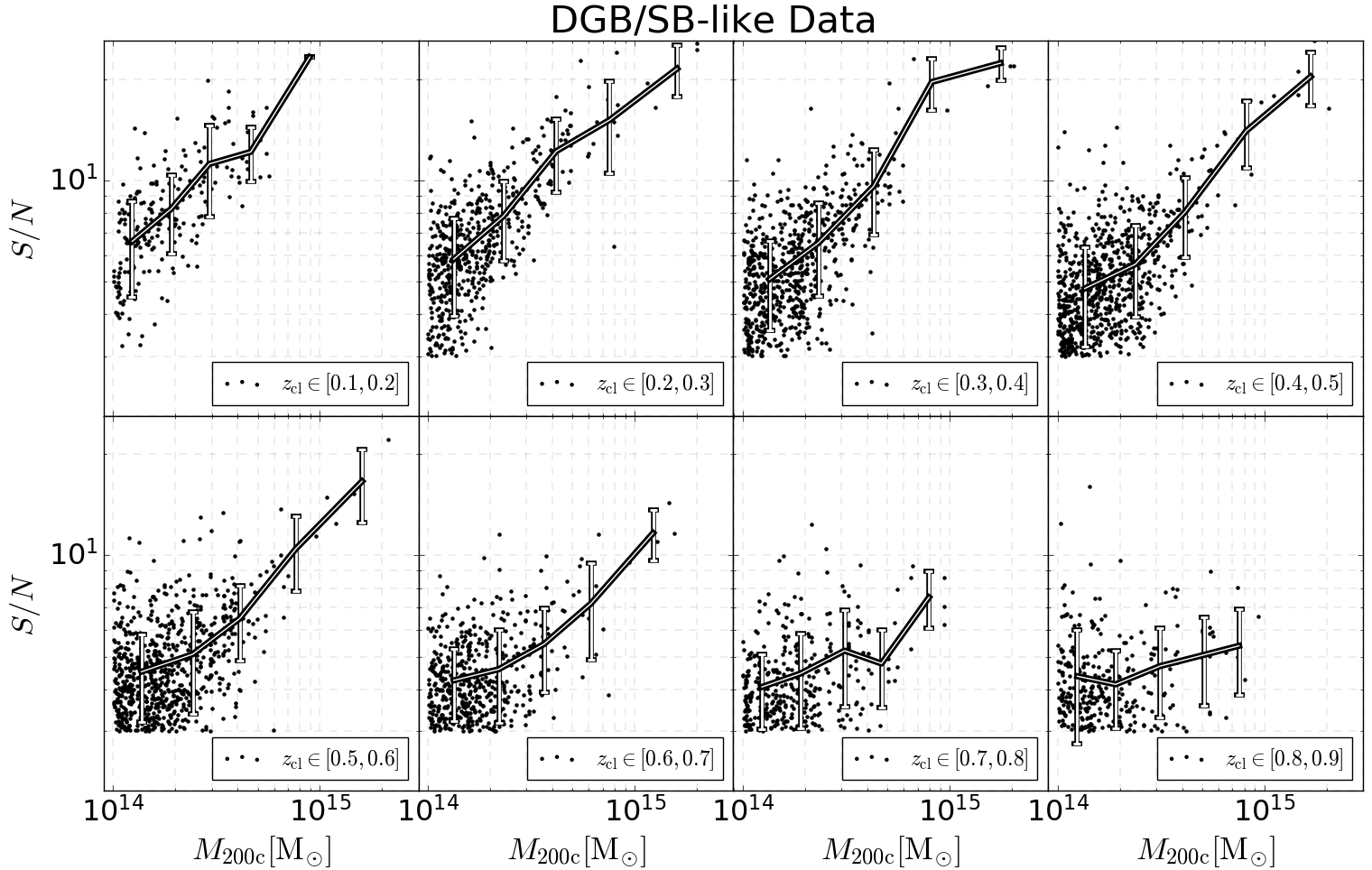}
         \caption{30 gal/arcmin$^2$}
         \label{fig:LSST_MvsSNR_binZ}
     \end{subfigure}
        \caption{The S/N peaks vs. $M_{200c}$ 
        for clusters in different redshift bins for KiDS and DGB/SB (left and right panels respectively). 
        The points are the S/N peak values ($\geq 3$) closest to the massive objects 
        within $2.0$ arcmin; the lines are the average of the S/N values in logarithmic mass bins and the error bars show the variance (note that there is a limit at S/N = 3).
        This plot is for $M_{\nu} = 0.06$ eV.
        The Pearson Correlation Coefficient values are ($z_{bin}$, PCC), for KiDS: ([0.1, 0.2], 0.62), ([0.2, 0.3], 0.71), ([0.3, 0.4], 0.66), ([0.4, 0.5], 0.54), ([0.5, 0.6], 0.21), ([0.6, 0.7], -0.09), ([0.7, 0.8], 0.26), ([0.8, 0.9], -0.20); and for DGB/SB: ([0.1, 0.2], 0.70), ([0.2, 0.3], 0.75), ([0.3, 0.4], 0.74), ([0.4, 0.5], 0.70), ([0.5, 0.6], 0.61), ([0.6, 0.7], 0.56), ([0.7, 0.8], 0.34), ([0.8, 0.9], 0.16).
        }
        \label{fig:survey_MvsSNR_binZ}
\end{figure*}

Figure \ref{fig:survey_MvsSNR_binZ} shows the S/N values vs. $M_{200c}$ for clusters segregated into different redshift bins for KiDS and DGB/SB (left and right panels respectively). The simulations are for summed neutrino mass $M_{\nu} = 0.06$ eV.
The lines are the average of the S/N values in logarithmic mass bins and the error bars show the variance of S/N inside the mass bins (note that there is a limit at S/N = 3, which impacts on the error bars in particular for the higher redshifts).
The PCC values are ($z_{bin}$, PCC), for KiDS: ([0.1, 0.2], 0.62), ([0.2, 0.3], 0.71), ([0.3, 0.4], 0.66), ([0.4, 0.5], 0.54), ([0.5, 0.6], 0.21), ([0.6, 0.7], -0.09), ([0.7, 0.8], 0.26), ([0.8, 0.9], -0.20); and for DGB/SB: ([0.1, 0.2], 0.70), ([0.2, 0.3], 0.75), ([0.3, 0.4], 0.74), ([0.4, 0.5], 0.70), ([0.5, 0.6], 0.61), ([0.6, 0.7], 0.56), ([0.7, 0.8], 0.34), ([0.8, 0.9], 0.16).
We have compared the relationships between galaxy cluster mass and random fields for each synthetic survey and redshift bin, where the S/N peak positions are randomised, similar to the randomisation process above. The PCC values have means $\sim0$ for both KiDS and DGB/SB survey characteristics.
This figure shows that there is not only a greater correlation between S/N peaks and $M_{200c}$ for the lower redshift bins, but that increasing the number density of background sources also has a significant impact on the correlation. 
The redshifts of the clusters enter into the calculation of the lensing convergence, while increasing the number density of background sources tends to increase the aperture mass signal. 
In the highest redshift bins there are fewer sources that are background to the clusters, again decreasing the lensing signal.
The differences in cosmology between the different summed neutrino mass models also have a slight impact on cosmological distance measures, which enter into Equation \ref{eq:simulationConvergence}. 

A similar study can be done by keeping a fixed value of summed neutrino mass while varying baryonic physics prescriptions. In the case of BAHAMAS, the low and high AGN feedback models are extreme cases, but a broader range of mass density profiles could be explored using the halo model.


\subsection{Correlation Between High S/N Peaks and Massive Clusters for Various Source Population Characteristics}
\label{sub:surveySummary}
In this subsection we present results on the correspondence between high S/N peaks and clusters determined for different source population characteristics (KiDS, DGB/SB, and DSB) with $M_{\nu} = $ 0.06 eV. For particular survey characteristics, ray-tracing through simulations with a reasonable prescription for baryonic physics and summed neutrino mass can be used to establish the correlation between S/N peaks and clusters or other features in the LSS.


\begin{figure}
    \includegraphics[width=\columnwidth]{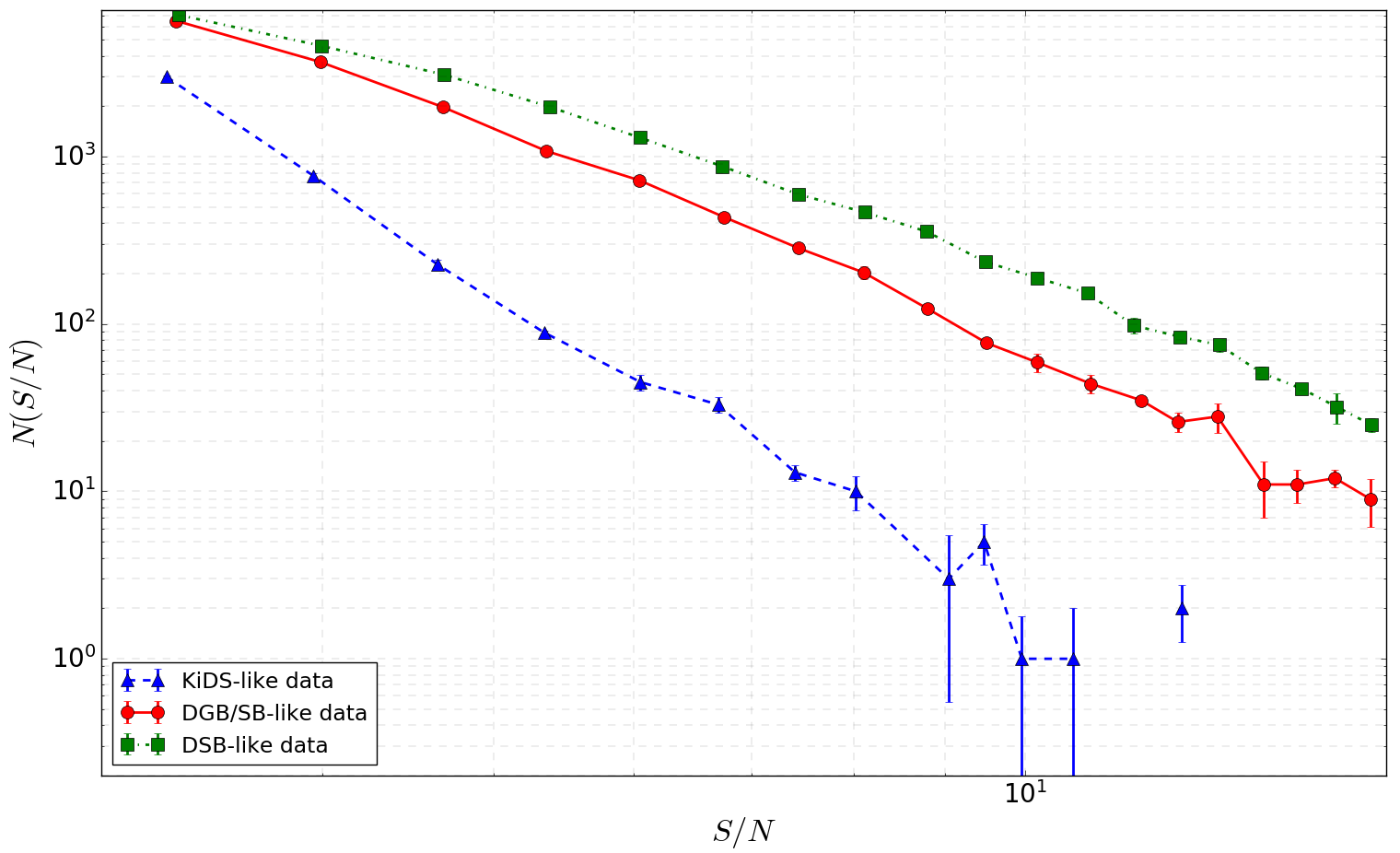}
    \caption{The high S/N peak distributions for $n_{\rm{eff}} = 9$, $30$, and $60$ gal/arcmin$^{2}$ (the triangle, circle, and square markers, respectively). 
    The error bars are the variance in the S/N bins for different noise realisations.
    Note that in some of the high S/N KiDS-like bins there are no detected peaks.
    }
    \label{fig:60gal_compareAllCountSNRs}
\end{figure}
Figure \ref{fig:60gal_compareAllCountSNRs} is the S/N peak distribution for KiDS, DGB/SB, and DSB (the triangle, circle, and square markers, respectively). 
The error bars are the variance in the S/N bins for different noise realisations. Note that the error bars are small.
As $n_{\rm{eff}}$ increases, the number of S/N peaks is significantly increased for S/N $\geq 3$. 
Furthermore, an increase in the number density of background sources can produce extremely high S/N peak values. 
Note that the filter function that we use downweights the strong lensing regime when centred on a cluster (Equation \ref{eq:NFWFilterFunction} and Figure \ref{fig:Qnfw}). 
In practice we would also apply corrections for factors such as intrinsic alignment of galaxies, and the boost factor in dense regions, as in for example \citet{Kacprzak2016, Martinet2018}.

\begin{figure}
    \includegraphics[width=\columnwidth]{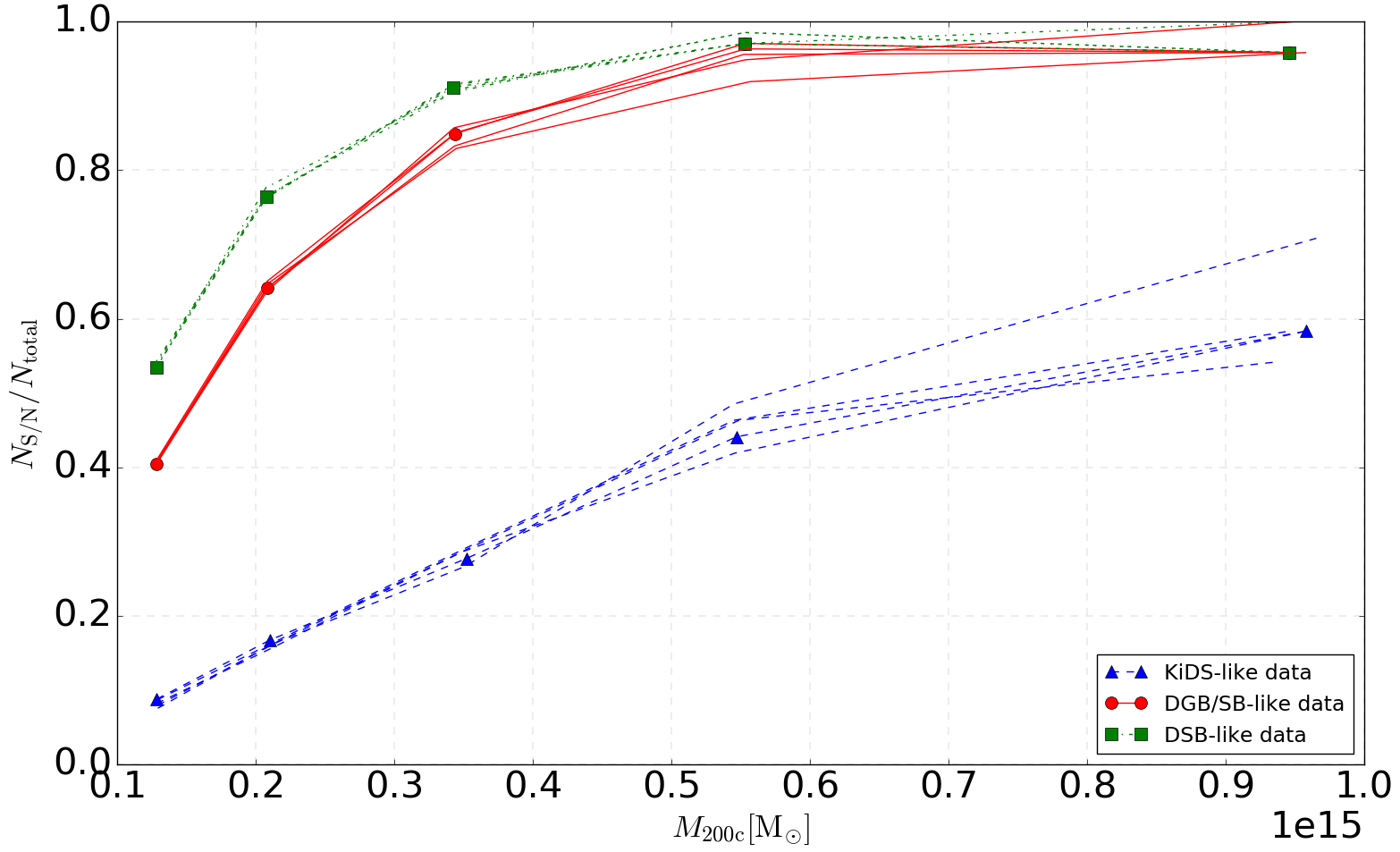}
    \caption{The fraction of clusters that have an associated S/N peak, 
    within a radius of $2.0$ arcmin from the cluster centre, as a function of cluster mass. Source number densities $n_{\rm{eff}} = 9$, $30$, and $60$ gal/arcmin$^{2}$ are represented by the triangle, circle, and square markers, respectively. Here we show the results for five different shape noise realisations of background galaxies to indicate the variation that would be expected over surveys of the same area.
    }
    \label{fig:fullM200s}
\end{figure}

Figure \ref{fig:fullM200s} shows the fraction of cluster-mass objects in our catalogue that enclose a S/N peak ($\geq 3$) within radius $2.0$ arcmin, or $N_{\rm S/N}/N_{\rm total}$.
KiDS, DGB/SB, and DSB are represented by the triangle, circle, and square markers, respectively.

Here we show the results for five different noise realisations of background galaxies.

Increasing source number density increases the ability to detect clusters with weak lensing peaks and the overall number of S/N peaks (see Figure \ref{fig:60gal_compareAllCountSNRs}). Therefore there are more high S/N peaks being produced around galaxy clusters, i.e. the number of S/N peaks for massive haloes tends to increase (see Figure \ref{fig:MvsSNR}, $M_{\nu} = $ 0.06 eV). In the Appendix we check the statistical significance of these results by relating galaxy cluster masses and S/N peaks at random positions, in Figure \ref{fig:fullRandomM200s}.

\begin{figure}
    \includegraphics[width=\columnwidth]{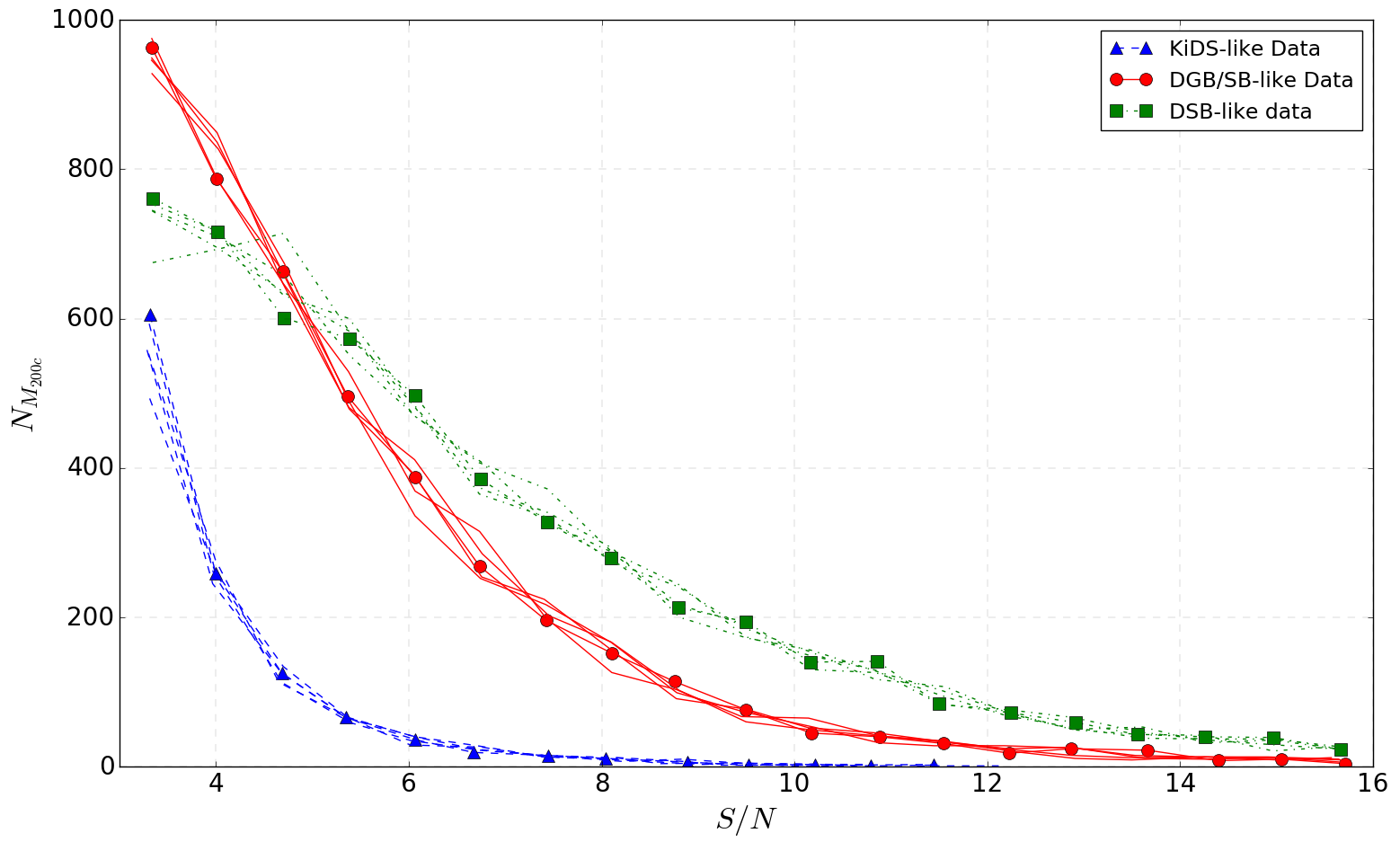}
    \caption{The number of S/N peaks ($\geq 3$) that have a nearby cluster within radius $2.0$ arcmin, for different background galaxy number densities $n_{\rm{eff}} = 9$, $30$, and $60$ gal/arcmin$^{2}$ (represented by the triangle, circle, and square markers, respectively). Here we show the results for five different shape noise realisations of background galaxies to indicate the variation that would be expected over surveys of the same area.
    }
    \label{fig:fullSNRs}
\end{figure}

Figure \ref{fig:fullSNRs} is the number of S/N peaks ($\geq 3$) that have a nearby halo ($\geq 10^{14} \rm{M_{\odot}}$) within $2.0$ arcmin, divided into S/N bins. The different galaxy number densities $n_{\rm{eff}} = 9$, $30$, and $60$ gal/arcmin$^{2}$ are represented by the triangle, circle, and square markers, respectively. We show the results for five different noise realisations of background galaxies to indicate the variation that would be expected over surveys of the same area. 

For the deeper data massive objects below the mass cutoff threshold could produce S/N peaks at the lower end of the range. 
We also consider a lower mass cutoff threshold and the number of galaxy clusters detected using a blind weak lensing survey increases with background galaxy number density for high S/N bins.


\section{Conclusions}
\label{s:discussions}
In this paper we have quantified the impact of baryonic physics and massive neutrinos on weak lensing peak statistics.
We have considered a range of prescriptions for the baryonic physics (with zero neutrino mass) and summed neutrino mass ($M_\nu = 0.06, 0.12, 0.24, 0.48$ eV, for the fiducial baryonic physics model) implemented in the BAHAMAS simulations \citet{McCarthy2018}. Our results for baryonic physics and massive neutrinos can guide the error budget when deriving cosmological parameters from WL peak statistics.
We have also considered the correspondence between high S/N peaks and galaxy clusters for a fiducial baryonic physics prescription, while varying summed neutrino mass.

The WL peak statistics were determined from synthetic aperture mass S/N maps calculated from shape noise realisations of simulated WL data fields. Calculation of the convergence assumed different source redshift distributions from \citet{Chang2013} for LSST and from \citet{Hildebrandt2017} and \citet{deJong2017} for KiDS. We use source number densities of $9$ and $30$ gal/arcmin$^{2}$, roughly corresponding to the KiDS data and the expectation for LSST and \textit{Euclid}. 
We also considered a higher source number density of $60$ gal/arcmin$^{2}$ corresponding to space-based observations, such as \textit{HST}.
Aperture mass S/N maps were determined with the publicly available code from \citet{Bard2012}, using the aperture mass filter function optimised to the NFW profile \citep{Schirmer2007}. We use a fiducial filter size of 12.5 arcmin for consistency with \citet{Martinet2015}.

In summary we find that:\begin{itemize}
    \item Considering the \textit{WMAP}~9 cosmology, the impact of baryonic physics (in addition to the gravity of DM) boosts (suppresses) the peak counts for low (high) S/N. The low S/N peaks are consistently boosted by less than a few percent with the KiDS survey, whereas for deeper DGB/SB data the low S/N peaks are boosted by about 2-6 percent relative to DMONLY case. With DGB/SB number density, the lower S/N peak statistics are more sensitive to baryonic physics. 
    The high S/N peaks become more suppressed with increasing S/N value. We explain this suppression using the fact that AGN feedback changes the shape of the mass density profiles of massive clusters. Baryonic physics is roughly degenerate with the impact of $S_8$ seen in \citet{Martinet2018} as well as with massive neutrinos.
    
    \item During the early epoch of the Universe, free-streaming neutrinos hinder formation of LSS and result in decreasing $S_8$ (measured today) and, therefore, impact the WL peak distribution. For the \textit{WMAP}~9 cosmology 0.06 and 0.12 eV summed neutrino mass models, the impact of massive neutrinos (at fixed fiducial baryonic physics) on the peak counts compared with DMONLY (collisionless dynamics) tends to be less significant or similar to that of baryonic physics. For higher 0.24 and 0.48 eV summed neutrino mass models, the impact of massive neutrinos tends to be greater than that of baryonic physics.
    The lowest source density (9 gal/arcmin$^2$) peak distributions for the 0.06 and 0.12 eV models are not substantially different (see discussion of $S_8$ above and summary item below), but higher summed neutrino mass models become more distinguishable from the DMONLY model, even at this low source number density. 
    For deeper surveys the peak distributions have greater power to differentiate between summed neutrino mass models.
    
    \item Different cosmological parameters based on \textit{WMAP}~9 and \textit{Planck}~2015 surveys were compared, using models with summed neutrino mass 0.06, 0.12, 0.24, and 0.48 eV, all with fiducial baryonic physics. In comparison with \textit{WMAP}~9 fiducial baryonic physics with massless neutrinos, the peak distributions show clear differences for the \textit{WMAP}~9 and \textit{Planck}~2015 cosmologies if restricted to the 0.06 and 0.12 eV models; at low S/N the peak distributions of \textit{Planck}~2015 are consistently suppressed by less than $\sim5$ percent for the 30 gal/arcmin$^2$ source number density case, and the high S/N peaks are boosted by $\sim$5 to $\sim$10 percent, compared to the \textit{WMAP}~9 peak distributions. Overall the trends in the weak lensing peak counts are ordered with respect to $S_8$, see Table \ref{tab:cosmological_parameters}. For example, in models across cosmologies but with similar $S_8$ values the peak counts are difficult to distinguish. The \textit{WMAP}~9 $M_\nu=$0.00 (0.24) eV model has similar peak counts to the \textit{Planck}~2015 $M_\nu=$0.24 (0.48) eV model, respectively. 
    
    Note that since neutrinos change the formation of structure, for example impacting on the measured $\sigma_8$, whereas for massive haloes baryons primarily alter the mass density profiles, ordering in weak lensing peak counts as a function of $S_8$ is not seen for models with different baryonic physics. For example, although our \textit{WMAP}~9 AGN low, fiducial, and high models have the same $S_8$ they do not have the same weak lensing peak distributions.

    \item Impact of massive neutrinos on high S/N peaks and massive clusters: consistent with \citet{Costanzi2013, Castorina2014, Mummery2017, Hagstotz2018}, the cluster mass function is reduced in amplitude when including baryonic physics and non-zero neutrino mass. For a fixed prescription of baryonic physics in the \textit{Planck}~2015 cosmology, higher neutrino mass reduces the number of high S/N peaks. 
    Increasing the summed neutrino mass typically, although not always, reduces the masses and S/N peak heights of individual galaxy clusters.

    \item Detecting peaks around galaxy clusters within a radius of $2.0$ arcmin, for $M_{\nu} = 0.06$ eV, and with fiducial baryonic physics: even for the most massive clusters ($\geq 10^{14} \rm{M_{\odot}}$), a weak lensing S/N peak is not always present at the lowest source number density, $9$ gal/arcmin$^{2}$. For example, in Figure \ref{fig:fullM200s}, at $M_{200c} \approx  6 \times 10^{14} \rm{M_{\odot}}$ there is $\approx 50$ percent likelihood that a cluster has a corresponding weak lensing peak, with S/N $\geq 3$. At higher source number density ($30$ gal/arcmin$^{2}$) we find that more than $90$ percent of clusters have a detected weak lensing peak. For the highest number density $60$ gal/arcmin$^{2}$, we find that nearly all clusters have a detected weak lensing peak. Note that the initially random distribution of intrinsic background galaxy shapes from which the shear is measured has an impact on the detection of a foreground cluster using aperture mass peaks. The ability to study the peak statistics for intermediate redshift galaxy clusters relies on deeper lensing data, as can be seen in the differences between Figures \ref{fig:KiDS_MvsSNR_binZ} and \ref{fig:LSST_MvsSNR_binZ}.
    Increasing the source number density increases the overall number of higher S/N peaks, and for deeper data other features such as projections of large scale structure and combinations of lower mass haloes can also be significantly detected. With higher source number density the overall number of detected galaxy clusters increases. These statements on cluster detection with weak lensing peaks for different source number densities hold for the range of summed neutrino masses considered. However, the number of clusters decreases as a function of summed neutrino mass.
    
    \item Figure \ref{fig:percentDiff} and Table \ref{tab:varyMnu_SNRpercentDiff} encapsulate the main findings of this paper. For the \textit{WMAP}~9 cosmology, the percentage difference between the S/N peak counts in comparison with the DMONLY (collisionless dynamics) model is shown for summed neutrino mass 0.00, 0.06, 0.12, 0.24 and 0.48 eV models (all with fiducial baryonic physics). Assuming that baryonic physics and massive neutrinos act independently \citep{Mummery2017}: for lower (non-zero) summed neutrino mass models 0.06 and 0.12 eV, baryonic physics tends to have a greater or similar impact on the results than massive neutrinos; for higher summed neutrino mass models 0.24 and 0.48 eV, massive neutrinos can change the peak counts by more than fiducial baryonic physics. We have considered a range of models for baryonic physics and summed neutrino mass; the precise impact will depend on the true baryonic physics and summed neutrinos mass, as well as on the characteristics of the survey itself. A detailed study of baryonic physics and massive neutrinos (impacting on mass density profiles and halo formation) would be possible in the framework of the halo model. 
\end{itemize}

We look forward to applying the results of this theoretical work to observational data, such as the LSST and \textit{Euclid} surveys.
Here we have considered the synthetic peak counts on the sky, independent of redshift. More precise predictions for the peak statistics can also be made accounting for tomographic information in surveys, and the ability to break degeneracies between baryonic physics, massive neutrinos, and other cosmological parameters can be studied.

\section*{Acknowledgements}
MWF, MC, and LJK acknowledge support for this work from the National Science Foundation, grant number 1517954.
BEL acknowledges support for this work by National Aeronautics and Space Administration Grant No. NNX16AF53G.
VC was supported during part of this work by the Clark Summer Program at The University of Texas at Dallas.
RLB acknowledges support from the National Science Foundation Research Experiences for Undergraduates program at the Maria Mitchell Observatory.
We thank Regina Jorgenson for advising RLB at Maria Mitchell Observatory's REU program and for helpful discussions.
We thank the Texas Advanced Computer Center for use of resources and NVIDIA for donating a GPU for our computations.
IGM thanks Simeon Bird and Joop Schaye for their contributions to the BAHAMAS simulations.  
This project has received funding from the European Research Council (ERC) under the European Union's Horizon 2020 research and innovation programme (grant agreement No 769130).
AP acknowledges support from an Enhanced Eurotalents Fellowship, a Marie Sk\l{}odowska-Curie Actions Programme co-funded by the European Commission and Commissariat {\`a} l'{\'e}nergie atomique et aux {\'e}nergies alternatives (CEA).
We thank the referee for a detailed and insightful report that has improved the manuscript.

This research made use of \textsc{Astropy},\footnote{http://www.astropy.org} a community-developed core Python package for Astronomy \citep{astropy2013, astropy2018}.
Plots in this paper were produced with \textsc{python} and its \textsc{matplotlib} \citep{Hunter2007}.





\bibliographystyle{mnras}
\bibliography{citation}

\begin{thebibliography}{}
\makeatletter
\relax
\def\mn@urlcharsother{\let\do\@makeother \do\$\do\&\do\#\do\^\do\_\do\%\do\~}
\def\mn@doi{\begingroup\mn@urlcharsother \@ifnextchar [ {\mn@doi@}
  {\mn@doi@[]}}
\def\mn@doi@[#1]#2{\def\@tempa{#1}\ifx\@tempa\@empty \href
  {http://dx.doi.org/#2} {doi:#2}\else \href {http://dx.doi.org/#2} {#1}\fi
  \endgroup}
\def\mn@eprint#1#2{\mn@eprint@#1:#2::\@nil}
\def\mn@eprint@arXiv#1{\href {http://arxiv.org/abs/#1} {{\tt arXiv:#1}}}
\def\mn@eprint@dblp#1{\href {http://dblp.uni-trier.de/rec/bibtex/#1.xml}
  {dblp:#1}}
\def\mn@eprint@#1:#2:#3:#4\@nil{\def\@tempa {#1}\def\@tempb {#2}\def\@tempc
  {#3}\ifx \@tempc \@empty \let \@tempc \@tempb \let \@tempb \@tempa \fi \ifx
  \@tempb \@empty \def\@tempb {arXiv}\fi \@ifundefined
  {mn@eprint@\@tempb}{\@tempb:\@tempc}{\expandafter \expandafter \csname
  mn@eprint@\@tempb\endcsname \expandafter{\@tempc}}}

\bibitem[\protect\citeauthoryear{Addison, Huang, Watts, Bennett, Halpern,
  Hinshaw  \& Weiland}{Addison et~al.}{2016}]{Addison2016}
Addison G.~E.,  Huang Y.,  Watts D.~J.,  Bennett C.~L.,  Halpern M.,  Hinshaw
  G.,   Weiland J.~L.,  2016, \mn@doi [The Astrophysical Journal]
  {10.3847/0004-637x/818/2/132}, 818, 132

\bibitem[\protect\citeauthoryear{{Allen}, {Evrard}  \& {Mantz}}{{Allen}
  et~al.}{2011}]{Allen2011}
{Allen} S.~W.,  {Evrard} A.~E.,   {Mantz} A.~B.,  2011, \mn@doi [ARA\&A]
  {10.1146/annurev-astro-081710-102514}, \href
  {http://adsabs.harvard.edu/abs/2011ARA%26A..49..409A} {49, 409}

\bibitem[\protect\citeauthoryear{{Amendola} et~al.,}{{Amendola}
  et~al.}{2018}]{Amendola2018}
{Amendola} L.,  et~al., 2018, \mn@doi [Living Reviews in Relativity]
  {10.1007/s41114-017-0010-3}, \href
  {https://ui.adsabs.harvard.edu/#abs/2018LRR....21....2A} {21, 2}

\bibitem[\protect\citeauthoryear{{Bard}, {Bellis}, {Allen}, {Yepremyan}  \&
  {Kratochvil}}{{Bard} et~al.}{2012}]{Bard2012}
{Bard} D.,  {Bellis} M.,  {Allen} M.~T.,  {Yepremyan} H.,   {Kratochvil} J.~M.,
   2012, arXiv e-prints, \href
  {https://ui.adsabs.harvard.edu/\#abs/2012arXiv1208.3658B} {p.
  arXiv:1208.3658}

\bibitem[\protect\citeauthoryear{{Bartelmann}}{{Bartelmann}}{1996}]{Bartelmann1996}
{Bartelmann} M.,  1996, A\&A, \href
  {http://adsabs.harvard.edu/abs/1996A%26A...313..697B} {313, 697}

\bibitem[\protect\citeauthoryear{{Bartelmann} \& {Schneider}}{{Bartelmann} \&
  {Schneider}}{2001}]{Bartelmann2001}
{Bartelmann} M.,  {Schneider} P.,  2001, \mn@doi [Physics Reports]
  {10.1016/S0370-1573(00)00082-X}, \href
  {http://adsabs.harvard.edu/abs/2001PhR...340..291B} {340, 291}

\bibitem[\protect\citeauthoryear{{Bashinsky} \& {Seljak}}{{Bashinsky} \&
  {Seljak}}{2004}]{Bashinsky2004}
{Bashinsky} S.,  {Seljak} U.,  2004, \mn@doi [\prd]
  {10.1103/PhysRevD.69.083002}, \href
  {https://ui.adsabs.harvard.edu/\#abs/2004PhRvD..69h3002B} {69, 083002}

\bibitem[\protect\citeauthoryear{{Battye} \& {Moss}}{{Battye} \&
  {Moss}}{2014}]{Battye2014}
{Battye} R.~A.,  {Moss} A.,  2014, \mn@doi [Physical Review Letters]
  {10.1103/PhysRevLett.112.051303}, \href
  {http://adsabs.harvard.edu/abs/2014PhRvL.112e1303B} {112, 051303}

\bibitem[\protect\citeauthoryear{{Bird}, {Viel}  \& {Haehnelt}}{{Bird}
  et~al.}{2012}]{Bird2012}
{Bird} S.,  {Viel} M.,   {Haehnelt} M.~G.,  2012, \mn@doi [\mnras]
  {10.1111/j.1365-2966.2011.20222.x}, \href
  {https://ui.adsabs.harvard.edu/\#abs/2012MNRAS.420.2551B} {420, 2551}

\bibitem[\protect\citeauthoryear{Blumenthal, Faber, Primack  \&
  Rees}{Blumenthal et~al.}{1984}]{Blumenthal1984}
Blumenthal G.~R.,  Faber S.~M.,  Primack J.~R.,   Rees M.~J.,  1984, Nature,
  311, 517

\bibitem[\protect\citeauthoryear{Bond, Efstathiou  \& Silk}{Bond
  et~al.}{1980}]{Bond1980}
Bond J.~R.,  Efstathiou G.,   Silk J.,  1980, Phys. Rev. Lett., 45

\bibitem[\protect\citeauthoryear{{Castorina}, {Sefusatti}, {Sheth},
  {Villaescusa-Navarro}  \& {Viel}}{{Castorina} et~al.}{2014}]{Castorina2014}
{Castorina} E.,  {Sefusatti} E.,  {Sheth} R.~K.,  {Villaescusa-Navarro} F.,
  {Viel} M.,  2014, \mn@doi [Journal of Cosmology and Astro-Particle Physics]
  {10.1088/1475-7516/2014/02/049}, \href
  {https://ui.adsabs.harvard.edu/\#abs/2014JCAP...02..049C} {2014, 049}

\bibitem[\protect\citeauthoryear{{Castro}, {Quartin}, {Giocoli}, {Borgani}  \&
  {Dolag}}{{Castro} et~al.}{2018}]{Castro2018}
{Castro} T.,  {Quartin} M.,  {Giocoli} C.,  {Borgani} S.,   {Dolag} K.,  2018,
  \mn@doi [\mnras] {10.1093/mnras/sty1117}, \href
  {https://ui.adsabs.harvard.edu/abs/2018MNRAS.478.1305C} {478, 1305}

\bibitem[\protect\citeauthoryear{{Chang} et~al.,}{{Chang}
  et~al.}{2013}]{Chang2013}
{Chang} C.,  et~al., 2013, \mn@doi [MNRAS] {10.1093/mnras/stt1156}, \href
  {http://adsabs.harvard.edu/abs/2013MNRAS.434.2121C} {434, 2121}

\bibitem[\protect\citeauthoryear{{Clowe}, {De Lucia}  \& {King}}{{Clowe}
  et~al.}{2004}]{Clowe2004}
{Clowe} D.,  {De Lucia} G.,   {King} L.,  2004, \mn@doi [MNRAS]
  {10.1111/j.1365-2966.2004.07723.x}, \href
  {http://adsabs.harvard.edu/abs/2004MNRAS.350.1038C} {350, 1038}

\bibitem[\protect\citeauthoryear{{Costanzi}, {Villaescusa-Navarro}, {Viel},
  {Xia}, {Borgani}, {Castorina}  \& {Sefusatti}}{{Costanzi}
  et~al.}{2013}]{Costanzi2013}
{Costanzi} M.,  {Villaescusa-Navarro} F.,  {Viel} M.,  {Xia} J.-Q.,  {Borgani}
  S.,  {Castorina} E.,   {Sefusatti} E.,  2013, \mn@doi [Journal of Cosmology
  and Astro-Particle Physics] {10.1088/1475-7516/2013/12/012}, \href
  {https://ui.adsabs.harvard.edu/\#abs/2013JCAP...12..012C} {2013, 012}

\bibitem[\protect\citeauthoryear{{Cusworth}, {Kay}, {Battye}  \&
  {Thomas}}{{Cusworth} et~al.}{2014}]{Cusworth2014}
{Cusworth} S.~J.,  {Kay} S.~T.,  {Battye} R.~A.,   {Thomas} P.~A.,  2014,
  \mn@doi [MNRAS] {10.1093/MNRAS/stu105}, \href
  {http://adsabs.harvard.edu/abs/2014MNRAS.439.2485C} {439, 2485}

\bibitem[\protect\citeauthoryear{{Davies}, {Cautun}  \& {Li}}{{Davies}
  et~al.}{2019}]{Davies2019}
{Davies} C.~T.,  {Cautun} M.,   {Li} B.,  2019, arXiv e-prints, \href
  {https://ui.adsabs.harvard.edu/abs/2019arXiv190501710D} {p. arXiv:1905.01710}

\bibitem[\protect\citeauthoryear{Di~Valentino, Melchiorri, Linder  \&
  Silk}{Di~Valentino et~al.}{2017}]{Valentino2017}
Di~Valentino E.,  Melchiorri A.,  Linder E.~V.,   Silk J.,  2017, \mn@doi
  [Phys. Rev. D] {10.1103/PhysRevD.96.023523}, 96, 023523

\bibitem[\protect\citeauthoryear{{Dietrich} \& {Hartlap}}{{Dietrich} \&
  {Hartlap}}{2010}]{Dietrich2010}
{Dietrich} J.~P.,  {Hartlap} J.,  2010, \mn@doi [\mnras]
  {10.1111/j.1365-2966.2009.15948.x}, \href
  {http://adsabs.harvard.edu/abs/2010MNRAS.402.1049D} {402, 1049}

\bibitem[\protect\citeauthoryear{{Dolag}, {Borgani}, {Murante}  \&
  {Springel}}{{Dolag} et~al.}{2009}]{Dolag2009}
{Dolag} K.,  {Borgani} S.,  {Murante} G.,   {Springel} V.,  2009, \mn@doi
  [MNRAS] {10.1111/j.1365-2966.2009.15034.x}, \href
  {http://adsabs.harvard.edu/abs/2009MNRAS.399..497D} {399, 497}

\bibitem[\protect\citeauthoryear{{Dubois} et~al.,}{{Dubois}
  et~al.}{2014}]{Dubois2014}
{Dubois} Y.,  et~al., 2014, \mn@doi [MNRAS] {10.1093/mnras/stu1227}, \href
  {http://adsabs.harvard.edu/abs/2014MNRAS.444.1453D} {444, 1453}

\bibitem[\protect\citeauthoryear{{Fan}, {Shan}  \& {Liu}}{{Fan}
  et~al.}{2010}]{Fan2010}
{Fan} Z.,  {Shan} H.,   {Liu} J.,  2010, \mn@doi [ApJ]
  {10.1088/0004-637X/719/2/1408}, \href
  {https://ui.adsabs.harvard.edu/\#abs/2010ApJ...719.1408F} {719, 1408}

\bibitem[\protect\citeauthoryear{{Feeney}, {Mortlock}  \& {Dalmasso}}{{Feeney}
  et~al.}{2018}]{Feeney2018}
{Feeney} S.~M.,  {Mortlock} D.~J.,   {Dalmasso} N.,  2018, \mn@doi [\mnras]
  {10.1093/mnras/sty418}, \href
  {https://ui.adsabs.harvard.edu/\#abs/2018MNRAS.476.3861F} {476, 3861}

\bibitem[\protect\citeauthoryear{{Gratton}, {Lewis}  \& {Efstathiou}}{{Gratton}
  et~al.}{2008}]{Gratton2008}
{Gratton} S.,  {Lewis} A.,   {Efstathiou} G.,  2008, \mn@doi [\prd]
  {10.1103/PhysRevD.77.083507}, \href
  {https://ui.adsabs.harvard.edu/\#abs/2008PhRvD..77h3507G} {77, 083507}

\bibitem[\protect\citeauthoryear{{Hagstotz}, {Costanzi}, {Baldi}  \&
  {Weller}}{{Hagstotz} et~al.}{2018}]{Hagstotz2018}
{Hagstotz} S.,  {Costanzi} M.,  {Baldi} M.,   {Weller} J.,  2018, arXiv
  e-prints, \href {https://ui.adsabs.harvard.edu/\#abs/2018arXiv180607400H} {p.
  arXiv:1806.07400}

\bibitem[\protect\citeauthoryear{{Hamana}, {Takada}  \& {Yoshida}}{{Hamana}
  et~al.}{2004}]{Hamana2004}
{Hamana} T.,  {Takada} M.,   {Yoshida} N.,  2004, \mn@doi [MNRAS]
  {10.1111/j.1365-2966.2004.07691.x}, \href
  {http://adsabs.harvard.edu/abs/2004MNRAS.350..893H} {350, 893}

\bibitem[\protect\citeauthoryear{{Hamana}, {Oguri}, {Shirasaki}  \&
  {Sato}}{{Hamana} et~al.}{2012}]{Hamana2012}
{Hamana} T.,  {Oguri} M.,  {Shirasaki} M.,   {Sato} M.,  2012, \mn@doi [\mnras]
  {10.1111/j.1365-2966.2012.21582.x}, \href
  {http://adsabs.harvard.edu/abs/2012MNRAS.425.2287H} {425, 2287}

\bibitem[\protect\citeauthoryear{{Hannestad}, {Tu}  \& {Wong}}{{Hannestad}
  et~al.}{2006}]{Hannestad2006}
{Hannestad} S.,  {Tu} H.,   {Wong} Y.~Y.,  2006, \mn@doi [Journal of Cosmology
  and Astro-Particle Physics] {10.1088/1475-7516/2006/06/025}, \href
  {https://ui.adsabs.harvard.edu/\#abs/2006JCAP...06..025H} {2006, 025}

\bibitem[\protect\citeauthoryear{{Hennawi} \& {Spergel}}{{Hennawi} \&
  {Spergel}}{2005}]{Hennawi2005}
{Hennawi} J.~F.,  {Spergel} D.~N.,  2005, \mn@doi [\apj] {10.1086/428749},
  \href {https://ui.adsabs.harvard.edu/\#abs/2005ApJ...624...59H} {624, 59}

\bibitem[\protect\citeauthoryear{{Henson}, {Barnes}, {Kay}, {McCarthy}  \&
  {Schaye}}{{Henson} et~al.}{2017}]{Henson2017}
{Henson} M.~A.,  {Barnes} D.~J.,  {Kay} S.~T.,  {McCarthy} I.~G.,   {Schaye}
  J.,  2017, \mn@doi [MNRAS] {10.1093/mnras/stw2899}, \href
  {http://adsabs.harvard.edu/abs/2017MNRAS.465.3361H} {465, 3361}

\bibitem[\protect\citeauthoryear{{Hetterscheidt}, {Erben}, {Schneider},
  {Maoli}, {van Waerbeke}  \& {Mellier}}{{Hetterscheidt}
  et~al.}{2005}]{Hetterscheidt2005}
{Hetterscheidt} M.,  {Erben} T.,  {Schneider} P.,  {Maoli} R.,  {van Waerbeke}
  L.,   {Mellier} Y.,  2005, \mn@doi [\aap] {10.1051/0004-6361:20053339}, \href
  {http://adsabs.harvard.edu/abs/2005A%26A...442...43H} {442, 43}

\bibitem[\protect\citeauthoryear{{Heymans} et~al.,}{{Heymans}
  et~al.}{2012}]{Heymans2012}
{Heymans} C.,  et~al., 2012, \mn@doi [MNRAS]
  {10.1111/j.1365-2966.2012.21952.x}, \href
  {http://adsabs.harvard.edu/abs/2012MNRAS.427..146H} {427, 146}

\bibitem[\protect\citeauthoryear{{Hildebrandt}, {Viola}, {Heymans}, {Joudaki},
  {Kuijken}, {Blake}, {Erben}  \& {Joachimi}}{{Hildebrandt}
  et~al.}{2017}]{Hildebrandt2017}
{Hildebrandt} H.,  {Viola} M.,  {Heymans} C.,  {Joudaki} S.,  {Kuijken} K.,
  {Blake} C.,  {Erben} T.,   {Joachimi} B.,  2017, \mn@doi [MNRAS]
  {10.1093/mnras/stw2805}, \href
  {http://adsabs.harvard.edu/abs/2017MNRAS.465.1454H} {465, 1454}

\bibitem[\protect\citeauthoryear{{Hinshaw} et~al.,}{{Hinshaw}
  et~al.}{2013}]{Hinshaw2013}
{Hinshaw} G.,  et~al., 2013, \mn@doi [ApJ] {10.1088/0067-0049/208/2/19}, \href
  {http://adsabs.harvard.edu/abs/2013ApJS..208...19H} {208, 19}

\bibitem[\protect\citeauthoryear{{Hu}, {Eisenstein}  \& {Tegmark}}{{Hu}
  et~al.}{1998}]{Hu1998}
{Hu} W.,  {Eisenstein} D.~J.,   {Tegmark} M.,  1998, \mn@doi [\prl]
  {10.1103/PhysRevLett.80.5255}, \href
  {https://ui.adsabs.harvard.edu/\#abs/1998PhRvL..80.5255H} {80, 5255}

\bibitem[\protect\citeauthoryear{Jain \& Waerbeke}{Jain \&
  Waerbeke}{2000}]{Jain2000}
Jain B.,  Waerbeke L.~V.,  2000, \mn@doi [The Astrophysical Journal]
  {10.1086/312480}, 530, L1

\bibitem[\protect\citeauthoryear{{Kacprzak} et~al.,}{{Kacprzak}
  et~al.}{2016}]{Kacprzak2016}
{Kacprzak} T.,  et~al., 2016, \mn@doi [\mnras] {10.1093/mnras/stw2070}, \href
  {http://adsabs.harvard.edu/abs/2016MNRAS.463.3653K} {463, 3653}

\bibitem[\protect\citeauthoryear{{King} et~al.,}{{King}
  et~al.}{2016}]{King2016}
{King} L.~J.,  et~al., 2016, \mn@doi [MNRAS] {10.1093/mnras/stw507}, \href
  {http://adsabs.harvard.edu/abs/2016MNRAS.459..517K} {459, 517}

\bibitem[\protect\citeauthoryear{{Kratochvil}, {Haiman}  \& {May}}{{Kratochvil}
  et~al.}{2010}]{Kratochvil2010}
{Kratochvil} J.~M.,  {Haiman} Z.,   {May} M.,  2010, \mn@doi [\prd]
  {10.1103/PhysRevD.81.043519}, \href
  {https://ui.adsabs.harvard.edu/\#abs/2010PhRvD..81d3519K} {81, 043519}

\bibitem[\protect\citeauthoryear{{Kravtsov} \& {Borgani}}{{Kravtsov} \&
  {Borgani}}{2012}]{Kravtsov2012}
{Kravtsov} A.~V.,  {Borgani} S.,  2012, \mn@doi [Annual Review of Astronomy and
  Astrophysics] {10.1146/annurev-astro-081811-125502}, \href
  {https://ui.adsabs.harvard.edu/\#abs/2012ARA&A..50..353K} {50, 353}

\bibitem[\protect\citeauthoryear{{Kruse} \& {Schneider}}{{Kruse} \&
  {Schneider}}{1999}]{Kruse1999}
{Kruse} G.,  {Schneider} P.,  1999, \mn@doi [\mnras]
  {10.1046/j.1365-8711.1999.02195.x}, \href
  {http://adsabs.harvard.edu/abs/1999MNRAS.302..821K} {302, 821}

\bibitem[\protect\citeauthoryear{{Kruse} \& {Schneider}}{{Kruse} \&
  {Schneider}}{2000}]{Kruse2000}
{Kruse} G.,  {Schneider} P.,  2000, \mn@doi [\mnras]
  {10.1046/j.1365-8711.2000.03389.x}, \href
  {http://adsabs.harvard.edu/abs/2000MNRAS.318..321K} {318, 321}

\bibitem[\protect\citeauthoryear{{Lahav}, {Kiakotou}, {Abdalla}  \&
  {Blake}}{{Lahav} et~al.}{2010}]{Lahav2010}
{Lahav} O.,  {Kiakotou} A.,  {Abdalla} F.~B.,   {Blake} C.,  2010, \mn@doi
  [\mnras] {10.1111/j.1365-2966.2010.16472.x}, \href
  {https://ui.adsabs.harvard.edu/\#abs/2010MNRAS.405..168L} {405, 168}

\bibitem[\protect\citeauthoryear{{Laureijs} et~al.,}{{Laureijs}
  et~al.}{2011}]{Laureijs2011}
{Laureijs} R.,  et~al., 2011, preprint, \href
  {https://ui.adsabs.harvard.edu/#abs/2011arXiv1110.3193L} {p. arXiv:1110.3193}
  (\mn@eprint {arXiv} {1110.3193})

\bibitem[\protect\citeauthoryear{{Le Brun}, {McCarthy}, {Schaye}  \&
  {Ponman}}{{Le Brun} et~al.}{2014}]{LeBrun2014}
{Le Brun} A.~M.~C.,  {McCarthy} I.~G.,  {Schaye} J.,   {Ponman} T.~J.,  2014,
  \mn@doi [MNRAS] {10.1093/mnras/stu608}, \href
  {http://adsabs.harvard.edu/abs/2014MNRAS.441.1270L} {441, 1270}

\bibitem[\protect\citeauthoryear{{Lee}, {Le Brun}, {Haq}, {Deering}, {King},
  {Applegate}  \& {McCarthy}}{{Lee} et~al.}{2018}]{Lee2018}
{Lee} B.~E.,  {Le Brun} A.~M.~C.,  {Haq} M.~E.,  {Deering} N.~J.,  {King}
  L.~J.,  {Applegate} D.,   {McCarthy} I.~G.,  2018, \mn@doi [\mnras]
  {10.1093/mnras/sty1377}, \href
  {http://adsabs.harvard.edu/abs/2018MNRAS.479..890L} {479, 890}

\bibitem[\protect\citeauthoryear{{Leonard}, {Pires}  \& {Starck}}{{Leonard}
  et~al.}{2012}]{Leonard2012}
{Leonard} A.,  {Pires} S.,   {Starck} J.-L.,  2012, \mn@doi [MNRAS]
  {10.1111/j.1365-2966.2012.21133.x}, \href
  {https://ui.adsabs.harvard.edu/#abs/2012MNRAS.423.3405L} {423, 3405}

\bibitem[\protect\citeauthoryear{{Lesgourgues} \& {Pastor}}{{Lesgourgues} \&
  {Pastor}}{2006}]{Lesgourgues2006}
{Lesgourgues} J.,  {Pastor} S.,  2006, \mn@doi [\physrep]
  {10.1016/j.physrep.2006.04.001}, \href
  {http://adsabs.harvard.edu/abs/2006PhR...429..307L} {429, 307}

\bibitem[\protect\citeauthoryear{{Lesgourgues} \& {Pastor}}{{Lesgourgues} \&
  {Pastor}}{2012}]{Lesgourgues2012}
{Lesgourgues} J.,  {Pastor} S.,  2012, arXiv e-prints, \href
  {http://adsabs.harvard.edu/abs/2012arXiv1212.6154L} {}

\bibitem[\protect\citeauthoryear{{Li}, {Liu}, {Matilla}  \& {Coulton}}{{Li}
  et~al.}{2019}]{Li2019}
{Li} Z.,  {Liu} J.,  {Matilla} J. M.~Z.,   {Coulton} W.~R.,  2019, \mn@doi
  [\prd] {10.1103/PhysRevD.99.063527}, \href
  {https://ui.adsabs.harvard.edu/abs/2019PhRvD..99f3527L} {99, 063527}

\bibitem[\protect\citeauthoryear{{Lin} \& {Kilbinger}}{{Lin} \&
  {Kilbinger}}{2015}]{Lin2015}
{Lin} C.-A.,  {Kilbinger} M.,  2015, \mn@doi [\aap]
  {10.1051/0004-6361/201526659}, \href
  {https://ui.adsabs.harvard.edu/\#abs/2015A&A...583A..70L} {583, A70}

\bibitem[\protect\citeauthoryear{{Liu} \& {Haiman}}{{Liu} \&
  {Haiman}}{2016}]{Liu2016a}
{Liu} J.,  {Haiman} Z.,  2016, \mn@doi [\prd] {10.1103/PhysRevD.94.043533},
  \href {https://ui.adsabs.harvard.edu/\#abs/2016PhRvD..94d3533L} {94, 043533}

\bibitem[\protect\citeauthoryear{{Liu} \& {Madhavacheril}}{{Liu} \&
  {Madhavacheril}}{2019}]{Liu2019}
{Liu} J.,  {Madhavacheril} M.~S.,  2019, \mn@doi [\prd]
  {10.1103/PhysRevD.99.083508}, \href
  {https://ui.adsabs.harvard.edu/abs/2019PhRvD..99h3508L} {99, 083508}

\bibitem[\protect\citeauthoryear{{Liu} et~al.,}{{Liu} et~al.}{2016}]{Liu2016b}
{Liu} X.,  et~al., 2016, \mn@doi [\prl] {10.1103/PhysRevLett.117.051101}, \href
  {https://ui.adsabs.harvard.edu/\#abs/2016PhRvL.117e1101L} {117, 051101}

\bibitem[\protect\citeauthoryear{{Martinet}, {Bartlett}, {Kiessling}  \&
  {Sartoris}}{{Martinet} et~al.}{2015}]{Martinet2015}
{Martinet} N.,  {Bartlett} J.~G.,  {Kiessling} A.,   {Sartoris} B.,  2015,
  \mn@doi [\aap] {10.1051/0004-6361/201425164}, \href
  {http://adsabs.harvard.edu/abs/2015A%26A...581A.101M} {581, A101}

\bibitem[\protect\citeauthoryear{{Martinet} et~al.,}{{Martinet}
  et~al.}{2018}]{Martinet2018}
{Martinet} N.,  et~al., 2018, \mn@doi [MNRAS] {10.1093/mnras/stx2793}, \href
  {http://adsabs.harvard.edu/abs/2018MNRAS.474..712M} {474, 712}

\bibitem[\protect\citeauthoryear{{Massara}, {Villaescusa-Navarro}  \&
  {Viel}}{{Massara} et~al.}{2014}]{Massara2014}
{Massara} E.,  {Villaescusa-Navarro} F.,   {Viel} M.,  2014, \mn@doi [Journal
  of Cosmology and Astro-Particle Physics] {10.1088/1475-7516/2014/12/053},
  \href {https://ui.adsabs.harvard.edu/abs/2014JCAP...12..053M} {2014, 053}

\bibitem[\protect\citeauthoryear{{Maturi}, {Angrick}, {Pace}  \&
  {Bartelmann}}{{Maturi} et~al.}{2010}]{Maturi2010}
{Maturi} M.,  {Angrick} C.,  {Pace} F.,   {Bartelmann} M.,  2010, \mn@doi
  [\aap] {10.1051/0004-6361/200912866}, \href
  {http://adsabs.harvard.edu/abs/2010A%26A...519A..23M} {519, A23}

\bibitem[\protect\citeauthoryear{{McCarthy}, {Schaye}, {Bird}  \& {Le
  Brun}}{{McCarthy} et~al.}{2017}]{McCarthy2017}
{McCarthy} I.~G.,  {Schaye} J.,  {Bird} S.,   {Le Brun} A.~M.~C.,  2017,
  \mn@doi [MNRAS] {10.1093/mnras/stw2792}, \href
  {http://adsabs.harvard.edu/abs/2017MNRAS.465.2936M} {465, 2936}

\bibitem[\protect\citeauthoryear{{McCarthy}, {Bird}, {Schaye},
  {Harnois-Deraps}, {Font}  \& {van Waerbeke}}{{McCarthy}
  et~al.}{2018}]{McCarthy2018}
{McCarthy} I.~G.,  {Bird} S.,  {Schaye} J.,  {Harnois-Deraps} J.,  {Font}
  A.~S.,   {van Waerbeke} L.,  2018, \mn@doi [MNRAS] {10.1093/mnras/sty377},
  \href {http://adsabs.harvard.edu/abs/2018MNRAS.476.2999M} {476, 2999}

\bibitem[\protect\citeauthoryear{{Mead}, {Heymans}, {Lombriser}, {Peacock},
  {Steele}  \& {Winther}}{{Mead} et~al.}{2016}]{Mead2016}
{Mead} A.~J.,  {Heymans} C.,  {Lombriser} L.,  {Peacock} J.~A.,  {Steele}
  O.~I.,   {Winther} H.~A.,  2016, \mn@doi [\mnras] {10.1093/mnras/stw681},
  \href {https://ui.adsabs.harvard.edu/abs/2016MNRAS.459.1468M} {459, 1468}

\bibitem[\protect\citeauthoryear{Moscardini, Baldi  \& Giocoli}{Moscardini
  et~al.}{2018}]{Moscardini2018}
Moscardini L.,  Baldi M.,   Giocoli C.,  2018, \mn@doi [Monthly Notices of the
  Royal Astronomical Society] {10.1093/mnras/sty2465}, 481, 2813

\bibitem[\protect\citeauthoryear{{Mummery}, {McCarthy}, {Bird}  \&
  {Schaye}}{{Mummery} et~al.}{2017}]{Mummery2017}
{Mummery} B.~O.,  {McCarthy} I.~G.,  {Bird} S.,   {Schaye} J.,  2017, \mn@doi
  [\mnras] {10.1093/mnras/stx1469}, \href
  {http://adsabs.harvard.edu/abs/2017MNRAS.471..227M} {471, 227}

\bibitem[\protect\citeauthoryear{{Namikawa}, {Saito}  \& {Taruya}}{{Namikawa}
  et~al.}{2010}]{Namikawa2010}
{Namikawa} T.,  {Saito} S.,   {Taruya} A.,  2010, \mn@doi [Journal of Cosmology
  and Astro-Particle Physics] {10.1088/1475-7516/2010/12/027}, \href
  {https://ui.adsabs.harvard.edu/\#abs/2010JCAP...12..027N} {2010, 027}

\bibitem[\protect\citeauthoryear{{Osato}, {Shirasaki}  \& {Yoshida}}{{Osato}
  et~al.}{2015}]{Osato2015ApJ...806..186O}
{Osato} K.,  {Shirasaki} M.,   {Yoshida} N.,  2015, \mn@doi [\apj]
  {10.1088/0004-637X/806/2/186}, \href
  {https://ui.adsabs.harvard.edu/#abs/2015ApJ...806..186O} {806, 186}

\bibitem[\protect\citeauthoryear{{Peel}, {Lin}, {Lanusse}, {Leonard}, {Starck}
  \& {Kilbinger}}{{Peel} et~al.}{2017}]{Peel2017}
{Peel} A.,  {Lin} C.-A.,  {Lanusse} F.,  {Leonard} A.,  {Starck} J.-L.,
  {Kilbinger} M.,  2017, \mn@doi [\aap] {10.1051/0004-6361/201629928}, \href
  {http://adsabs.harvard.edu/abs/2017A%26A...599A..79P} {599, A79}

\bibitem[\protect\citeauthoryear{{Peel}, {Pettorino}, {Giocoli}, {Starck}  \&
  {Baldi}}{{Peel} et~al.}{2018}]{Peel2018}
{Peel} A.,  {Pettorino} V.,  {Giocoli} C.,  {Starck} J.-L.,   {Baldi} M.,
  2018, \mn@doi [\aap] {10.1051/0004-6361/201833481}, \href
  {https://ui.adsabs.harvard.edu/\#abs/2018A&A...619A..38P} {619, A38}

\bibitem[\protect\citeauthoryear{{Pillepich} et~al.,}{{Pillepich}
  et~al.}{2018}]{Pillepich2018}
{Pillepich} A.,  et~al., 2018, \mn@doi [MNRAS] {10.1093/mnras/stx2656}, \href
  {https://ui.adsabs.harvard.edu/\#abs/2018MNRAS.473.4077P} {473, 4077}

\bibitem[\protect\citeauthoryear{{Planck Collaboration} et~al.,}{{Planck
  Collaboration} et~al.}{2016a}]{PlanckXIII}
{Planck Collaboration} et~al., 2016a, \mn@doi [\aap]
  {10.1051/0004-6361/201525830}, \href
  {http://adsabs.harvard.edu/abs/2016A%26A...594A..13P} {594, A13}

\bibitem[\protect\citeauthoryear{{Planck Collaboration} et~al.,}{{Planck
  Collaboration} et~al.}{2016b}]{PlanckCollaboration2016}
{Planck Collaboration} et~al., 2016b, \mn@doi [A\&A]
  {10.1051/0004-6361/201525830}, \href
  {http://adsabs.harvard.edu/abs/2016A%26A...594A..13P} {594, A13}

\bibitem[\protect\citeauthoryear{{Planck Collaboration} et~al.,}{{Planck
  Collaboration} et~al.}{2016c}]{PlanckXVI}
{Planck Collaboration} et~al., 2016c, \mn@doi [\aap]
  {10.1051/0004-6361/201526681}, \href
  {http://adsabs.harvard.edu/abs/2016A%26A...594A..16P} {594, A16}

\bibitem[\protect\citeauthoryear{{Planck Collaboration} et~al.,}{{Planck
  Collaboration} et~al.}{2017}]{Ashdown2017}
{Planck Collaboration} et~al., 2017, \mn@doi [\aap]
  {10.1051/0004-6361/201629504}, \href
  {https://ui.adsabs.harvard.edu/\#abs/2017A&A...607A..95P} {607, A95}

\bibitem[\protect\citeauthoryear{{Planck Collaboration} et~al.,}{{Planck
  Collaboration} et~al.}{2018}]{Planck2018}
{Planck Collaboration} et~al., 2018, arXiv e-prints, \href
  {https://ui.adsabs.harvard.edu/\#abs/2018arXiv180706209P} {p.
  arXiv:1807.06209}

\bibitem[\protect\citeauthoryear{Poulin, Boddy, Bird  \& Kamionkowski}{Poulin
  et~al.}{2018}]{Poulin2018}
Poulin V.,  Boddy K.~K.,  Bird S.,   Kamionkowski M.,  2018, \mn@doi [Phys.
  Rev. D] {10.1103/PhysRevD.97.123504}, 97, 123504

\bibitem[\protect\citeauthoryear{{Riess} et~al.,}{{Riess}
  et~al.}{2018}]{Riess2018}
{Riess} A.~G.,  et~al., 2018, \mn@doi [\apj] {10.3847/1538-4357/aaadb7}, \href
  {https://ui.adsabs.harvard.edu/abs/2018ApJ...855..136R} {855, 136}

\bibitem[\protect\citeauthoryear{{Roncarelli}, {Carbone}  \&
  {Moscardini}}{{Roncarelli} et~al.}{2015}]{Roncarelli2015}
{Roncarelli} M.,  {Carbone} C.,   {Moscardini} L.,  2015, \mn@doi [\mnras]
  {10.1093/mnras/stu2546}, \href
  {https://ui.adsabs.harvard.edu/abs/2015MNRAS.447.1761R} {447, 1761}

\bibitem[\protect\citeauthoryear{{Sawala}, {Frenk}, {Crain}, {Jenkins},
  {Schaye}, {Theuns}  \& {Zavala}}{{Sawala} et~al.}{2013}]{Sawala2013}
{Sawala} T.,  {Frenk} C.~S.,  {Crain} R.~A.,  {Jenkins} A.,  {Schaye} J.,
  {Theuns} T.,   {Zavala} J.,  2013, \mn@doi [MNRAS] {10.1093/MNRAS/stt259},
  \href {http://adsabs.harvard.edu/abs/2013MNRAS.431.1366S} {431, 1366}

\bibitem[\protect\citeauthoryear{{Schaller} et~al.,}{{Schaller}
  et~al.}{2015}]{Schaller2015}
{Schaller} M.,  et~al., 2015, \mn@doi [\mnras] {10.1093/mnras/stv1341}, \href
  {http://adsabs.harvard.edu/abs/2015MNRAS.452..343S} {452, 343}

\bibitem[\protect\citeauthoryear{{Schaye} et~al.,}{{Schaye}
  et~al.}{2010}]{Schaye2010}
{Schaye} J.,  et~al., 2010, \mn@doi [MNRAS] {10.1111/j.1365-2966.2009.16029.x},
  \href {http://adsabs.harvard.edu/abs/2010MNRAS.402.1536S} {402, 1536}

\bibitem[\protect\citeauthoryear{{Schaye} et~al.,}{{Schaye}
  et~al.}{2015}]{Schaye2015}
{Schaye} J.,  et~al., 2015, \mn@doi [\mnras] {10.1093/mnras/stu2058}, \href
  {http://adsabs.harvard.edu/abs/2015MNRAS.446..521S} {446, 521}

\bibitem[\protect\citeauthoryear{{Schirmer}, {Erben}, {Hetterscheidt}  \&
  {Schneider}}{{Schirmer} et~al.}{2007}]{Schirmer2007}
{Schirmer} M.,  {Erben} T.,  {Hetterscheidt} M.,   {Schneider} P.,  2007,
  \mn@doi [\aap] {10.1051/0004-6361:20065955}, \href
  {http://adsabs.harvard.edu/abs/2007A%26A...462..875S} {462, 875}

\bibitem[\protect\citeauthoryear{{Schneider}}{{Schneider}}{1996}]{Schneider1996}
{Schneider} P.,  1996, \mn@doi [MNRAS] {10.1093/mnras/283.3.837}, \href
  {http://adsabs.harvard.edu/abs/1996MNRAS.283..837S} {283, 837}

\bibitem[\protect\citeauthoryear{{Schneider}}{{Schneider}}{2005}]{Schneider2005}
{Schneider} P.,  2005, ArXiv Astrophysics e-prints, \href
  {http://adsabs.harvard.edu/abs/2005astro.ph..9252S} {}

\bibitem[\protect\citeauthoryear{{Schneider} \& {Teyssier}}{{Schneider} \&
  {Teyssier}}{2015}]{Schneider2015}
{Schneider} A.,  {Teyssier} R.,  2015, \mn@doi [J. Cosmology Astropart. Phys.]
  {10.1088/1475-7516/2015/12/049}, \href
  {http://adsabs.harvard.edu/abs/2015JCAP...12..049S} {12, 049}

\bibitem[\protect\citeauthoryear{{Schneider}, {van Waerbeke}, {Jain}  \&
  {Kruse}}{{Schneider} et~al.}{1998}]{Schneider1998}
{Schneider} P.,  {van Waerbeke} L.,  {Jain} B.,   {Kruse} G.,  1998, \mn@doi
  [MNRAS] {10.1046/j.1365-8711.1998.01422.x}, \href
  {http://adsabs.harvard.edu/abs/1998MNRAS.296..873S} {296, 873}

\bibitem[\protect\citeauthoryear{{Schrabback} et~al.,}{{Schrabback}
  et~al.}{2018}]{Schrabback2018}
{Schrabback} T.,  et~al., 2018, \mn@doi [\mnras] {10.1093/mnras/stx2666}, \href
  {https://ui.adsabs.harvard.edu/abs/2018MNRAS.474.2635S} {474, 2635}

\bibitem[\protect\citeauthoryear{{Semboloni}, {Hoekstra}, {Schaye}, {van
  Daalen}  \& {McCarthy}}{{Semboloni} et~al.}{2011}]{Semboloni2011}
{Semboloni} E.,  {Hoekstra} H.,  {Schaye} J.,  {van Daalen} M.~P.,   {McCarthy}
  I.~G.,  2011, \mn@doi [\mnras] {10.1111/j.1365-2966.2011.19385.x}, \href
  {https://ui.adsabs.harvard.edu/abs/2011MNRAS.417.2020S} {417, 2020}

\bibitem[\protect\citeauthoryear{{Shan} et~al.,}{{Shan}
  et~al.}{2018}]{Shan2018}
{Shan} H.,  et~al., 2018, \mn@doi [\mnras] {10.1093/mnras/stx2837}, \href
  {https://ui.adsabs.harvard.edu/\#abs/2018MNRAS.474.1116S} {474, 1116}

\bibitem[\protect\citeauthoryear{{Velliscig}, {van Daalen}, {Schaye},
  {McCarthy}, {Cacciato}, {Le Brun}  \& {Dalla Vecchia}}{{Velliscig}
  et~al.}{2014}]{Velliscig2014}
{Velliscig} M.,  {van Daalen} M.~P.,  {Schaye} J.,  {McCarthy} I.~G.,
  {Cacciato} M.,  {Le Brun} A.~M.~C.,   {Dalla Vecchia} C.,  2014, \mn@doi
  [MNRAS] {10.1093/MNRAS/stu1044}, \href
  {http://adsabs.harvard.edu/abs/2014MNRAS.442.2641V} {442, 2641}

\bibitem[\protect\citeauthoryear{{Villaescusa-Navarro}, {Marulli}, {Viel},
  {Branchini}, {Castorina}, {Sefusatti}  \& {Saito}}{{Villaescusa-Navarro}
  et~al.}{2014}]{Villaescusa-Navarro2014}
{Villaescusa-Navarro} F.,  {Marulli} F.,  {Viel} M.,  {Branchini} E.,
  {Castorina} E.,  {Sefusatti} E.,   {Saito} S.,  2014, \mn@doi [Journal of
  Cosmology and Astro-Particle Physics] {10.1088/1475-7516/2014/03/011}, \href
  {https://ui.adsabs.harvard.edu/\#abs/2014JCAP...03..011V} {2014, 011}

\bibitem[\protect\citeauthoryear{{Vogelsberger}, {Genel}, {Sijacki}, {Torrey},
  {Springel}  \& {Hernquist}}{{Vogelsberger} et~al.}{2013}]{Vogelsberger2013}
{Vogelsberger} M.,  {Genel} S.,  {Sijacki} D.,  {Torrey} P.,  {Springel} V.,
  {Hernquist} L.,  2013, \mn@doi [MNRAS] {10.1093/mnras/stt1789}, \href
  {https://ui.adsabs.harvard.edu/abs/2013MNRAS.436.3031V} {436, 3031}

\bibitem[\protect\citeauthoryear{{Voit}}{{Voit}}{2005}]{Voit2005}
{Voit} G.~M.,  2005, \mn@doi [Reviews of Modern Physics]
  {10.1103/RevModPhys.77.207}, \href
  {http://adsabs.harvard.edu/abs/2005RvMP...77..207V} {77, 207}

\bibitem[\protect\citeauthoryear{{Wagner}, {Verde}  \& {Jimenez}}{{Wagner}
  et~al.}{2012}]{Wagner2012}
{Wagner} C.,  {Verde} L.,   {Jimenez} R.,  2012, \mn@doi [\apj]
  {10.1088/2041-8205/752/2/L31}, \href
  {https://ui.adsabs.harvard.edu/\#abs/2012ApJ...752L..31W} {752, L31}

\bibitem[\protect\citeauthoryear{{White} \& {Vale}}{{White} \&
  {Vale}}{2004}]{White2004}
{White} M.,  {Vale} C.,  2004, \mn@doi [Astroparticle Physics]
  {10.1016/j.astropartphys.2004.05.002}, \href
  {https://ui.adsabs.harvard.edu/\#abs/2004APh....22...19W} {22, 19}

\bibitem[\protect\citeauthoryear{Wyman, Rudd, Vanderveld  \& Hu}{Wyman
  et~al.}{2014}]{Wyman2014}
Wyman M.,  Rudd D.~H.,  Vanderveld R.~A.,   Hu W.,  2014, \mn@doi [Phys. Rev.
  Lett.] {10.1103/PhysRevLett.112.051302}, 112, 051302

\bibitem[\protect\citeauthoryear{{Yang}, {Kratochvil}, {Wang}, {Lim}, {Haiman}
  \& {May}}{{Yang} et~al.}{2011}]{Yang2011}
{Yang} X.,  {Kratochvil} J.~M.,  {Wang} S.,  {Lim} E.~A.,  {Haiman} Z.,   {May}
  M.,  2011, \mn@doi [\prd] {10.1103/PhysRevD.84.043529}, \href
  {https://ui.adsabs.harvard.edu/\#abs/2011PhRvD..84d3529Y} {84, 043529}

\bibitem[\protect\citeauthoryear{{Yang}, {Kratochvil}, {Huffenberger}, {Haiman}
   \& {May}}{{Yang} et~al.}{2013}]{Yang2013}
{Yang} X.,  {Kratochvil} J.~M.,  {Huffenberger} K.,  {Haiman} Z.,   {May} M.,
  2013, \mn@doi [Phys.~Rev.~D] {10.1103/PhysRevD.87.023511}, \href
  {https://ui.adsabs.harvard.edu/#abs/2013PhRvD..87b3511Y} {87, 023511}

\bibitem[\protect\citeauthoryear{{de Jong} et~al.,}{{de Jong}
  et~al.}{2017}]{deJong2017}
{de Jong} J. T.~A.,  et~al., 2017, \mn@doi [\aap]
  {10.1051/0004-6361/201730747}, \href
  {https://ui.adsabs.harvard.edu/abs/2017A&A...604A.134D} {604, A134}

\bibitem[\protect\citeauthoryear{{van Daalen}, {Schaye}, {Booth}  \& {Dalla
  Vecchia}}{{van Daalen} et~al.}{2011}]{vanDaalen2011}
{van Daalen} M.~P.,  {Schaye} J.,  {Booth} C.~M.,   {Dalla Vecchia} C.,  2011,
  \mn@doi [MNRAS] {10.1111/j.1365-2966.2011.18981.x}, \href
  {http://adsabs.harvard.edu/abs/2011MNRAS.415.3649V} {415, 3649}

\makeatother
\end{thebibliography}

\appendix
\section{Additional Peak Distributions}
\label{appendix}

Here, we present the other comparisons of weak lensing peak distributions, referred to in Section \ref{sub:miyoungResults}, including varying filter sizes. In the main text, our discussions focus on the plots with filter size $\theta_{\rm{ap}} = 12.5$ arcmin. The error bars are the standard deviation of the 5 different shape noise realisations.

\begin{figure}
     \centering
     \begin{subfigure}[b]{\columnwidth}
         \centering
         \includegraphics[width=\columnwidth]{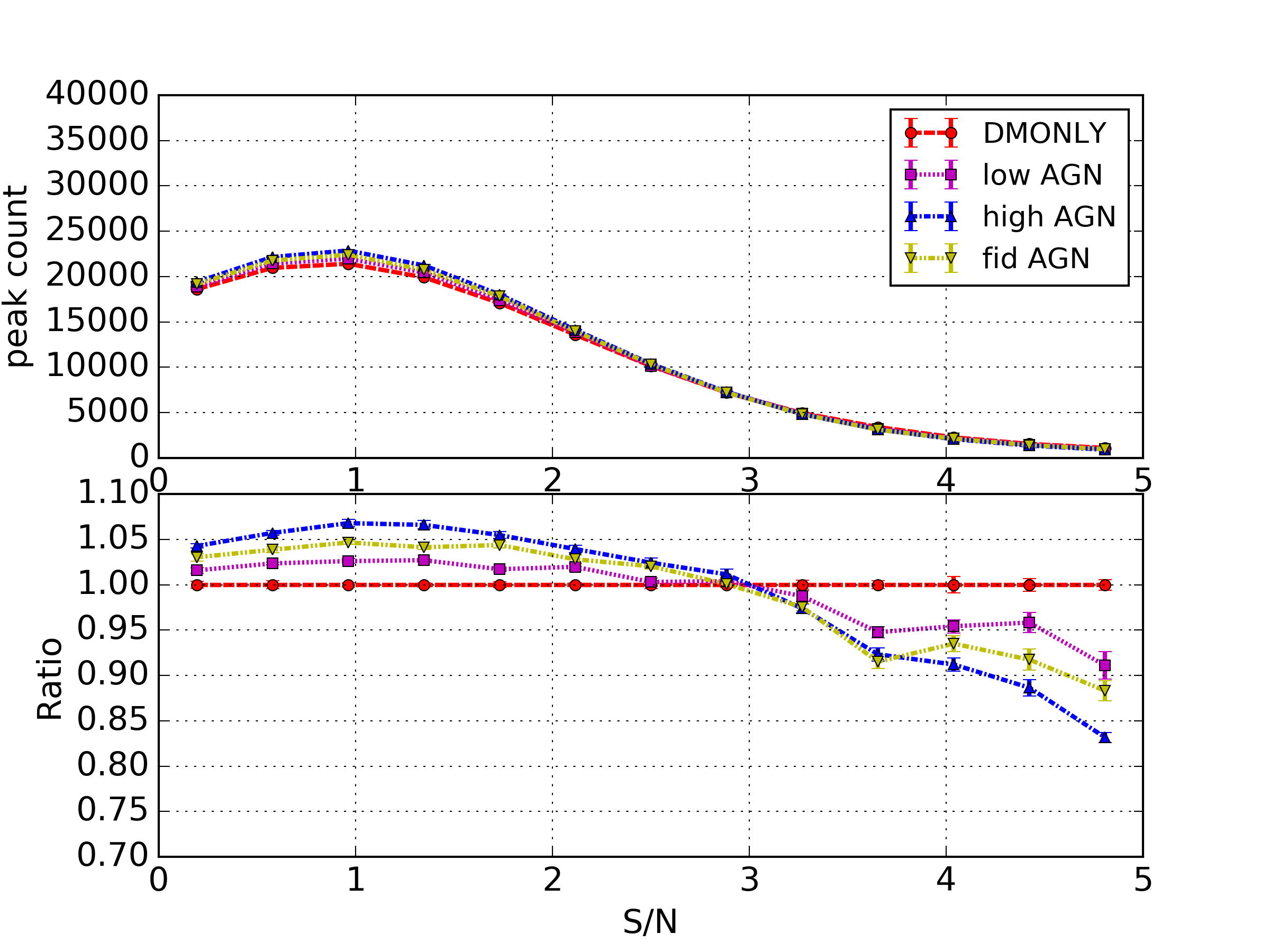}
         \caption{$\theta_{\rm{ap}} = 10.0$ arcmin}
         \label{fig:KiDS_agn_filt10_0}
     \end{subfigure}
     \hfill
     \begin{subfigure}[b]{\columnwidth}
         \centering
         \includegraphics[width=\columnwidth]{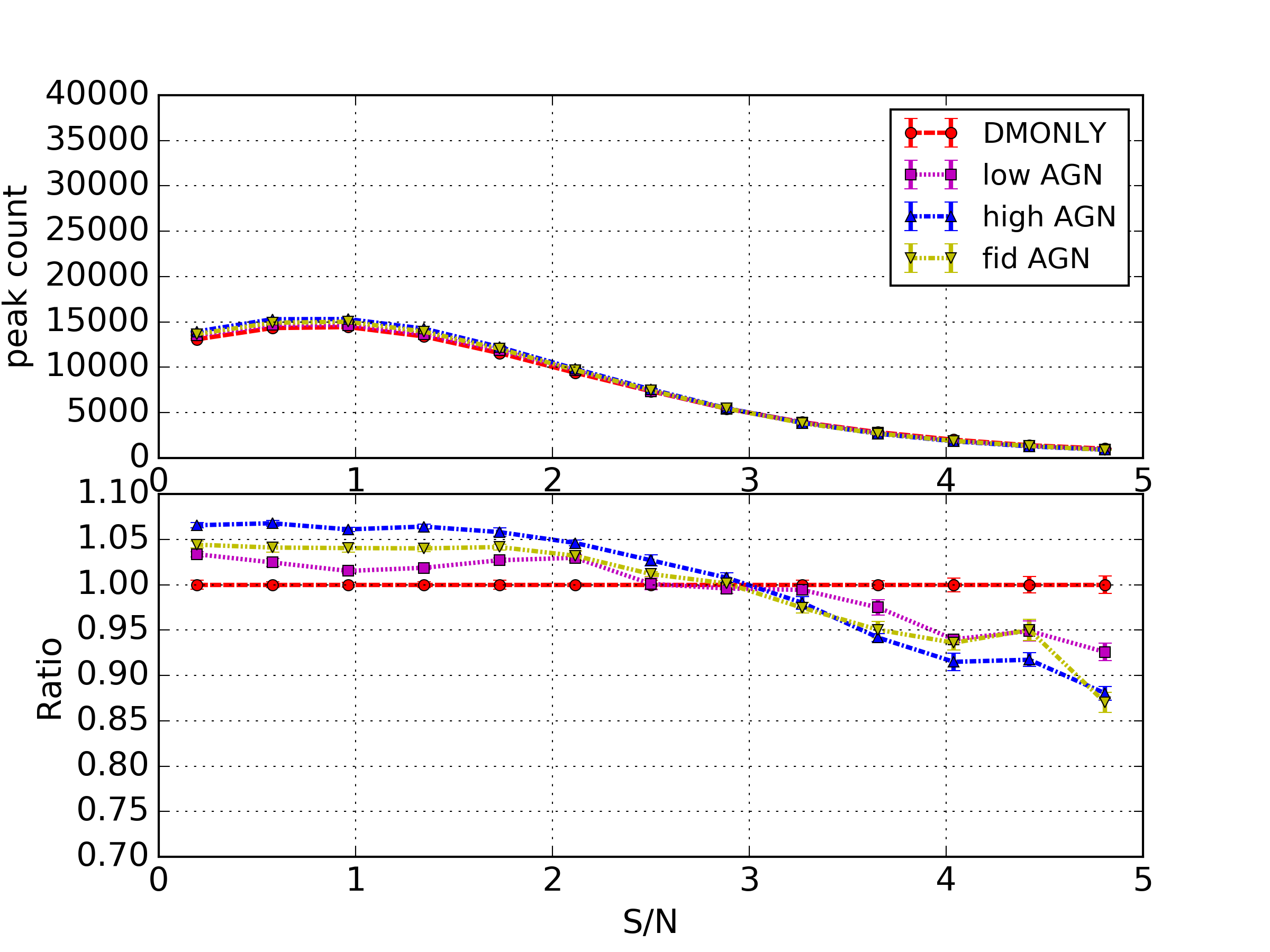}
         \caption{$\theta_{\rm{ap}} = 12.5$ arcmin}
         \label{fig:KiDS_agn_filt12_5}
     \end{subfigure}
     \hfill
      \begin{subfigure}[b]{\columnwidth}
         \centering
         \includegraphics[width=\columnwidth]{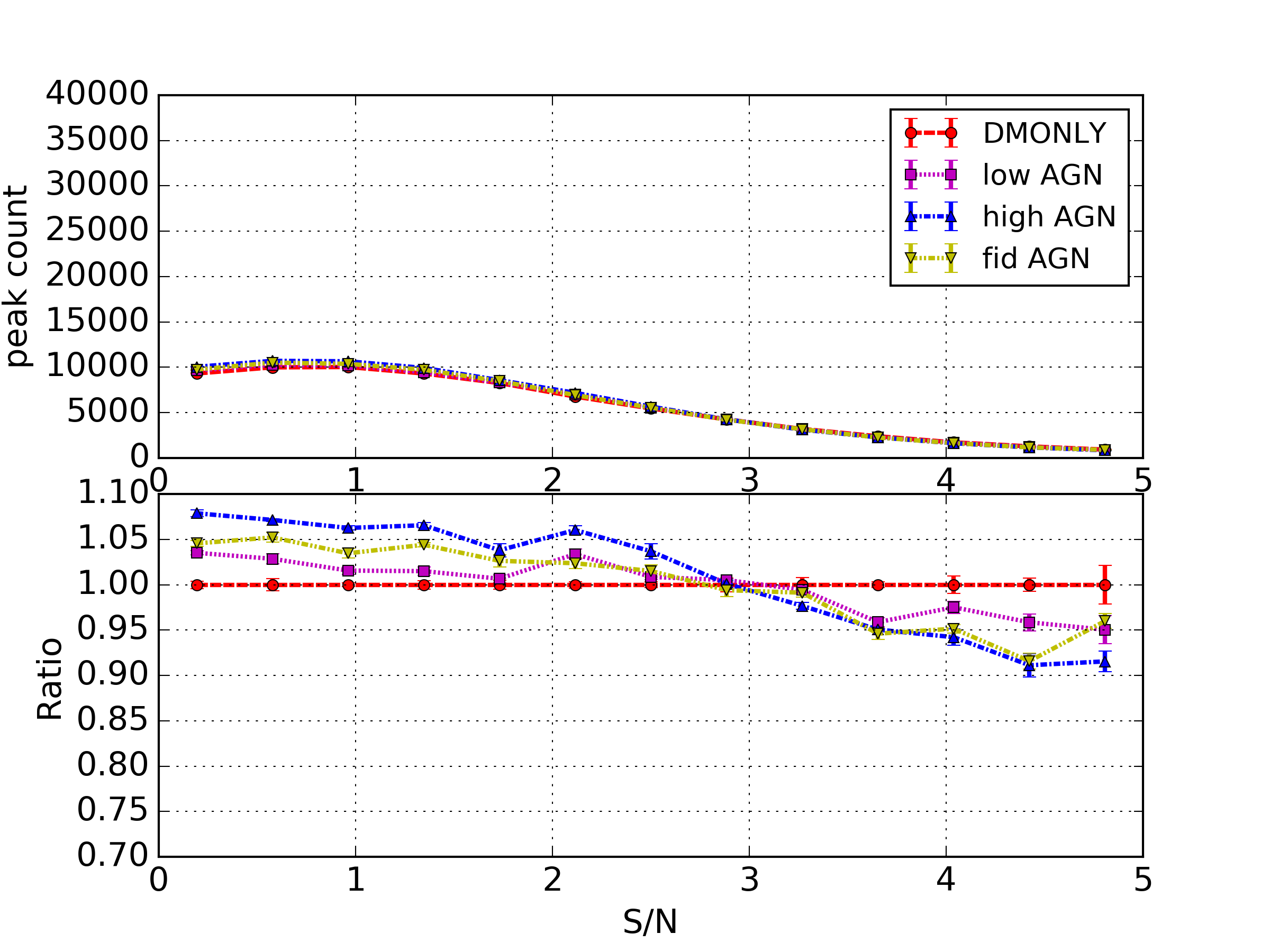}
         \caption{$\theta_{\rm{ap}} = 15.0$ arcmin}
         \label{fig:KiDS_agn_filt15_0}
     \end{subfigure}
        \caption{The impact of baryonic processes on the weak lensing peak statistics with different aperture filter sizes shown in the three subpanels. The source density is 9 gal/arcmin$^2$. The ratio is taken with respect to the DMONLY simulation in \textit{WMAP}~9.}
        \label{fig:result_peak_AGNs}
\end{figure}

\begin{figure}
     \centering
     \begin{subfigure}[b]{\columnwidth}
         \centering
         \includegraphics[width=\columnwidth]{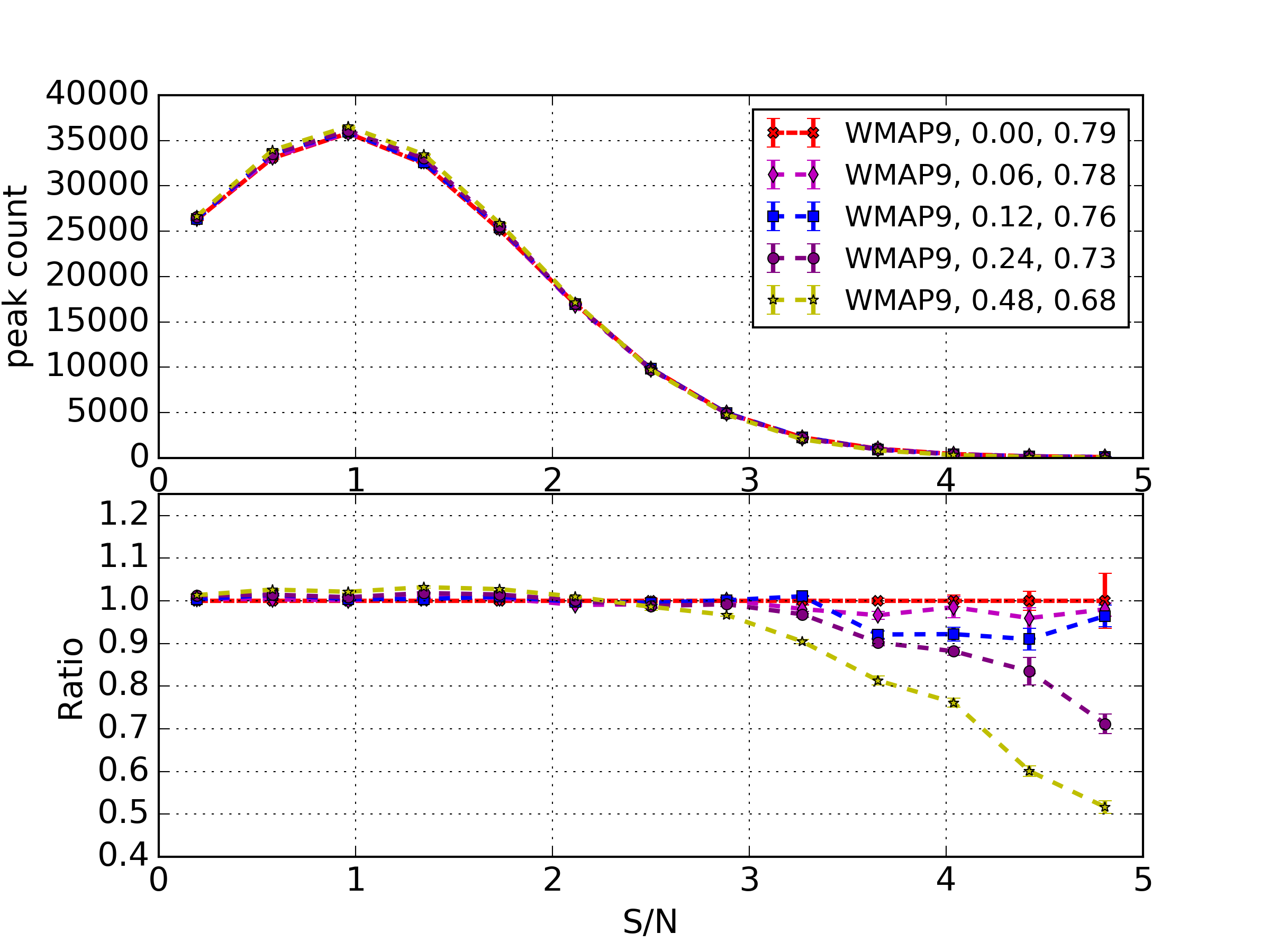}
         \caption{$\theta_{\rm{ap}} = 10.0$ arcmin}
         \label{fig:KiDS_mnu_filt10_0}
     \end{subfigure}
     \hfill
     \begin{subfigure}[b]{\columnwidth}
         \centering
         \includegraphics[width=\columnwidth]{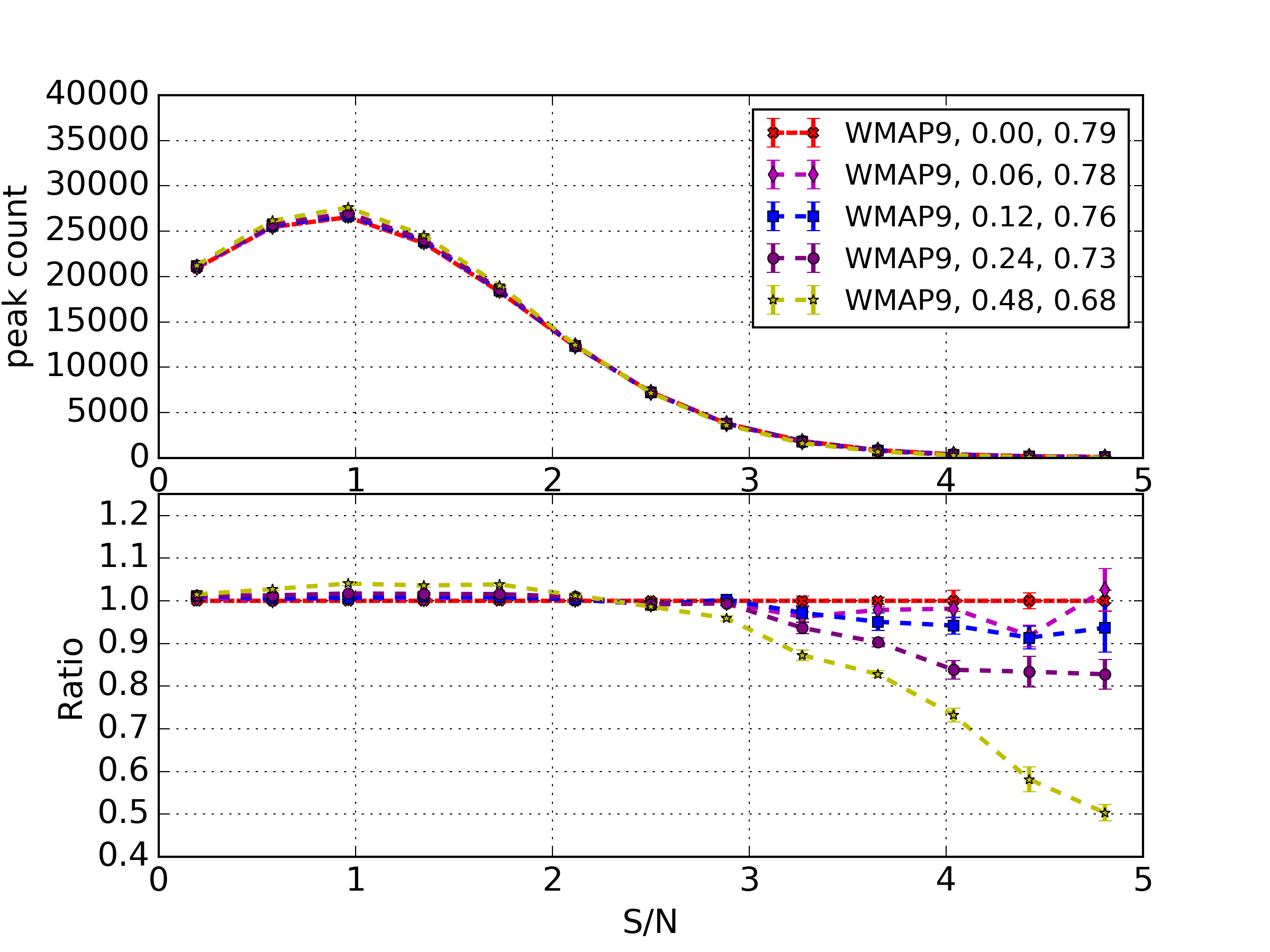}
         \caption{$\theta_{\rm{ap}} = 12.5$ arcmin}
         \label{fig:KiDS_mnu_filt12_5}
     \end{subfigure}
     \hfill
     \begin{subfigure}[b]{\columnwidth}
         \centering
         \includegraphics[width=\columnwidth]{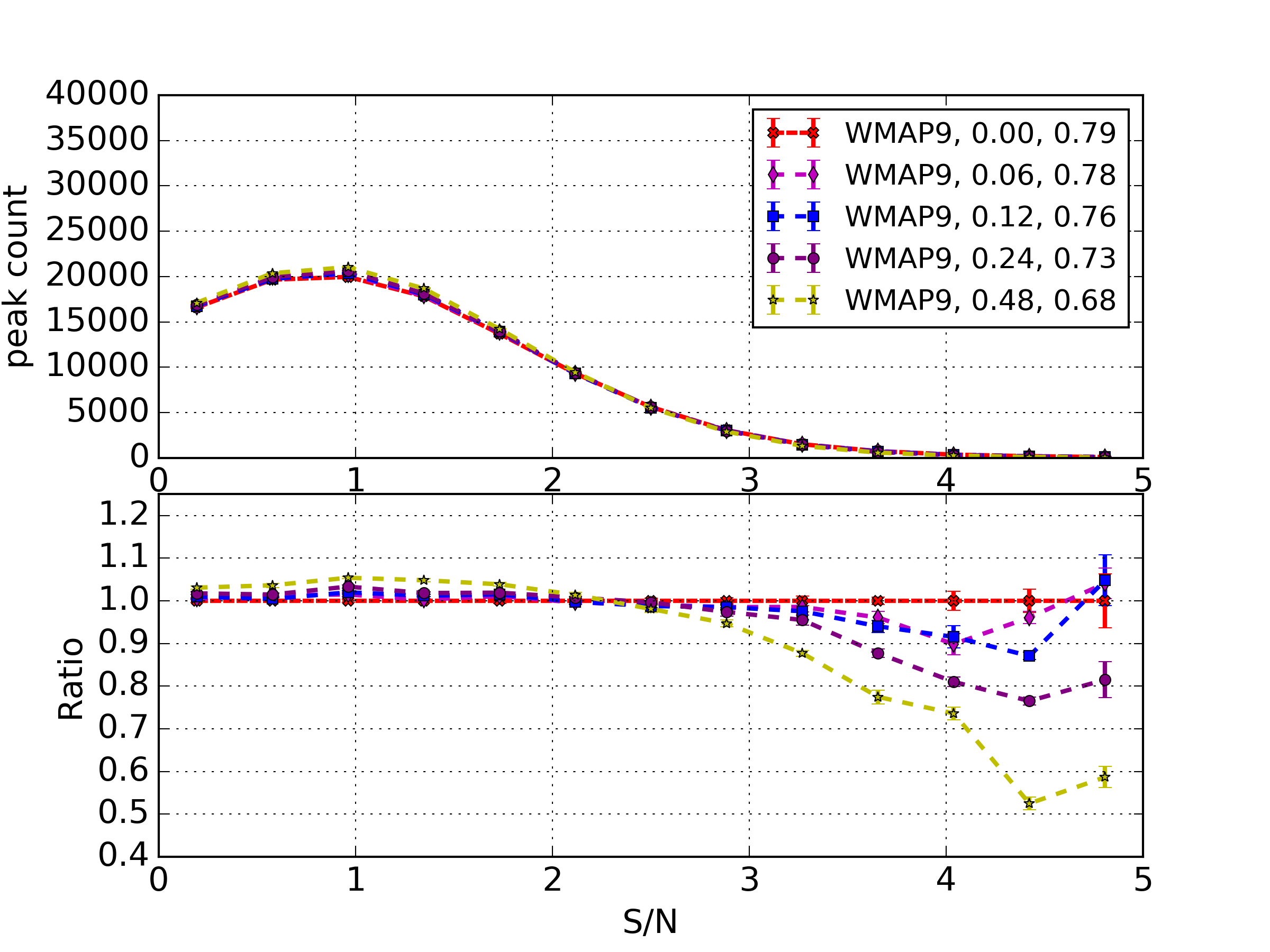}
         \caption{$\theta_{\rm{ap}} = 15.0$ arcmin}
         \label{fig:KiDS_mnu_filt15_0}
     \end{subfigure}
     \caption{The impact of varying neutrino mass on the weak lensing peak statistics for different filter sizes shown in the three subpanels. The source density is 9 gal/arcmin$^2$. The ratios are taken with respect to the fiducial baryonic physics simulation in \textit{WMAP}~9.}
     \label{fig:result_peak_Mnus}
\end{figure}

\begin{figure}
     \centering
     \begin{subfigure}[b]{\columnwidth}
         \centering
         \includegraphics[width=\columnwidth]{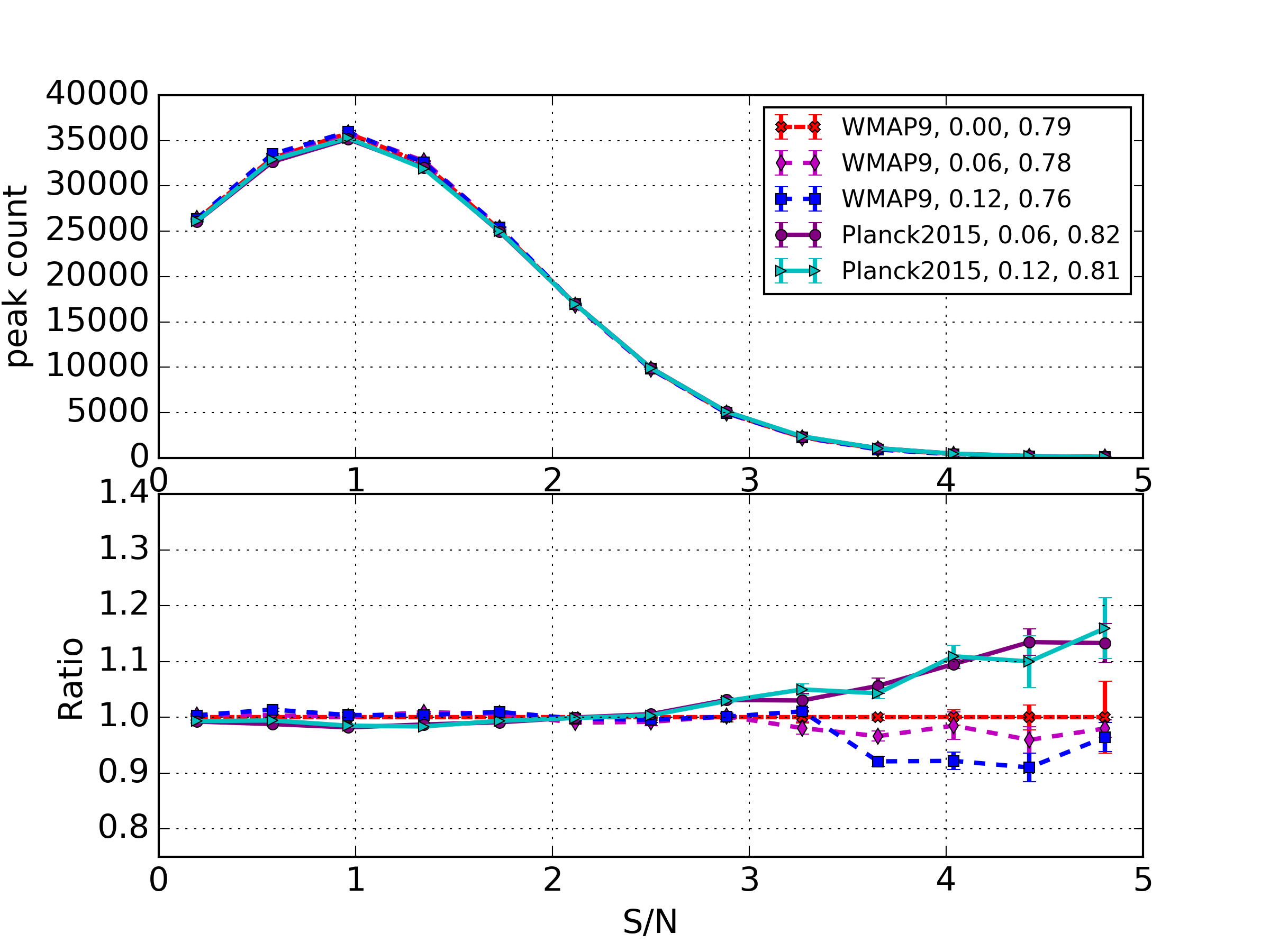}
         \caption{$\theta_{\rm{ap}} = 10.0$ arcmin}
         \label{fig:KiDS_agn_filt10_0}
     \end{subfigure}
     \hfill
     \begin{subfigure}[b]{\columnwidth}
         \centering
         \includegraphics[width=\columnwidth]{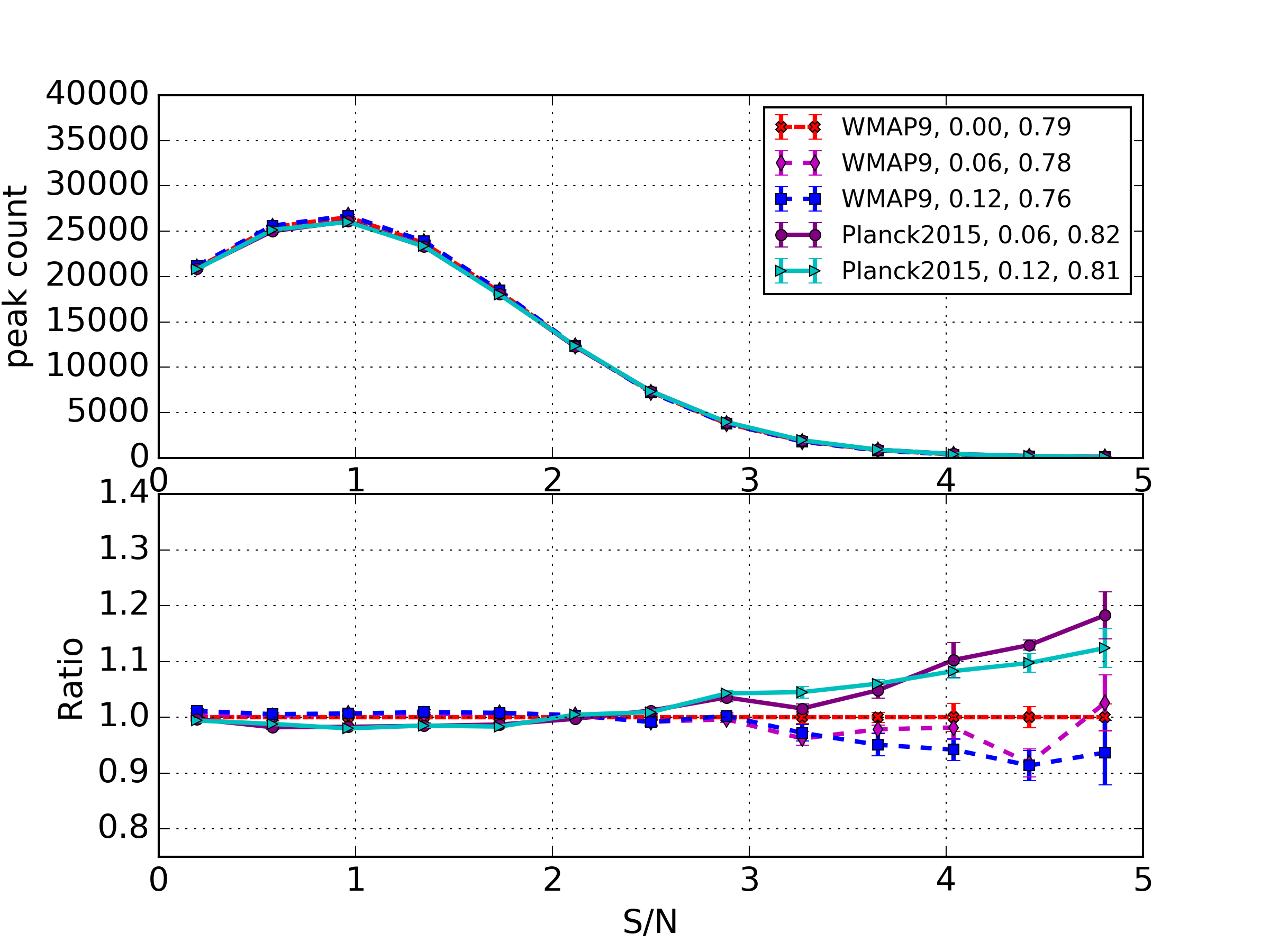}
         \caption{$\theta_{\rm{ap}} = 12.5$ arcmin}
         \label{fig:KiDS_agn_filt12_5}
     \end{subfigure}
     \hfill
      \begin{subfigure}[b]{\columnwidth}
         \centering
         \includegraphics[width=\columnwidth]{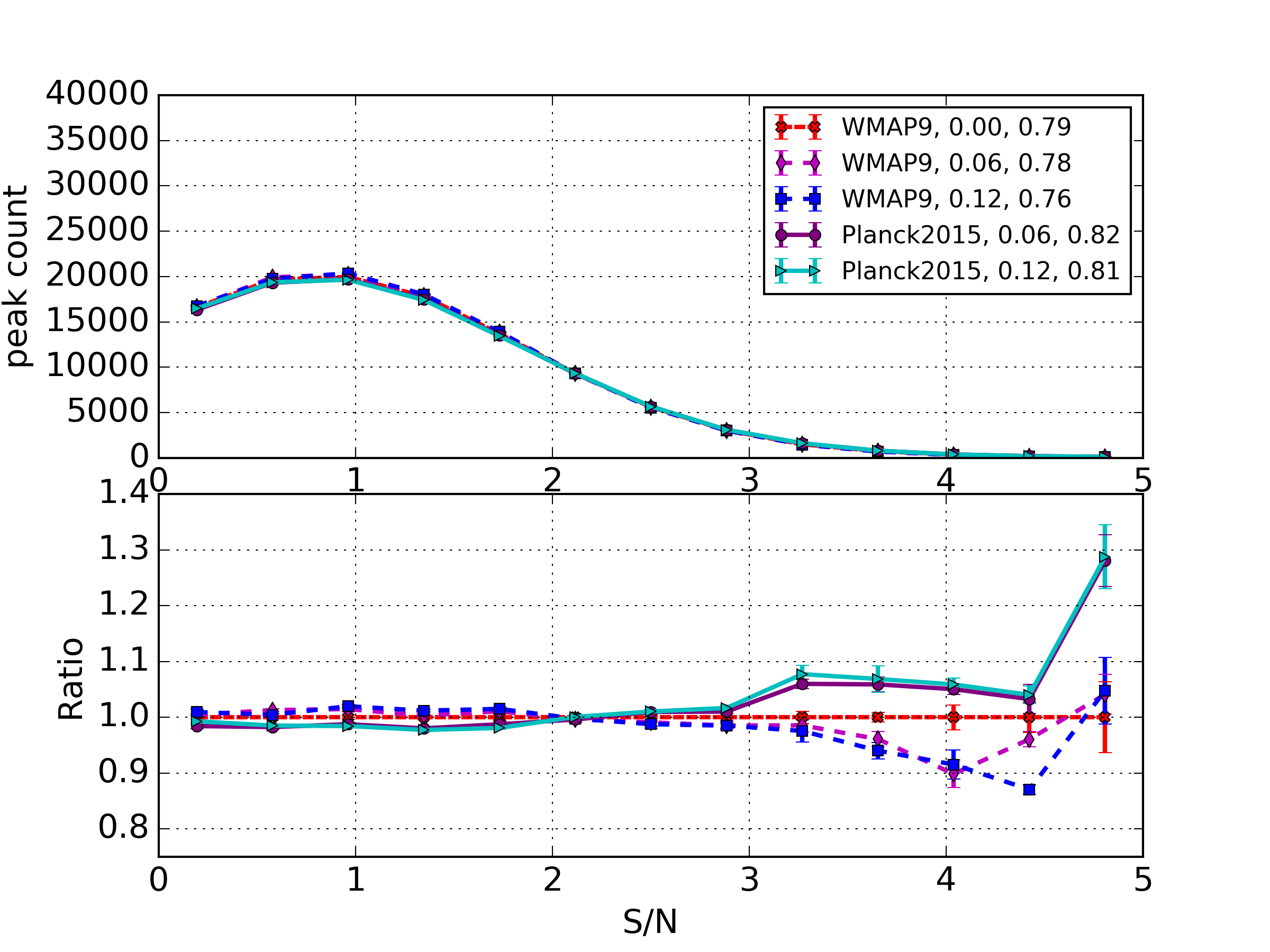}
         \caption{$\theta_{\rm{ap}} = 15.0$ arcmin}
         \label{fig:KiDS_agn_filt15_0}
     \end{subfigure}
        \caption{The impact of \textit{WMAP}~9 and \textit{Planck}~2015 cosmologies with summed neutrino mass of 0.06 and 0.12 eV on the weak lensing peak statistics for different filter sizes shown in the three subpanels. The source density is 9 gal/arcmin$^2$. The ratios are taken with respect to the fiducial baryonic physics simulation in \textit{WMAP}~9.}
        \label{fig:result_peak_CosmologiesAndNeutrinos}
\end{figure}

\begin{figure}
     \centering
     \begin{subfigure}[b]{\columnwidth}
         \centering
         \includegraphics[width=\columnwidth]{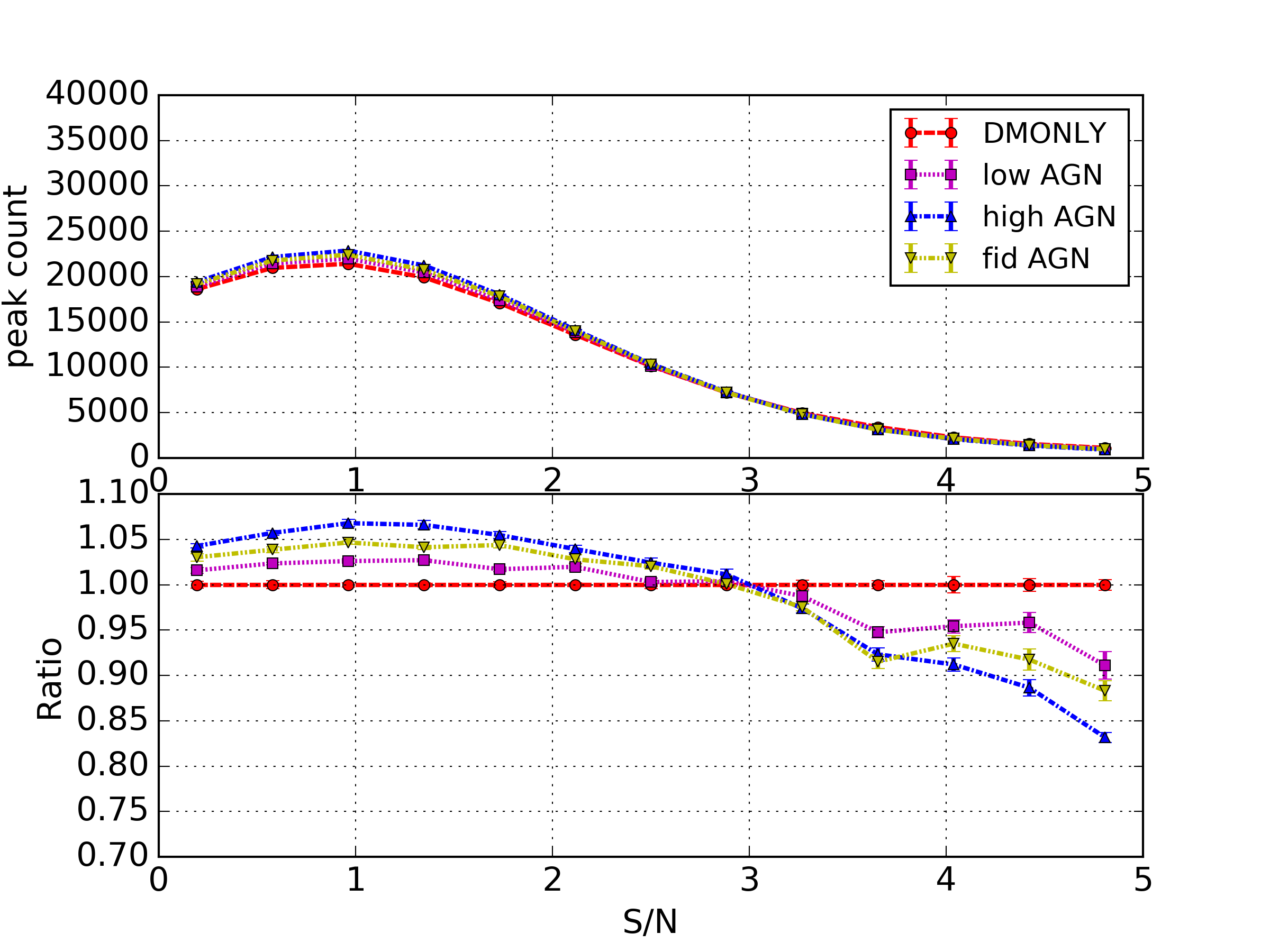}
         \caption{$\theta_{\rm{ap}} = 10.0$ arcmin}
         \label{fig:LSST_agn_filt10_0}
     \end{subfigure}
     \hfill
     \begin{subfigure}[b]{\columnwidth}
         \centering
         \includegraphics[width=\columnwidth]{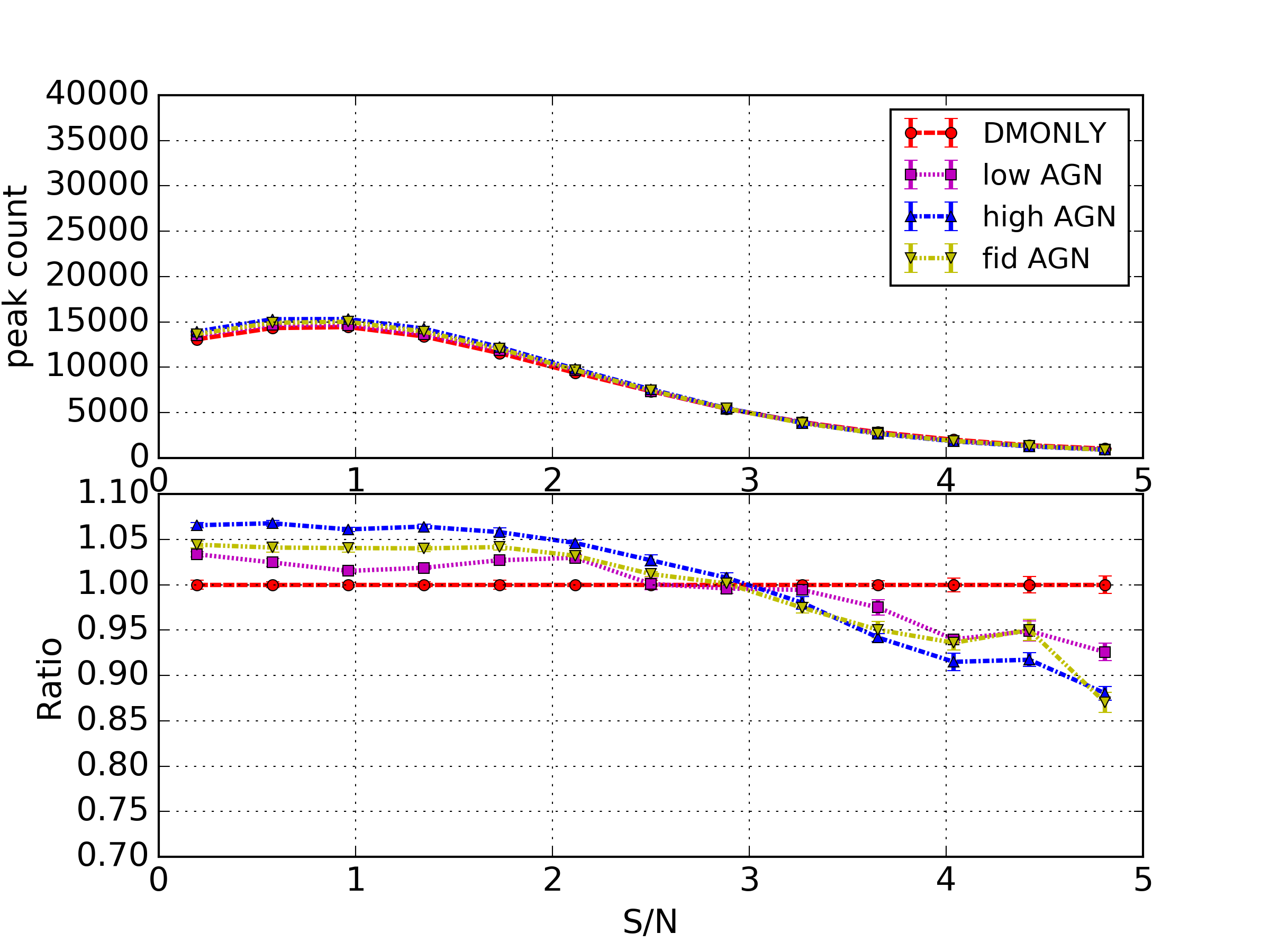}
         \caption{$\theta_{\rm{ap}} = 12.5$ arcmin}
         \label{fig:LSST_agn_filt12_5}
     \end{subfigure}
     \hfill
      \begin{subfigure}[b]{\columnwidth}
         \centering
         \includegraphics[width=\columnwidth]{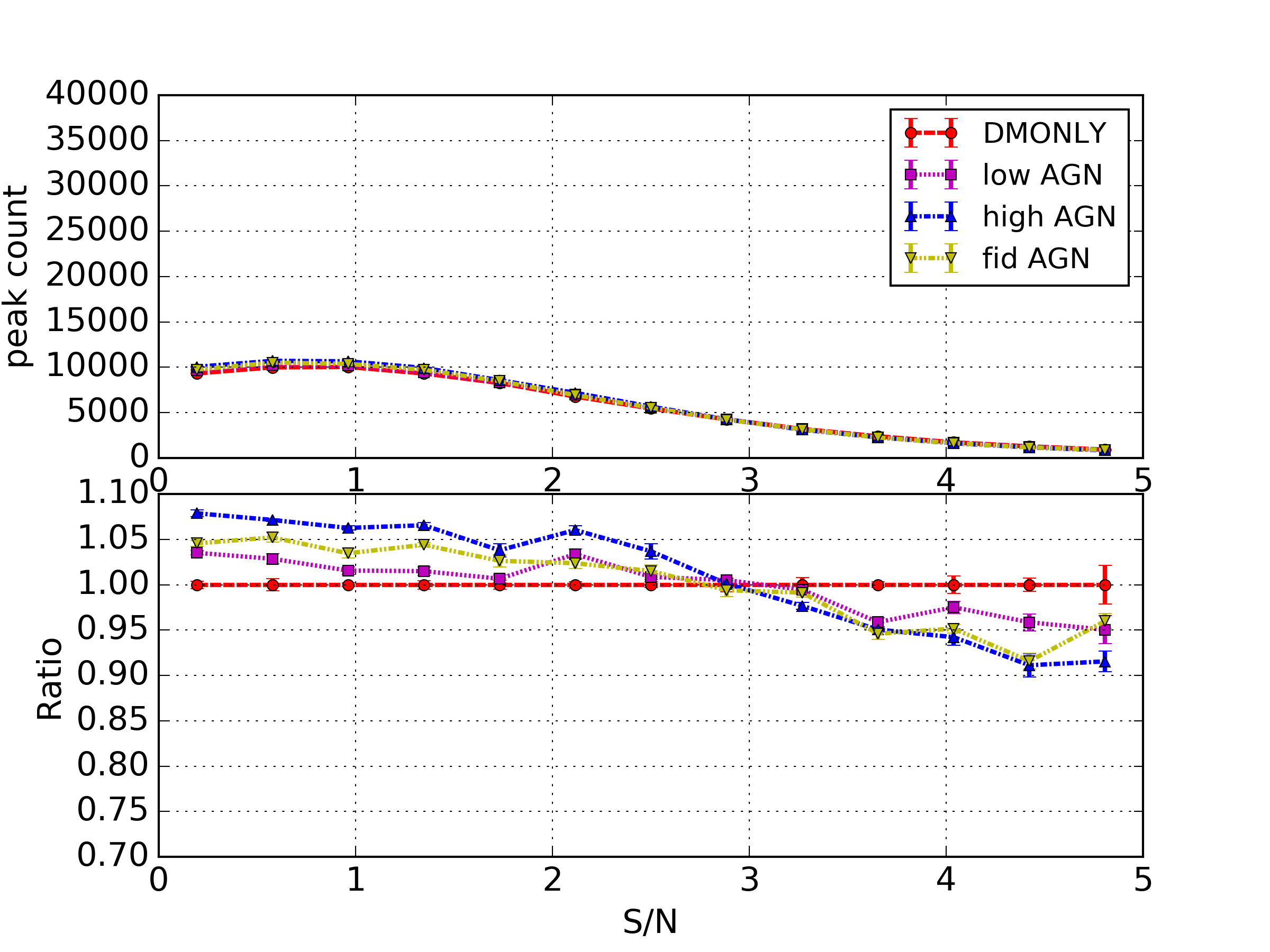}
         \caption{$\theta_{\rm{ap}} = 15.0$ arcmin}
         \label{fig:LSST_agn_filt15_0}
     \end{subfigure}
        \caption{The impact of baryonic processes on the weak lensing peak statistics for different filter sizes shown in the three subpanels. The source density is 30 gal/arcmin$^2$. The ratios are taken with respect to the DMONLY simulation in \textit{WMAP}~9.}
        \label{fig:result_peak_AGNs}
\end{figure}

\begin{figure}
     \centering
     \begin{subfigure}[b]{\columnwidth}
         \centering
         \includegraphics[width=\columnwidth]{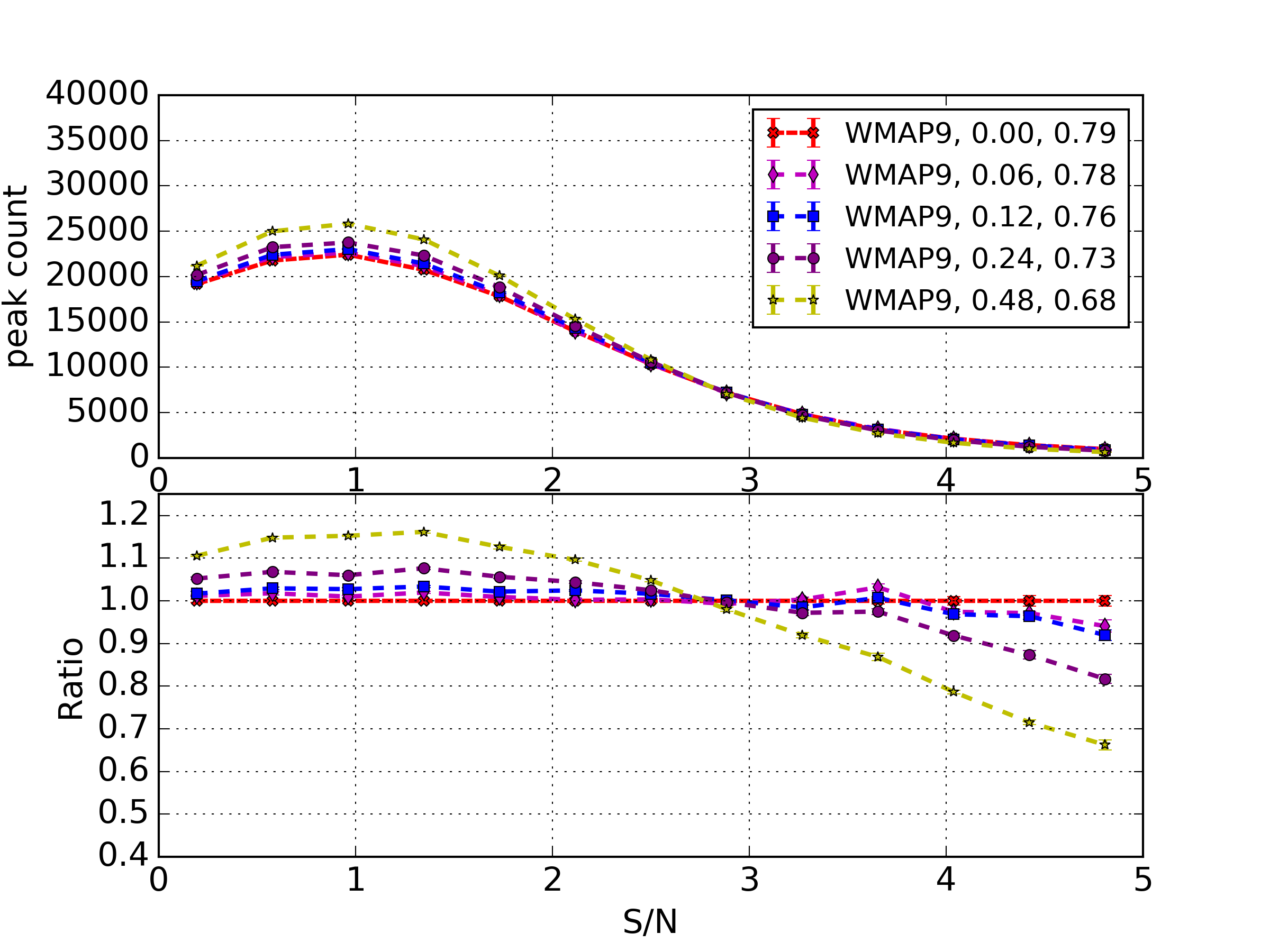}
         \caption{$\theta_{\rm{ap}} = 10.0$ arcmin}
         \label{fig:LSST_mnu_filt10_0}
     \end{subfigure}
     \hfill
     \begin{subfigure}[b]{\columnwidth}
         \centering
         \includegraphics[width=\columnwidth]{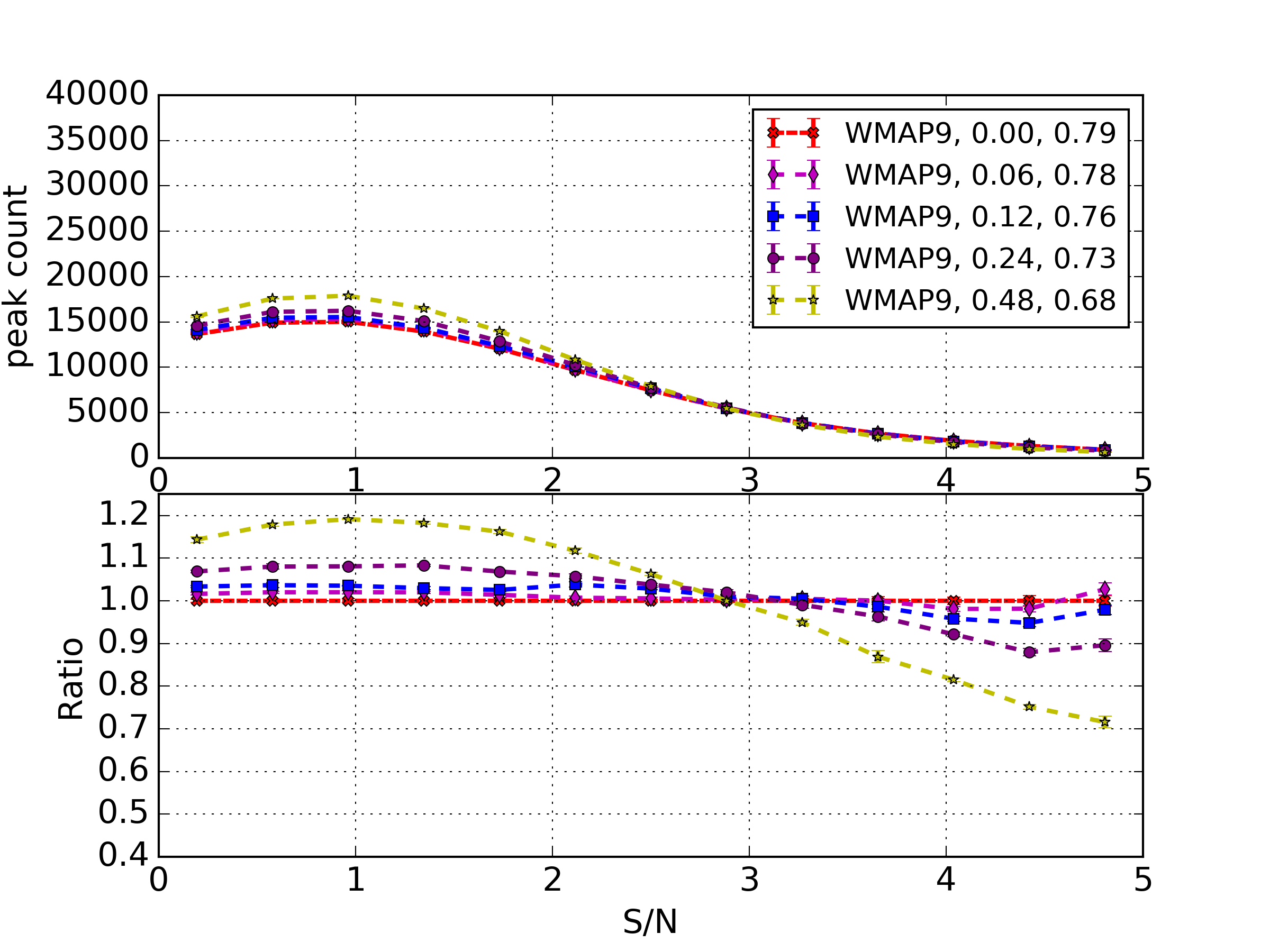}
         \caption{$\theta_{\rm{ap}} = 12.5$ arcmin}
         \label{fig:LSST_mnu_filt12_5}
     \end{subfigure}
     \hfill
     \begin{subfigure}[b]{\columnwidth}
         \centering
         \includegraphics[width=\columnwidth]{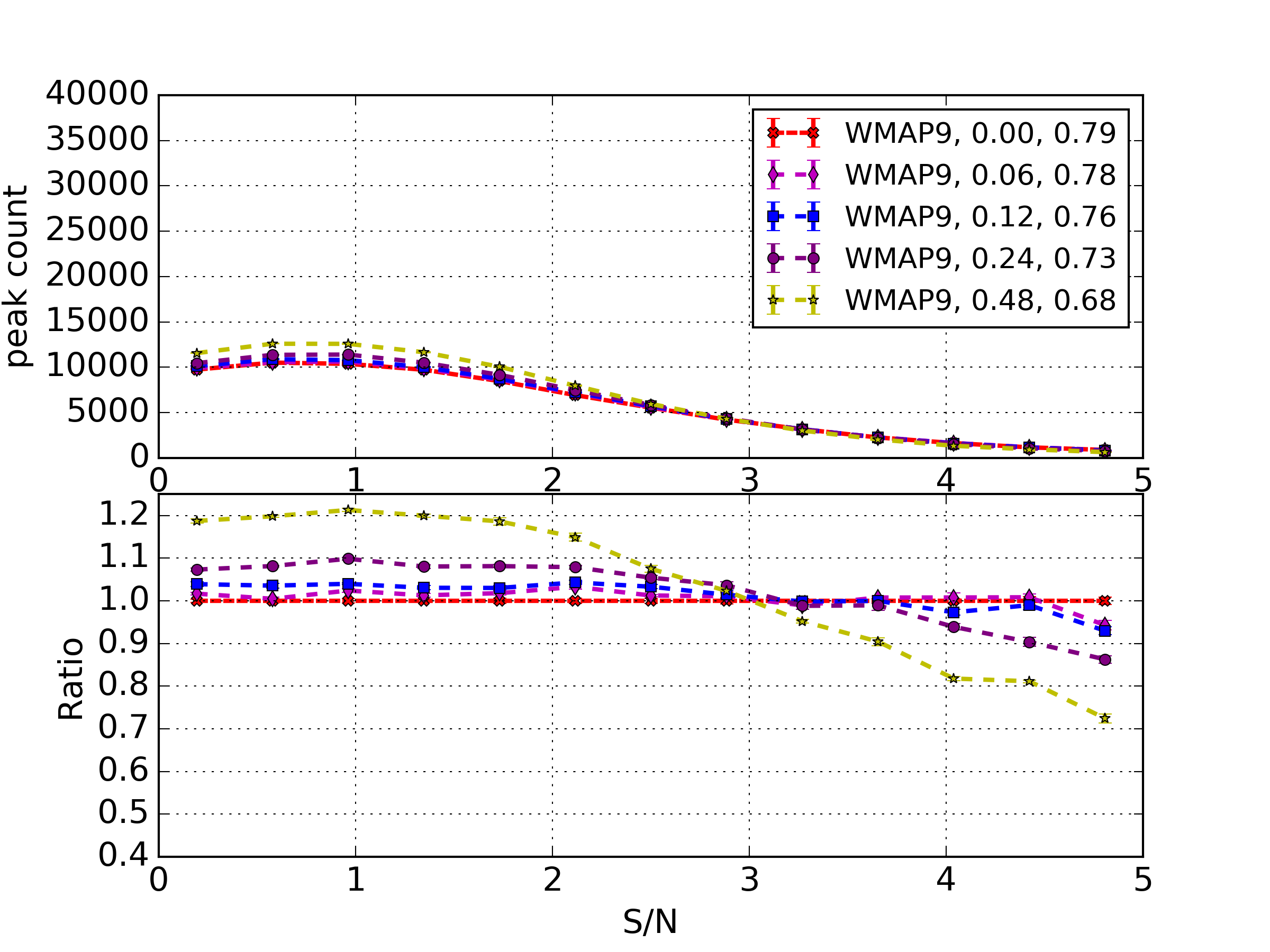}
         \caption{$\theta_{\rm{ap}} = 15.0$ arcmin}
         \label{fig:LSST_mnu_filt15_0}
     \end{subfigure}
     \caption{The impact of varying neutrino mass on the weak lensing peak statistics for different filter sizes shown in the three subpanels. The source density is 30 gal/arcmin$^2$. The ratios are taken with respect to the fiducial baryonic physics simulation in \textit{WMAP}~9.}
     \label{fig:result_peak_Mnus}
\end{figure}

\begin{figure}
     \centering
     \begin{subfigure}[b]{\columnwidth}
         \centering
         \includegraphics[width=\columnwidth]{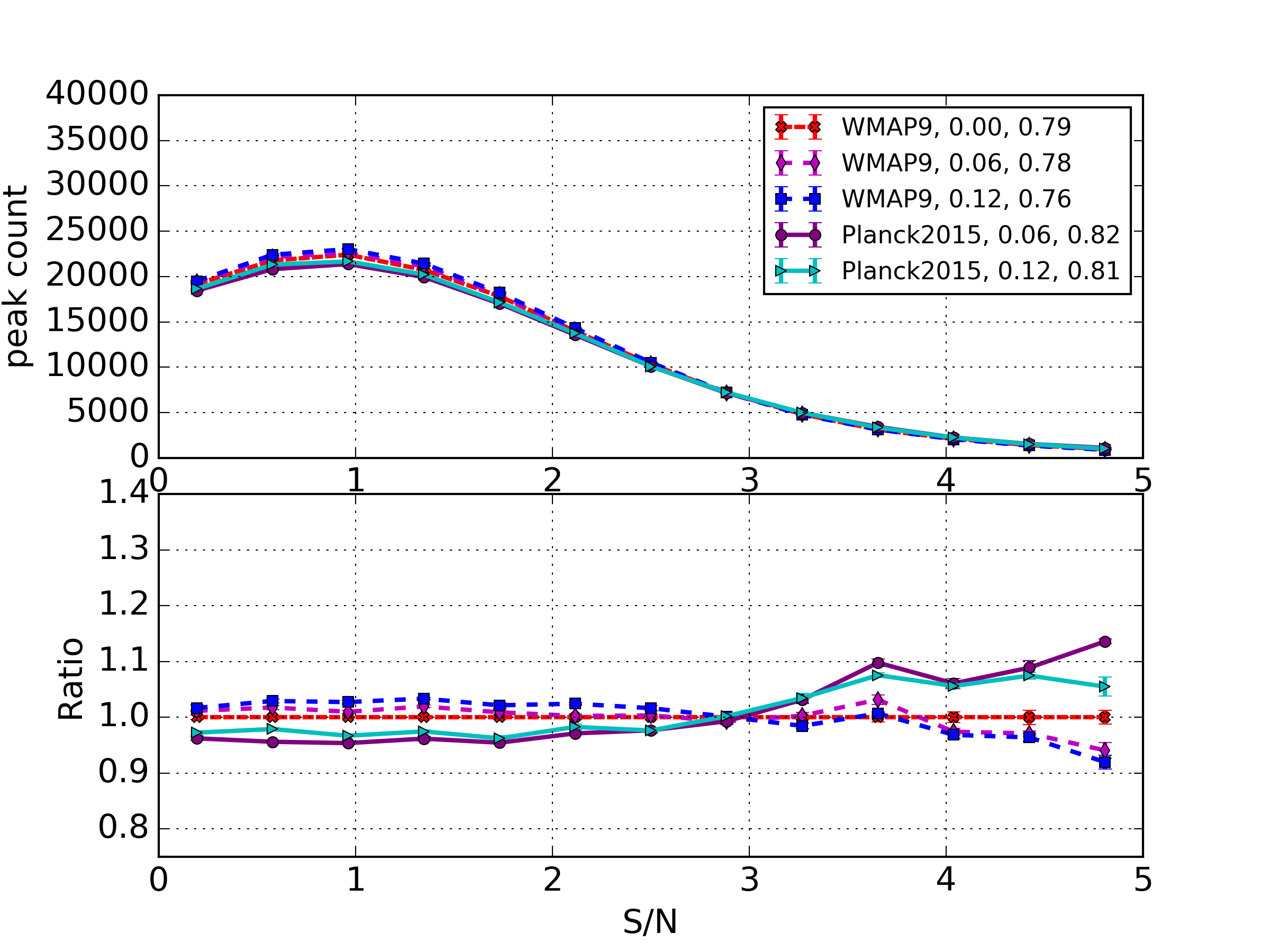}
         \caption{$\theta_{\rm{ap}} = 10.0$ arcmin}
         \label{fig:mnu_filt10_0}
     \end{subfigure}
     \hfill
     \begin{subfigure}[b]{\columnwidth}
         \centering
         \includegraphics[width=\columnwidth]{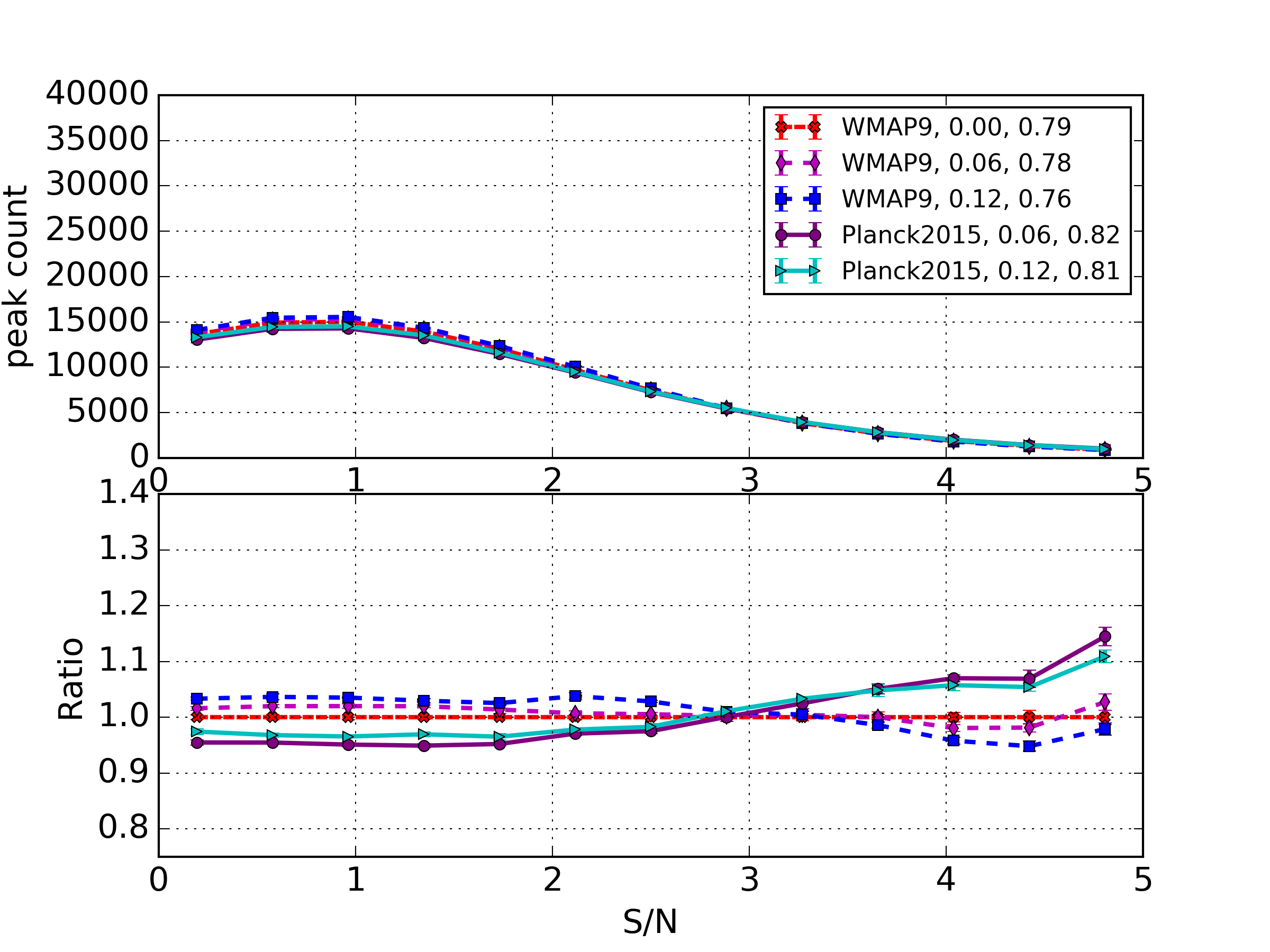}
         \caption{$\theta_{\rm{ap}} = 12.5$ arcmin}
         \label{fig:mnu_filt12_5}
     \end{subfigure}
     \hfill
     \begin{subfigure}[b]{\columnwidth}
         \centering
         \includegraphics[width=\columnwidth]{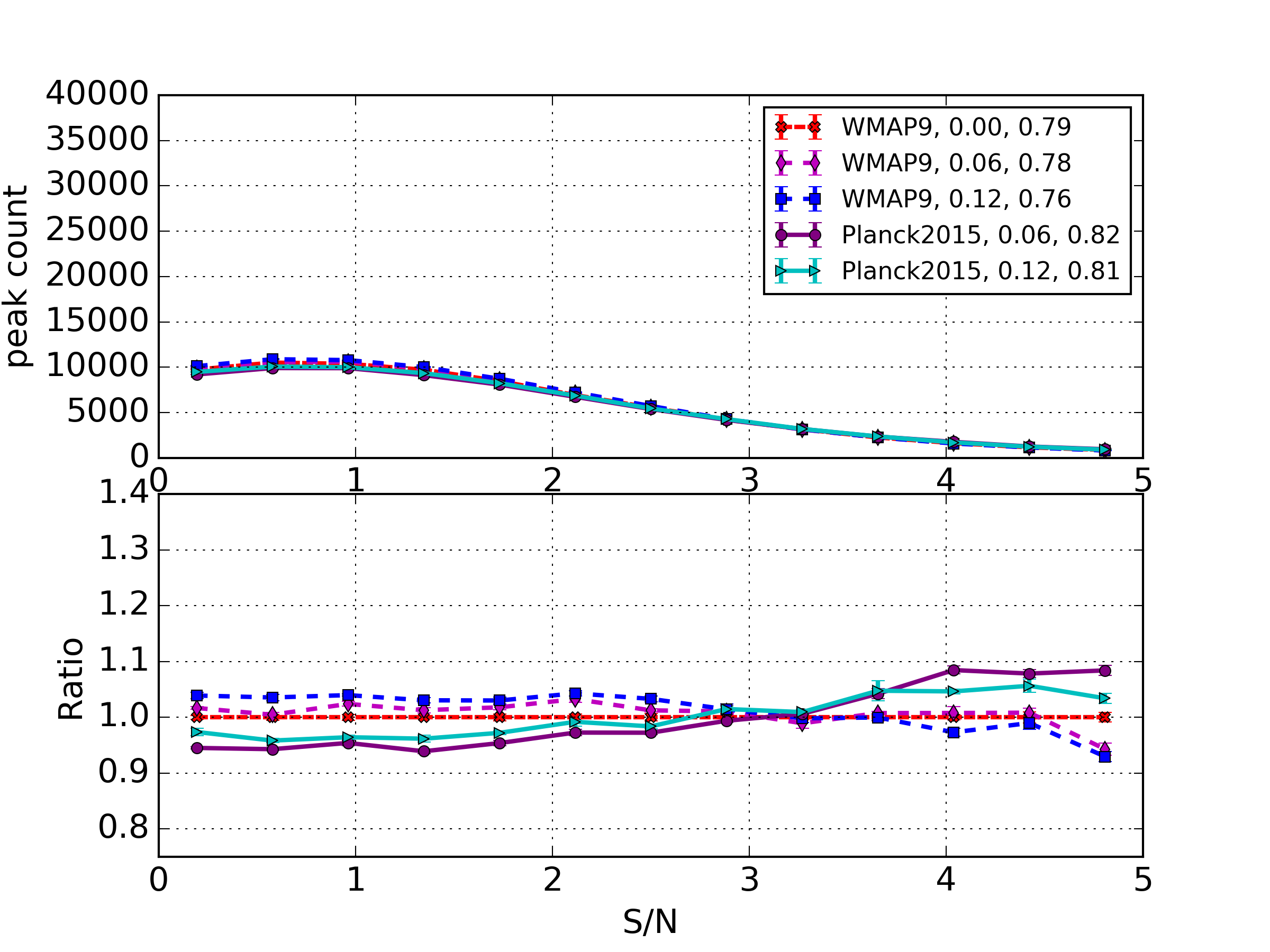}
         \caption{$\theta_{\rm{ap}} = 15.0$ arcmin}
         \label{fig:mnu_filt15_0}
     \end{subfigure}
     \caption{The impact \textit{WMAP}~9 and \textit{Planck}~2015 cosmologies with summed neutrino mass of 0.06 and 0.12 eV on the weak lensing peak statistics for different filter sizes shown in the three subpanels. The source density is 30 gal/arcmin$^2$. The ratios are taken with respect to the fiducial baryonic physics simulation in \textit{WMAP}~9.}
     \label{fig:result_peak_Cosmologies}
\end{figure}

\begin{figure}
    \includegraphics[width=\columnwidth]{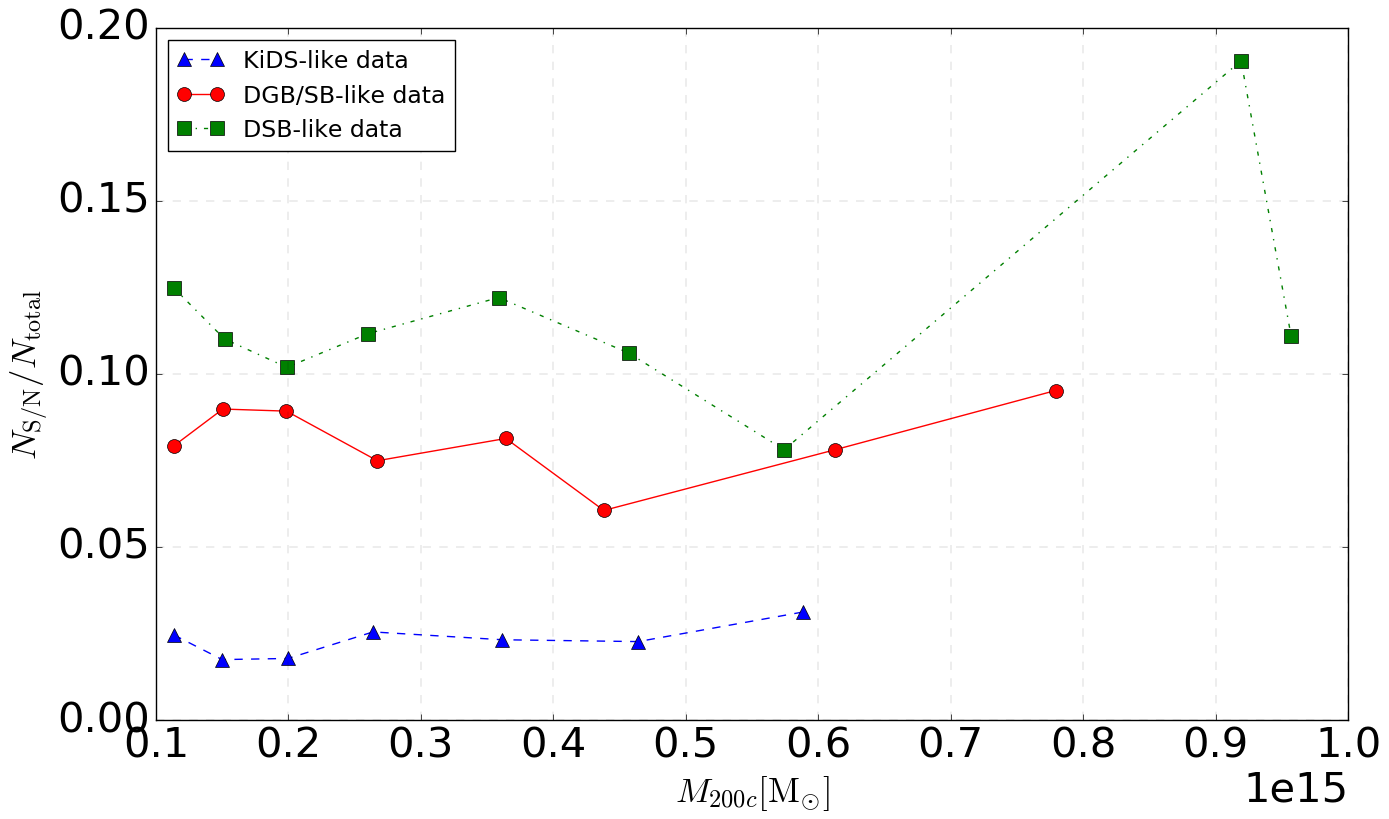}
    \caption{We show the fraction of haloes that have a randomised S/N peak position within $2.0$ arcmin, as a function of cluster mass. Source number densities $n_{\rm{eff}} = 9$, $30$, and $60$ gal/arcmin$^{2}$ are represented by the triangle, circle, and square markers respectively.
    }
    \label{fig:fullRandomM200s}
\end{figure}
For Figure \ref{fig:fullRandomM200s} we follow the same randomisation process mentioned in Section \ref{sub:surveySummary} for Figure \ref{fig:fullM200s}. This plot shows the relationship between galaxy cluster locations and a random noise field, with the same number of random positions as S/N peaks from the original synthetic survey data. The fraction $N_{\rm S/N}/N_{\rm total}$ increases with source number density because the total number of high S/N peaks are increased (see Figure \ref{fig:60gal_compareAllCountSNRs}). The Synthetic DGB/SB and DSB fractions extend to higher mass bins due to the fact that there are more high S/N peaks in these surveys, and that high galaxy cluster masses are more rare. The relatively flat relationships shown here illustrates that the results in Figure \ref{fig:fullM200s} are not due to random noise.







\bsp	
\label{lastpage}
\end{document}